\documentclass[11pt,letterpaper]{article}

\usepackage{amsmath}
\usepackage{amsfonts}
\usepackage{amssymb}
\usepackage[pdftex]{graphicx}
\usepackage{color}
\usepackage{rotating}
\usepackage{bm}
\usepackage{subcaption}
\usepackage[bottom]{footmisc}
\usepackage{amsthm}

\allowdisplaybreaks

\newtheorem{theo}{Theorem}
\newtheorem{lem}{Lemma}
\newtheorem{ass}{Assumption}
\newtheorem{defin}{Definition}

\newtheorem{ex}{Example}

\begin{document}

\title{\textsc{Rationalizable Screening and Disclosure under Unawareness}\thanks{We thank Pierpaolo Battigalli, Kym Pram, the AE, and two anonymous reviewers for detailed comments on an earlier draft. Moreover, we thank Antonio Penta and participants in seminars at UC Davis, UW Bothell, UW Seattle, UW Tacoma, Monash University, Universidad Torcuato Di Tella, and Stony Brook University as well as the Conference on ``Unawareness and Unintended Consequences,'' 2021, the Zoom Mini-Workshop on ``Contract theory with unawareness,'' 2021, the Canadian Economic Association Conference, 2021, and SAET 2023. Burkhard gratefully acknowledges financial support through ARO Contract W911NF2210282.}}

\author{Alejandro Francetich\thanks{School of Business, University of Washington, Bothell. Email: aletich@uw.edu} \and Burkhard C. Schipper\thanks{Department of Economics, University of California, Davis. Email: bcschipper@ucdavis.edu
}}

\date{October 16, 2025}

\maketitle

\begin{abstract} We analyze a principal-agent procurement problem in which the principal (she) is unaware some of the marginal cost types of the agent (he). Communication arises naturally as some types of the agent may have an incentive to raise the principal's awareness (totally or partially) before a contract menu is offered. The resulting menu must not only reflect the principal's change in awareness, but also her learning about types from the agent's decision to raise her awareness in the first place. We capture this reasoning in a discrete concave model via a rationalizability procedure in which marginal beliefs over types are restricted to log-concavity, ``reverse'' Bayesianism, and mild assumptions of caution. 

We show that if the principal is ex ante only unaware of high-cost types, all of these types have an incentive raise her awareness of them---otherwise, they would not be served. With three types, the two lower-cost types that the principal is initially aware of also want to raise her awareness of the high-cost type: Their quantities suffer no additional distortions and they both earn an extra information rent. Intuitively, the presence of an even higher cost type makes the original two look better. With more than three types, we show that this intuition may break down for types of whom the principal is initially aware of so that raising the principal's awareness could cease to be profitable for those types. When the principal is ex ante only unaware of more efficient (low-cost) types, then \textit{no type} raises her awareness, leaving her none the wiser. 

\end{abstract}

\newpage


\section{Introduction}

In many contracting situations, whether large or small, the agent (he) hired by a principal (she) is much more experienced than the principal; this is often the reason why an agent is hired in the first place. For instance, when procuring novel, complex, and tailor-made weapon systems, the government may not be aware of all the technological details impacting the contractors' cost structure. Similarly, for an economics professor hiring a contractor to remodel her house. 

The standard approach to contracting under hidden information poses that the principal envisions all of the types of the agent and offers a menu of contracts to screen types (Mirrlees, 1971; Mussa and Rosen, 1978; Baron and Myerson, 1982; Maskin and Riley, 1984). However, when the principal lacks experience (as with novel, complex, or tailor-made projects), she may be unable to conceive of all the relevant events that affect the agent's costs. Consequently, the menu of contracts designed by the principal may fail to provide appealing contracts to some types of the agent. For instance, a contractor may find himself expected to deliver output quantities at prices that fail to account for his actual cost structure; so, he may choose to misrepresent himself or downright reject the principal's offer. If allowed, the agent may have an incentive to raise the principal's awareness of types in the hope of being presented with a more suitable menu of contracts. Thus, communication arises naturally. However, such a voluntary disclosure of possible types may reveal information about the agent's actual type, based on his incentives to raise her awareness in the first place. 

In this paper, we study a principal-agent problem in which the principal is unaware of some of the agent's types. Before the principal offers a menu of contracts to screen for the agent's types of which she is aware, the agent can raise the principal's awareness (fully or partially) of additional possible types. The principal updates her belief over the larger set of types, taking into account the different type's incentives to raise her awareness to begin with. Consequently, she offers a menu of contracts that she deems optimal, from which the agent either selects a contract or takes his outside option. 

We show that if the principal is ex ante only unaware of higher-cost types, all of these types have an incentive raise her awareness---otherwise, they would not be served. With three types, the two lower-cost types that the principal is initially aware of also want to raise her awareness of the high-cost type: Their quantities suffer no additional distortions and they both earn an extra information rent. Intuitively, the presence of an even higher cost type makes the original two look better. However, for more than three types, we show that this intuition may break down for types of whom the principal is initially aware of so that raising the principal's awareness could cease to be profitable for those types. The reason is that after raising awareness of higher-cost types, the principal may focus her belief on higher-cost types. Consequently, the principal may not design a menu with a dedicated contract for such lower-cost type, who would then need to bunch and select a contract dedicated for another type, which might be less appealing than the contract selected by the type without raising awareness. When the principal is ex ante only unaware of more efficient (i.e., lower-cost) types, then \textit{no type} raises her awareness, leaving her none the wiser. Intuitively, raising awareness of lower-cost types makes all of us look worse in the eye of the principal, while said types can enjoy a higher surplus under the ``default'' offer.

Analyzing this contracting problem poses several challenges. First, how should we model this dynamic interaction under unawareness? Not only is the principal initially unaware of some types of the agent, but her awareness may change (endogenously) throughout the interaction. To cope, we make use of games in extensive form with unawareness (Heifetz, Meier, and Schipper, 2013; Halpern and R\^{e}go, 2014; Feinberg, 2021). We have a collection of game trees ordered by set inclusion of types selected by nature, each representing a possible worldview that the principal may hold after communication by the agent. 

Second, how do we update beliefs upon becoming aware? Here, we make use of \textit{reverse Bayesianism} (Karni and Vier{\o}, 2013; Hayashi, 2012; Dominiak and Tserenjigmid, 2018): Upon becoming aware of new types, the relative likelihood ratios for types that the principal was initially aware of are preserved. In our setting, however, it might be the case that upon becoming aware of new types, the principal can rule out some types she was already aware of---based on incentives. That is why we only apply reverse Bayesianism to types that she cannot rule out. It turns out that reverse Bayesianism fits extremely well with the standard assumption of information economics of monotone inverse hazard rates (i.e., the monotone Mill's ratio). Since reverse Bayesianism preserves relative likelihoods and the inverse hazard rate is a likelihood ratio, reverse Bayesianism facilitates the comparison of optimal contract quantities across awareness levels. This observation should turn out to be extremely useful in other areas of information economics with unawareness.

Third, what is the appropriate solution concept? As discussed by Heifetz, Meier, and Schipper (2013, 2021) and Schipper (2021), when players' awareness changes along the outcome path, rationalizability concepts are more appropriate than equilibrium concepts such as Perfect Bayesian Equilibrium or Sequential Equilibrium because, typically, there is no ready-made equilibrium convention available in novel situations. However, using a rationalizability concept means that we cannot make use of equilibrium tie-breaking conventions in the agent's incentive and participation constraints. In standard screening problems, it is implicitly assumed that when a type of the agent is indifferent between two different contracts, or between the menu and his outside option, he always makes the ``right'' choice. Such a brute-force behavioral convention is foreign to rationalizability. Indifference between contracts could be broken with strict inventive constraints. However, with real-valued transfers, there is no well-defined ``smallest'' increment of transfers to break the indifference. To circumvent these technicalities while reflecting the discreteness of the real world, we employ the discrete concave screening approach we develop in Francetich and Schipper (2025). 

In the companion paper, Francetich and Schipper (2025), we analyze a screening problem with discrete types and discrete contracts (with full awareness). We show that the solutions to the discrete first-order conditions need not be unique \textit{even under discrete strict concavity}, but we also show that there cannot be more than two optimal quantities for each type and---if there are two---they must be adjacent. Moreover, strictly monotone virtual costs need not imply strict monotonicity of the quantities, unless we limit the ``degree of concavity'' of the principal’s utility. Finally, we  introduce a rationalizability notion with robustness to slight perturbations of beliefs over types called $\Delta-O$ \textit{Rationalizability} and show that the set of $\Delta-O$ rationalizable menus coincides with the set of usual equilibrium contracts---possibly augmented to include irrelevant contracts. In the present work, we extend this solution concept to dynamic interaction under unawareness. $\Delta-O$ rationalizability features full-support log-concave marginal beliefs over types. In our dynamic context, the full-support assumption is too strong since the principal may also learn from the agent's disclosure decision about the agent's type and consequently may want to exclude some types. That's why we replace the full-support assumption with weaker assumptions modeling some sort of cautiousness such as not excluding extreme types that the principal became newly aware. In addition, we require reverse Bayesianism as discussed above. 

Although games with unawareness are relatively new, this is not the first paper to study contracting under unawareness. von Thadden and Zhao (2012) study a principal-agent moral hazard problem in which the principal is aware of actions of which the agent is unaware. When contemplating whether or not to make the agent aware of such actions, the principal faces a trade-off between getting a better action and saving on information rents due to additional incentive compatibility constraints. In Auster (2013), the principal is aware of contingencies of which the agent is unaware, but whose realization is informative about the agent's actions. In the optimal contract, the principal faces a trade-off between exploiting the agent's unawareness and using said contingencies in order to provide incentives. Filiz-Ozbay (2012) studies a risk neutral insurer who is aware of some contingencies of which the insured is unaware. The insurer has an incentive to mention only some contingencies in a contract, while remaining silent on others. Li and Schipper (2024) study optimal second-price auctions with participation fees in settings where the auctioneer is aware of payoff relevant components and may raise them (with and without disclosing information about them) to unaware bidders before second-price auctions. 

In all of the papers above, in contrast to ours, the principal has a higher awareness level than the agent. Auster and Pavoni (2024) and Lei and Zhao (2021) feature an agent with higher awareness level than the principal but in the context of optimal delegation. In Auster and Pavoni (2024), the agent is aware of both the set of his actions and their performance, and only reveals extreme actions. In Lei and Zhao (2021), the agent only partially reveals contingencies of which the principal is unaware. Principals who are unaware of more contingencies delegate a large set of projects. Herweg and Schmidt (2020) study a procurement problem in which the seller may be aware of some design flaws. They propose an efficient two-stage mechanism with a neutral arbitrator that does not require an ex ante description of design flaws. For their argument to work it is crucial that any design flaws are verifiable ex post. However, we know from empirical studies (e.g., Bajari, Houghton, and Tadelis, 2014) that there is often disagreement ex post. Grant, Kline, and Quiggin (2012) discuss disagreements arising from asymmetric awareness in contracting. Finally, in the field of value-based business strategy, Bryan, Ryall, and Schipper (2021) devise cooperative games with incomplete information and unawareness for studying surplus creation and surplus appropriation within contracting relationships. 

Pram and Schipper (2025) take a more general approach and study efficient mechanism design under unawareness. They show that standard VCG mechanisms fail to implement efficiently at the pooled awareness level, and devise a dynamic elaboration VCG mechanisms in which agents report payoff types, a mediator pools awareness encapsulated in those payoff types and communicates it back to agents, and agents subsequently elaborate on their prior reported types at the pooled awareness level. This might be repeated until no agent wants to elaborate any further, upon which an efficient outcome is implemented using transfers inspired by VCG mechanisms. Their mechanisms can be interpreted as complementing VCG mechanisms with awareness-changing negotiation in order to take advantage of the expertise of the agents. They show that these mechanisms lead to efficiency at the pooled awareness level and do not impose additional constraints on budget balance. In particular, in a procurement context, they show that the dynamic elaboration reverse second-price auction pools awareness, leads to efficiency, to budget balance, and satisfies participation constraints. The extension to optimal multi-agent mechanisms remains an open problem. 

Somewhat similar to a principal-agent problem, Piermont (2024) studies how a decision maker can incentivize an expert to reveal novel aspects about a decision problem via an iterated revelation mechanism in which, at each round, the expert decides on whether or not to raise awareness of a contingency that the decision maker can consider when proposing a new contract to the expert. 

Since communication arises naturally in our principal-agent problem due to some types of the agent possibly having an incentive to raise the principal's awareness, our paper is also related to disclosure in games with unawareness. Heifetz, Meier, and Schipper (2021) show that, in disclosure games \`{a} la Milgrom and Roberts (1986), the unraveling argument breaks down when receivers may be unaware of some signaling dimension. In such a case, the receiver is unable to infer anything about the sender's type from the absence of a signal. This has been experimentally tested in Li and Schipper (2025). Schipper and Woo (2019) apply this insight to electoral campaigning to discuss microtargeting of voters and negative campaigning.

The closest work to ours on adverse selection without unawareness are Pram (2021) and Ali, Lewis, and Vasserman (2023). Both papers consider a screening problem in which the agent can disclose verifiable evidence about his type before the principal commits to a mechanism. Pram (2021) characterizes environments in which verifiable evidence is welfare improving, namely whenever in the mechanism without disclosure some types would be excluded. Ali, Lewis, and Vasserman (2023) also find that the agent can benefit from prior disclosure of evidence, but it depends on that partial disclosure being feasible. Somewhat related is Sher and Vohra (2015), who consider disclosure in a monopolist's price discrimination problem. The seller commits to a mechanism with evidence-contingent prices; i.e., disclosure occurs \emph{after} commitment to the mechanism. Such timing would be much less compelling in the face of unawareness, however, as the principal is unaware of potential evidence that could be disclosed. Hidir and Vellodi (2021) study buyer's cheap-talk prior to trade and buyer-optimal market segmentation consistent with their information revelation.

Our solution concept, \textit{$\Delta$-O Prudent Rationalizability}, shares features with strong rationalizability aka extensive-form rationalizability (Pearce, 1984, Battigalli, 1997, Heifetz, Meier, and Schipper, 2013). It also has features of $\Delta$-rationalizability (see Battigalli, 2003, Battigalli and Siniscalchi, 2003; Battigalli and Friedenberg, 2012; Battigalli and Prestipino, 2013; Brandenburger, Friedenberg, and Keisler, 2008; Brandenburger and Friedenberg, 2010) because it involves restrictions on first-order beliefs such as log-concavity and reverse Bayesianism---although we are not aware of a notion of $\Delta$-rationalizability in the literature featuring log-concavity or reverse Bayesianism. Our robustness notion is related to rationalizability by sets of beliefs, introduced in the context of static games with complete information by Ziegler and Zuazo-Guarin (2020), and can be interpreted as a notion of caution, making our rationalizability concept reminiscent of prudent rationalizability (Heifetz, Meier, and Schipper, 2021). Our solution concept is meant to capture some version of common cautious (strong) belief in rationality (Battigalli and Siniscalchi, 2002; Guarino, 2020) and common belief in reverse Bayesianism and in log-concavity of marginals on types. 

The full-awareness benchmark of our principal's problem is analyzed in Francetich and Schipper (2025), who show that our solution concept yields standard equilibrium menus of contracts---except perhaps for the inclusion of irrelevant contracts. Since we use a rationalizability notion with belief restrictions as solution concept, our work is also related to the mechanism design literature on implementation in rationalizable strategies. For instance, Oll\'{a}r and Penta (2017), who study full implementation in rationalizable strategies with belief restrictions. A direct comparison is not straightforward, however: Oll\'{a}r and Penta (2017) work with general belief restrictions and smoothness assumptions while we focus on log-concave marginal beliefs but in a discrete setting and allowing for unawareness.

The paper is organized as follows. Section~\ref{model} lays out our model and our solution concept. Section \ref{high} analyzes the case where the principal is initially unaware of high costs only, while Section~\ref{low} looks at the case of initial unawareness of low cost types only. Section \ref{discussion} concludes. Some auxiliary results are stated and proved in the appendix.


\section{Model}\label{model}

\subsection{Discrete Concave Screening with Unawareness}\label{beliefs}

Our basic setup is an extension of the discrete concave screening problem by Francetich and Schipper (2025) to unawareness and disclosure of awareness. The principal ($P$, ``she'') wants to procure a quantity of an object from an agent  ($A$, ``he''). Given her awareness level after a possible disclosure from the agent, the principal offers him a menu $M$ of contracts $\bm{c} = (q, t)$ to procure $q \in D := \{0, 1, \ldots, b\}$ units of output, where $b$ is a positive integer, in exchange for payment $t\in D$. The principal chooses a contract menu from the set $\mathcal{M} := \{M \subseteq D^{2}: M \neq \emptyset\}$. Presented with menu $M \in \mathcal{M}$, the agent chooses either a contract $\bm{c} \in M$ or takes his outside option, $\bm{o}$, identified by the null-contract $(0, 0)$; the game then ends. 

The agent's payoff depends on his marginal-cost type, denoted by $\theta$, which is his private information. Types are drawn from $\bar{\Theta} := \left\{ j - \frac{1}{\gamma}: j = 1, \ldots, m \right\}$ for some $\gamma \in \mathbb{R}_{++}$ and $m \in \mathbb{N}$. The payoff for the agent with type $\theta \in \bar{\Theta}$ from signing contract $\bm{c} = (q, t)$, is given by $u_A(\bm{c}, \theta) := t - \theta q$; his payoff from the outside option is $u_{A}(\bm{o},\theta) = 0$ for all $\theta \in \bar{\Theta}$. We require that both $\gamma$ and $b$ are large compared to the positive integer $m$, so that the principal is able to design menus of contracts with as many contracts as types---if she desires to do so. We also require $\gamma > b$. Notice that both $q$ and $t$ are integer, while $\theta$ and $\theta q$ are not. 

As discussed by Francetich and Schipper (2025), non-integer types can reflect the fact that private costs can be nuanced, idiosyncratic, implicit, unverifiable and even cognitive and subjective while the terms of a contract are typically standardized by objective scales of measurements that contracts can explicitly be written on and that can be verified ex post. We believe that's one of the reasons why contracts cannot directly conditions on marginal costs. As we will see, the model choice achieves two simultaneous technical goals: (1) Replicating the usual constraint-simplification steps and thus obtaining formulas more-closely comparable to standard results; and (2) breaking indifferences in incentive and participation constraints due to \textit{round-up} rents stemming from the mismatch between integer transfers and non-integer total costs, simplifying the rationalizability analysis. If both total costs and transfers were integers, we could encounter indifferences in the relevant incentive and participation constraints in the contract-design problem. Since we employ a rationalizability notion as the solution concept, we cannot easily resolve these indifferences invoking the usual tie-break equilibrium conventions. Moreover, our model accommodates integer types by simply taking the limit as $\gamma \to \infty$. Finally, when equidistant grids of real numbers for contracts and marginal costs are chosen independently at random, then ending up with the same grid is a probability zero event. Having the same grid for contracts and costs would be non-generic. 

Departing from standard screening problems, we posit that the principal is unaware of some types. More precisely, the principal is only aware of types in a proper subset $\Theta_P \subsetneq \bar{\Theta}$, while the agent is aware of all types in $\bar{\Theta}$---regardless of his type.\footnote{Letting the agent's awareness level depend on his marginal-cost type would make it easier for the principal to infer the agent's type upon having her awareness raised. However, our goal is to focus on the informativeness of disclosing awareness based on incentives alone, so we do not allow for this. See Pram (2021) and Ali, Lewis, and Vasserman (2023) for disclosure of information only.} The agent knows that the principal is only aware of types in $\Theta_P$. Thus, we allow him to approach the principal to raise her awareness of additional types before designing the contract. Initially, the agent decides whether to disclose the existence of some (larger) subset of types to the principal by sending her a message $\Theta \subseteq \bar{\Theta}$. We require the message $\Theta$ to be a \textit{truncation} of $\bar{\Theta}$ that includes $\Theta_P$: $$\Theta \in \mathcal{T} := \left\{\Theta' \subseteq \bar{\Theta}: \Theta' \supseteq \Theta_{P} \mbox{ and } \forall \theta\in \bar{\Theta} \ ((\min(\Theta') \leq \theta \leq \max (\Theta'))\Rightarrow \theta \in \Theta') \right\}.$$ 

Although our representation of cost types is unidimensional, this structure of messages allows us to think of the agent's type as a score that aggregates the impact of various events affecting marginal costs, reminiscent of ``pseudotypes'' in scoring auctions; see, for instance, Asker and Cantillon (2006) or Bajari, Houghton, and Tadelis (2014). The agent's marginal cost may be result of two components, with the principal being only aware of one of them. For instance, let $\theta$ be the sum of two components, $\theta_{1} \in \{0.99, 1.99\}$ and $\theta_{2} \in \{0, 1\}$, the latter of which is not on the principal's radar; if the agent makes the principal aware of $\theta_{2}$, knowing that $\theta = \theta_{1} + \theta_{2}$ allows her to figure out that $\theta\in\{0.99, 1.99, 2.99\}$.

Notice that, by raising the principal's awareness, he is not disclosing information to her in a standard sense. When the agent sends message $\Theta\in\mathcal{T}$, he is not saying ``I am type $\theta\in\Theta\setminus\Theta_{P}$.'' (This message would not necessarily be credible anyway.) Rather, he is saying ``Types in $\Theta\setminus\Theta_{P}$ also exist, so my type is in $\Theta$.'' This enables the principal to form beliefs over types in $\Theta$ while previously she was able to form beliefs only about types in $\Theta_P$. 

The principal's utility of output $q$ is given by a function $v(q)$, and her net utility or payoff from a contract $\bm{c} = (q, t)$ is $u_P(\bm{c}) = v(q) - t$. In order to state the assumptions imposed on $v$, denote by $\Delta^- v(q) := v(q) - v(q - 1)$ for $q \in \{1, ..., b\}$ and $\Delta^+ v(q) := v(q + 1) - v(q)$ for $q \in \{0, ..., b - 1\}$ the discrete backward and forward derivatives of $v(q)$, respectively. Let $\Delta^+ \Delta^- v(q) := \Delta^+(\Delta^- v(q))$ denote the second discrete derivative.  

\begin{ass}\label{properties_v} 
The function $v(q)$ satisfies the following properties:
\begin{enumerate}
\item\label{normalization}
$v(0) = 0$;
\item\label{sincreasing}
$v(q)$ is strictly increasing;
\item\label{sdc}
$v(q)$ is discrete strictly concave: For every $q \in \{1, \ldots, b - 1\}$, $v(q+1)+v(q-1)<2v(q)$;
\item\label{not_too_sdc} 
$\Delta^+ \Delta^- v(q)\geq -1$;
\item\label{efficient_unique} 
For every $q \in \{0, ..., b - 1\}$ and $n \in \mathbb{N}$, $\Delta^+ v(q) \neq n$. 
\end{enumerate}
\end{ass} 

Assumptions \ref{properties_v}.\ref{normalization} and \ref{properties_v}.\ref{sincreasing} are standard in the screening literature, while Assumption \ref{properties_v}.\ref{sdc} is the discrete counterpart of the typical assumption of strict concavity. The remaining two assumptions warrant further discussion. 

Discrete concavity introduces two technical complications: First, the menu of optimal contracts obtained by ignoring the monotonicity constraints may fail to satisfy strict monotonicity (thereby allowing for indifferences) \textit{even if virtual costs are strictly monotone}.
As explained in Francetich and Schipper (2025), this can happen if $v(q)$ is ``too concave.'' Assumption \ref{properties_v}.\ref{not_too_sdc} allows us to circumvent the problem by limiting the ``degree of concavity'' of $v(q)$. To see why, notice that discrete strict concavity can be rewritten using the second discrete derivative as $0 > \Delta^+ \Delta^- v(q)$, so Assumptions \ref{properties_v}.\ref{sdc} and \ref{properties_v}.\ref{not_too_sdc} collectively imply $0 > \Delta^+ \Delta^- v(q) \geq -1$.

Second, the optimal menu need not be unique \textit{even under strict concavity}. The discrete F.O.C. for the contract catered to any given type may have at most two adjacent solutions (Francetich and Schipper, 2025). As non-uniqueness is a knife-edge case, our prudence-inspired robustness notion with respect to beliefs over types ensures uniqueness for all types but one: the lowest type of which the principal is aware (and to which she assigns positive probability). The allocation for the latter type involves no distortion, so it is independent of beliefs. Assumption \ref{properties_v}.\ref{efficient_unique} tackles this case, ensuring uniqueness when paired with our solution concept. 

Our game is a dynamic game with unawareness (Heifetz, Meier, and Schipper, 2013). Dynamic games with unawareness consist of an ``ordered forest of game trees'' rather than a single game tree. In our case, there is exactly one tree for each subset of type-space messages/moves of nature $\Theta \in \mathcal{T}$. This is because when the principal is unaware of types in $\bar{\Theta}\setminus\Theta$, she cannot conceive that the agent could have disclosed some other $\Theta' \supsetneq \Theta$. That is, unawareness of types implies unawareness of disclosure actions of the agent that involve such types.

\begin{figure}[h!]
\begin{center}
\includegraphics[scale=.16]{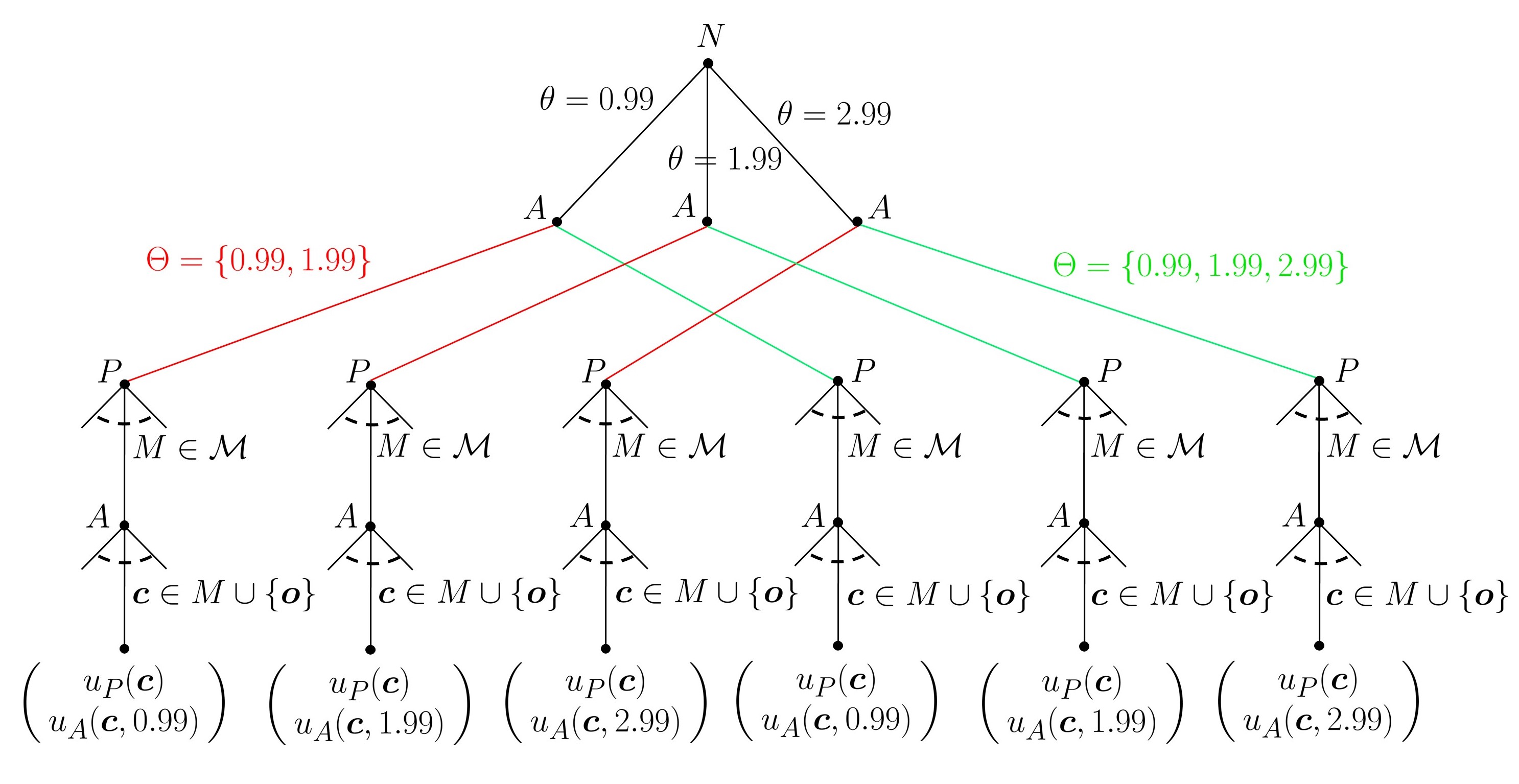}
\end{center}
\caption{Full tree of objective moves\label{tree1}}
\end{figure}

\begin{figure}[h!]
\begin{center}
\includegraphics[scale=.16]{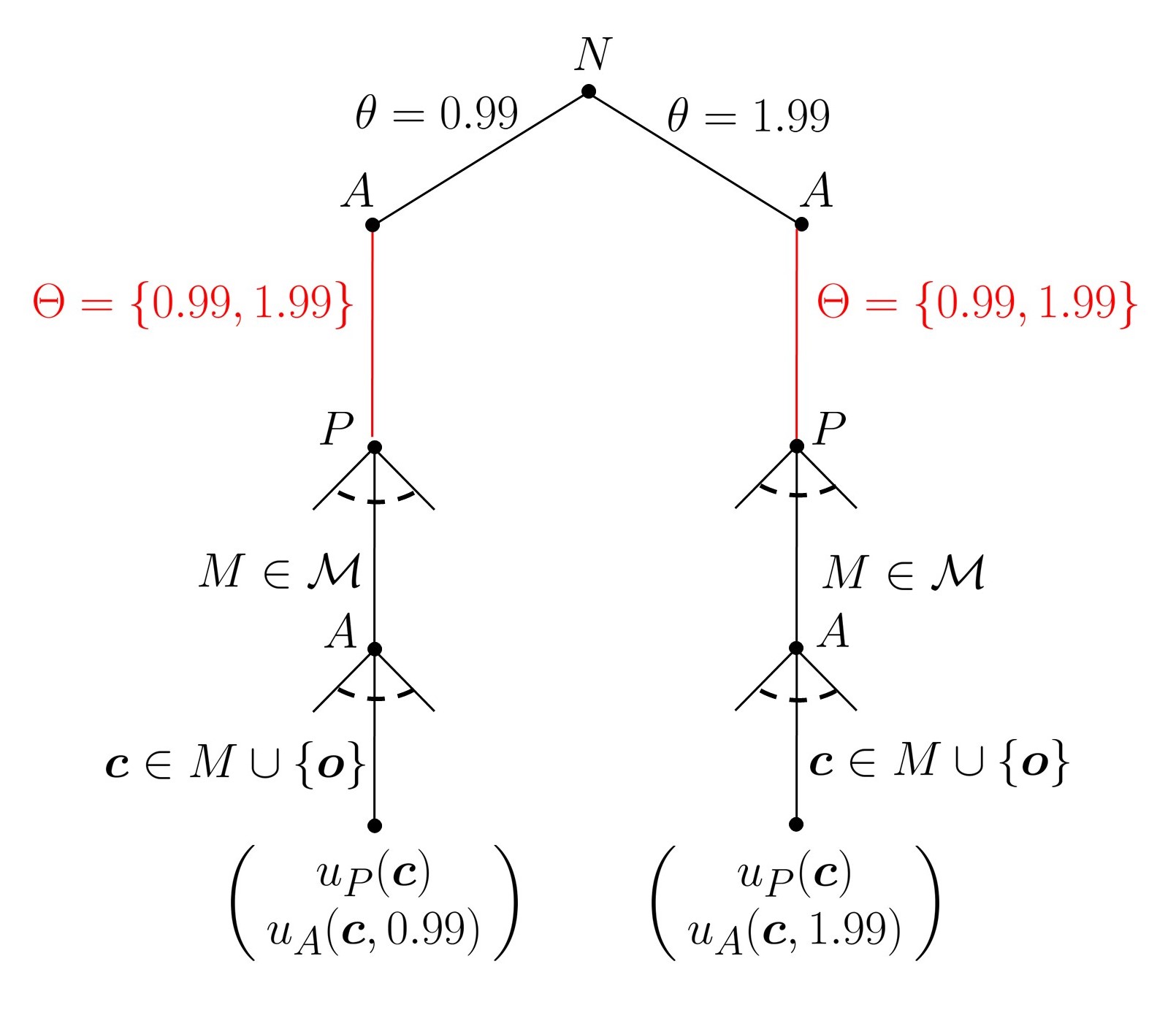}
\end{center}
\caption{The tree initially perceived by the principal\label{subtree1}}
\end{figure}

Consider an example where $D=\{0,1,\ldots,99\}$, $\bar{\Theta} = \{0.99, 1.99, 2.99\}$, and the principal is initially unaware of type $\theta=2.99$, so that $\Theta_{P}=\{0.99,1.99\}$. The agent can send two messages: the uninformative message $\Theta=\{0.99, 1.99\}$ and raising awareness of type $2.99$ with the message $\Theta=\{0.99, 1.99, 2.99\}$. Figure~\ref{tree1} represents the full tree that the agent conceives, but not the principal---at least not initially. Her initial perception of the strategic situation is given by the game tree in Figure~\ref{subtree1}, where cost type $2.99$ and the disclosure thereof are missing.

The two separate trees are not enough to represent the interaction between our principal and agent, however, because the principal's awareness is dynamic and depends on the agent's disclosure action. To model such dynamic and strategic change of the principal's perception, dynamic games with unawareness feature information sets at a history in one game tree which may be located in a subtree thereof. A player is aware of the tree (and all subtrees) in which their information set is located. 

\begin{figure}[h!]
\vspace{4mm}
\begin{center}
\includegraphics[scale=.155]{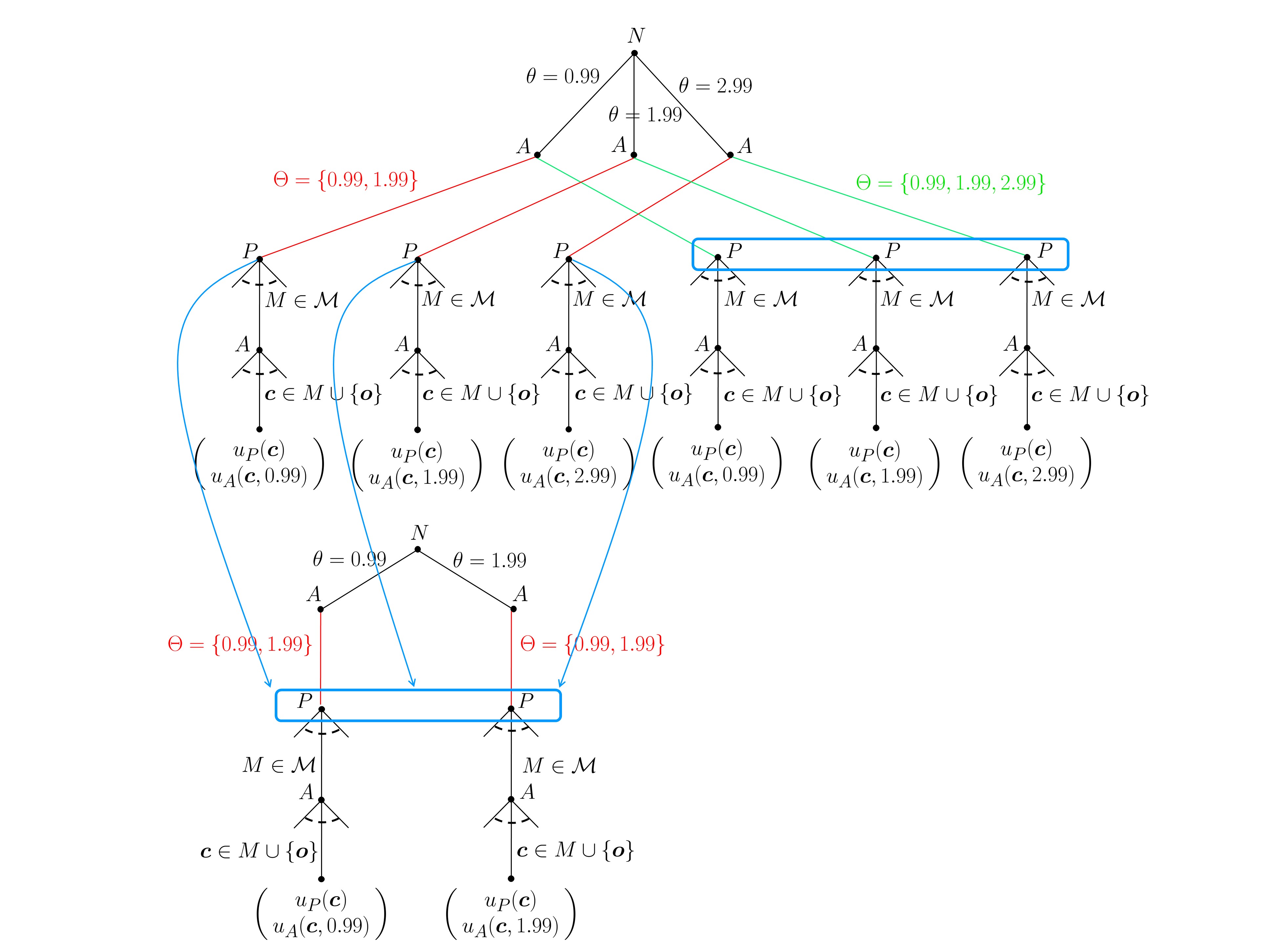}
\end{center}
\caption{Screening game with unawareness 
\label{gen_game}}
\vspace{4mm}
\end{figure}

The interaction in our game is illustrated in Figure~\ref{gen_game}. While the agent's information sets are trivial, i.e., there is a singleton information set at every history at which the agent moves, the principal's information sets are depicted in Figure~\ref{gen_game} with blue rectangles and arrows. When the agent does not disclose the existence of type $2.99$ (e.g., the orange actions in both trees), then the principal remains unaware of it and her information set is located in the bottom game tree. When the agent discloses everything (i.e., green actions in the upmost game tree), the principal becomes aware of everything and her information set is located in the upmost game tree. Moreover, she can now reason that the agent could have remained silent about type $2.99$ and chose not to, which may tell her something about his actual type. 

\subsection{Solution Concept}\label{solution}

The primitives of our solution concept comprise of strategies and belief systems. A player's strategy assigns an action to each of the player's information sets. It is easy to see that the principal's information sets are identified by $\mathcal{T}$. Thus, a strategy for the principal is a mapping $s_P: \mathcal{T} \longrightarrow \mathcal{M}$, and her set of strategies is $S_P := \mathcal{M}^{\mathcal{T}}$. 

The agent moves twice along each path. First, he decides which set of types to disclose (``first round''). Second, he either picks a contract from the menu presented to him by the principal or he goes for his outside option  (``second-round''). However, we cannot simply identify first-round information sets with the (full) type space $\bar{\Theta}$. To see why, consider type $\theta=0.99$ following the message $\Theta=\{0.99,1.99,2.99\}$. Here, the principal knows that Nature draws types from $\{0.99,1.99,2.99\}$, and that type $0.99$ could have sent her the message $\Theta=\{0.99,1.99\}$ and chose not to; thus, she can reason about the type's disclosure incentives. But if this type were to send the message $\Theta=\{0.99,1.99\}$, the principal would remain none the wiser about type $2.99$. She would conceive of Nature drawing types from $\{0.99,1.99\}$ only, and she could not reason about the agent being aware of type $\theta=2.99$---let alone about him raising her awareness about it---when she is not aware of it herself. 

Thus, the agent's first-round information sets depend on the type and the game tree. Since each game tree is identified by a subset of types, we identify them with the tuples $(\theta, \Theta)$ with $\theta \in \Theta$ and $\Theta \in \mathcal{T}$. When type $\theta$ in the $\Theta$ tree decides on disclosure, from the principal's viewpoint he could only disclose subsets of $\Theta$ (that contain $\Theta_P$). To model this, we define: $$\mathcal{T}(\Theta) := \left\{\Theta'\in\mathcal{T}:\Theta'\subseteq \Theta\right\}$$
for each $\Theta \in \mathcal{T}$. Hence, a first-round strategy for the agent is a mapping: $$s_{A1}: \bigcup_{\Theta \in \mathcal{T}} \bigcup_{\theta \in \Theta} \{(\theta, \Theta)\} \longrightarrow \mathcal{T},$$ with $s_{A1}(\theta, \Theta) \in \mathcal{T}(\Theta)$ for every $(\theta, \Theta)$ with $\theta \in \Theta$ and $\Theta \in \mathcal{T}$. 

For second-round strategies, note that second-round information sets are identified by type-tree profiles (i.e., first-round information sets), messages, and contract menus. That is, a second-round strategy for the agent is a mapping $s_{A2}: \bigcup_{\Theta \in \mathcal{T}} \bigcup_{\theta \in \Theta} \bigcup_{\Theta' \in \mathcal{T}(\Theta)} \{(\theta, \Theta, \Theta')\} \times \mathcal{M} \longrightarrow D^2$ with $s_{A2}(\theta, \Theta, \Theta', M) \in M \cup \{\bm{o}\}$ for all $\theta \in \Theta$, $\Theta \in \mathcal{T}$, $\Theta' \in \mathcal{T}(\Theta)$, and $M \in \mathcal{M}$. Note that $\Theta$ refers here to the game tree associated with the set of types $\Theta$, while $\Theta'\in \mathcal{T}(\Theta)$ refers to the set of types disclosed by the agent in the first round. We denote the agent's set of strategies $s_A = (s_{A1}, s_{A2})$ by $S_A$. 

While we may think of the agent's strategies as his own plan on how to act in every possible contingency, the interpretation is necessarily different for the principal's strategy. Initially, the principal can only conceive of $\Theta_{P}$; she cannot conceive of receiving messages containing types that she is unaware of, let alone plan for such types before becoming aware of them. Strategies for the principal are more accurately interpreted as objects of belief of the aware agent. They summarize what the principal might do in various situations; the agent, when deciding on his optimal disclosure, has to form beliefs about them. 

When a player is unaware, they still form beliefs about other player's moves; but since they may be unable to perceive the entire strategy of their opponents, they actually form beliefs about \textit{partial} strategies. For any $\Theta \in \mathcal{T}$ and any principal's strategy $s_P \in S_P$, denote by $s_P^{\Theta}$ a principal's $\Theta$-partial strategy, defined as the restriction of $s_P$ to information sets in the game tree identified by $\Theta$ and all ``less expressive'' game trees identified by $\Theta' \in \mathcal{T}(\Theta)$; it is a mapping $s_P^{\Theta}: \mathcal{T}(\Theta) \longrightarrow \mathcal{M}$. Denote by $S_P^{\Theta}$ her set of $\Theta$-partial strategies. Similarly, let $s_{A1}^{\Theta}$ and $s_{A2}^{\Theta}$ be agent's first-round and second-round $\Theta$-partial strategies, respectively, and denote by $S_A^{\Theta}$ his set of $\Theta$-partial strategies. In general, we use the notational convention that the superscript $\Theta$ on any strategy or any subset of strategies (of any player) indicates that they are the $\Theta$-partial strategies.  
 
A belief system of a player assigns to each of the player's information sets a belief about the other player's partial strategies and possibly moves of nature, with the provision of assigning probability one to the set of the other player's partial strategies and possibly moves of nature that reach said information set. 

The agent's second-round beliefs are trivial, since by then he has observed everything. Nonetheless, he forms first-round beliefs about the principal's menu offer: $$\beta_{A} \in B_A := \prod_{\Theta \in \mathcal{T}} \left(\Delta(S_P^{\Theta})^{\bigcup_{\theta \in \Theta} \{(\theta, \Theta)\}}\right).$$ More precisely, the agent of type $\theta$ in game tree $\Theta$ forms a belief $\beta_A(\theta, \Theta) \in \Delta(S_P^{\Theta})$ about $\Theta$-partial strategies of the principal. 

At each of her information sets, the principal forms beliefs both about moves of nature and about the strategies of the agent. Since not all information sets can be reached with all strategies of the agent, the principal's belief at an information set should be consistent with the agent's strategies that reach it. To make this statement precise, define: $$S_A(\Theta) := \left\{(s^{\Theta}_{A1}, s^{\Theta}_{A2}) \in S^{\Theta}_{A}: \exists \theta \in \Theta \  \left(s_{A1}(\theta, \Theta) = \Theta \right) \right\}$$ for every $\Theta \in \mathcal{T}$; $S_A(\Theta)$ is the set of $\Theta$-partial strategies of the agent for which there exists a type in $\Theta$ at the game tree identified by $\Theta$ who discloses $\Theta$. 


A belief system of the principal is defined by $\beta_P \in B_P \subset \prod_{\Theta \in \mathcal{T}} \Delta(\Theta \times S_A^{\Theta})$ that satisfies:
\begin{itemize} 
\item[(i)] \textit{Consistency}: For any information set $\Theta \in \mathcal{T}$, the marginal of the belief $\beta_P(\Theta)$ on $\Theta$-partial strategies of the agent, $\text{marg}_{S_A^{\Theta}} \ \beta_P(\Theta)$, satisfies $\text{marg}_{S_A^{\Theta}} \ \beta_P(\Theta)(S_A(\Theta)) = 1$.
\item[(ii)] \textit{Log-concavity}: For any information set $\Theta \in \mathcal{T}$, denote the marginal of belief $\beta_P(\Theta)$ on $\Theta$ by $p_{\Theta}(\theta):= \left(\text{marg}_{\Theta} \ \beta_P(\Theta)\right)(\{\theta\})$.\footnote{Note that $\Theta$ appears twice in $\text{marg}_{\Theta}(\beta_{P}(\Theta))$, which we can link to its dual role. In the argument of $\beta_{P}$, $\Theta$ represents the message received by the principal, which leads to a belief $\beta_{P}(\Theta)$ over types and the agent's strategies; $\text{marg}_{\Theta}(\beta_{P}(\Theta))$ is the marginal belief over the former of these two.} Let $\theta^{(1)}$ denote the largest element of $\Theta$, let $\theta^{(2)}$ be its second-largest element, and so on. If $\Theta$ has at least three elements, then for every $i = 2, \ldots, \left|\Theta\right| - 1$, 
$$p_{\Theta}\left(\theta^{(i)}\right) \cdot p_{\Theta}\left(\theta^{(i)}\right) \geq p_{\Theta}\left(\theta^{(i-1)}\right)\cdot p_{\Theta}\left(\theta^{(i+1)}\right).$$
\item[(iii)] \textit{Reverse Bayesianism}: For every $\Theta, \Theta'\in\mathcal{T}$ and every $\theta, \theta' \in \text{supp}(p_\Theta) \cap \text{supp}(p_{\Theta'})$,
$$\frac{p_{\Theta'}(\theta')}{p_{\Theta'}(\theta)} = \frac{p_{\Theta}(\theta')}{p_{\Theta}(\theta)}.$$
\item[(iv)] \textit{Wariness}: For any $\Theta \in \mathcal{T}$, 
\begin{itemize}
\item[a.] if $\min \Theta \notin \Theta_P$, then $\min \Theta \in \text{supp}(p_{\Theta})$,
\item[b.] if $\max \Theta \notin \Theta_P$, then $\max \Theta \in \text{supp}(p_{\Theta})$.
\end{itemize}
\item[(v)] \textit{Monotone Supports}: For any $\Theta, \Theta' \in \mathcal{T}$, $\min \Theta \leq \min \Theta'$ and $\max \Theta \leq \max \Theta'$ implies 
$$\min \left( \text{supp}\left(p_{\Theta}\right) \cup \text{supp}\left(p_{\Theta'}\right) \right) \in \text{supp}\left(p_{\Theta}\right) \mbox{ and } \max \left( \text{supp}\left(p_{\Theta}\right) \cup \text{supp}\left(p_{\Theta'}\right)\right) \in \text{supp}\left(p_{\Theta'}\right).$$
\end{itemize}

Property (i), consistency, simply requires that at every information set $\Theta \in \mathcal{T}$, the principal is certain that the agent disclosed $\Theta$ to her. 

Property (ii), log-concavity, means that the marginal beliefs of the principal on types are unimodal with respect to all types that she is aware of. This condition implies non-decreasing hazard rates (see for instance, Francetich and Schipper, 2025, Lemma B1 (ii)), a standard assumption in screening problems and other problems of information economics (Mussa and Rosen, 1978; Spence, 1980; Baron and Myerson, 1982; Maskin and Riley, 1984; Matthews and Moore, 1986, 1987; Bagnoli and Bergstrom, 2005). Although insufficient by themselves, monotone hazard rates facilitate solving for the optimal menu of contracts subject to the agent's incentive constraints---which, together with Property (iii), facilitates the comparison of agent payoffs from contract menus across awareness levels. 

Property (iii), reverse Bayesianism, addresses how the principal changes her beliefs about types with changes in her awareness level. Thus, it plays a central role in the comparison of contract menus across awareness levels. Reverse Bayesianism says that, after becoming aware of additional types in $\Theta \supseteq \Theta_{P}$, the relative likelihood of types in $\Theta_{P}$---i.e. those that the principal was initially aware of---remains the same as long as they are not ruled out by reasoning about the agent's incentives. For example, let $\Theta_{P} = \{\theta_{1},\theta_{2}\}$ and imagine the principal learns about type $\theta_{3}$: $\Theta = \{\theta_{1},\theta_{2},\theta_{3}\}$. If $\theta_{1}$ and $\theta_{2}$ were equally likely under $\beta_{P}(\Theta_{P})$, so that $p_{\Theta_{P}}(\theta_{1})=p_{\Theta_{P}}(\theta_{2})=\frac{1}{2}$, and if $\beta_{P}(\Theta)$ assigns positive probability to both types of the agent raising her awareness about $\theta_{3}$, then $\theta_{1}$ and $\theta_{2}$ must remain equally likely under $\beta_{P}(\Theta)$---so that for instance $p_{\Theta}(\theta_{1})=p_{\Theta}(\theta_{2}) = \frac{1}{3}$. If the principal rules out  $\theta_{1}$ from raising her awareness, so that $\theta_{1}\notin\text{supp}\left(\text{marg}_{\Theta} \ \beta_{P}(\Theta)\right)$, then reverse Bayesianism imposes no restrictions on the relative likelihood of $\theta_{2}$ and $\theta_{3}$. Note that reverse Bayesianism preserves log-concavity---and thus monotone hazard rates---across awareness levels. In our setting, reverse Bayesianism allows us to compare virtual costs, hence optimal allocations, across different awareness levels for the principal.

Hayashi (2012), Karni and Vier{\o} (2013), and Dominiak and Tserenjigmid (2018) axiomatize reverse Bayesianism. One of the key axioms is invariant risk preferences with respect to awareness, which is implicitly assumed in the literature on games with unawareness. One potential argument against applying reverse Bayesianism to strategic settings is that the expansion of the state space may be strategic, so expanding the support of the unaware player's beliefs accordingly could be incompatible with their opponent's incentives. We avoid this pitfall altogether by imposing reverse Bayesianism only for types that the ``enlightened'' principal cannot rule out. This is not really a restriction of reverse Bayesianism since even in decision theory it applies only to non-null events. 

Property (iv), wariness, requires that when the principal becomes aware of types $\Theta \setminus \Theta_P \neq \emptyset$, she assigns at least some probability to some of these new types. In particular, if she was initially unaware of $\max \Theta$ and/or $\min \Theta$, respectively, then she must assign positive probability to them, respectively. Not excluding these types is a form of caution or wariness, hence the name of the property. In general, property (iv) may clash with updating of beliefs; namely, when made aware of some types, the principal might be able to rule out the extreme ones. This is not the case in our screening problem, however.

The last property, monotone supports, implies that supports of beliefs move monotonically with awareness: When the principal is made aware of higher cost types, the support of her revised beliefs weakly ascends in the sense of the Veinott set order. The order on the principal's supports preserves the order on the agent's messages.

Properties (iv) and (v) are (collectively) weaker than requiring the principal to assign positive probability to all of the ``new'' types: $\text{supp}\left(\text{marg}_{\Theta} \beta_P(\Theta)\right) \supseteq \Theta \setminus \Theta_P$. In our setting, this condition would be too strong---our principal would never be able to \textit{effectively} rule out any types. We explain this point after introducing our solution concept.


In line with the literature on $\Delta$-rationalizability, properties (ii) to (v) constitute our $\Delta$-restrictions.

Given belief systems and strategies, we can define best responses. The principal's best response depends on the message received from the agent (her information set) and her belief about the agent's pick from contract menus, i.e., her joint belief about the agent's type and second-round strategy. Define the principal's conditional expected utility (i.e., conditional on her information set $\Theta \in \mathcal{T}$) from offering menu $M \in \mathcal{M}$ given her belief system $\beta_P \in B_P$ by: 
$$U_P(M \mid \beta_P, \Theta) := \sum_{(\theta, s_{A}^{\Theta}) \in \Theta \times S_{A}^{\Theta}} \beta_P(\Theta)(\{(\theta, s_{A}^{\Theta})\}) \cdot u_P(s_{A2}^{\Theta}(\theta, \Theta, \Theta, M)).$$  
The principal's set of best responses given her belief system $\beta_P \in B_P$ and her information set $\Theta \in \mathcal{T}$ is: 
$$BR_{P}(\beta_{P}, \Theta) := \arg \max_{M \in \mathcal{M}} U_{P}(M \mid \beta_{P}, \Theta).$$ Since the optimization problem is finite, the best-response correspondence is non-empty valued. 

The agent moves twice, once at his first-round information sets and once at his second-round information sets. Thus, we will consider his weak sequential best responses. Given belief system $\beta_A$, define the expected payoff of the agent with type $\theta$ in the game tree $\Theta$ from strategy $(s_{A1}, s_{A2})$ by: 
$$U_A(s_{A1}, s_{A2} \mid \beta_A, \theta, \Theta) := \sum_{s_P^{\Theta} \in S_P^{\Theta}} \beta_A(\theta, \Theta)(\{s_P^{\Theta}\}) \cdot u_A\left(s_{A2}(\theta, \Theta, s_{A1}(\theta, \Theta), s_P^{\Theta}(s_{A1}(\theta, \Theta))), \theta\right).$$ The set of weak sequential best responses of the agent with type $\theta$ in game tree $\Theta$ given the belief system $\beta_A \in B_A$ is: 
$$BR_A(\beta_A, \theta, \Theta) := \left\{ (s_{A1}, s_{A2}) \in S_A : \begin{array}{l} (i) \ \forall M \in \mathcal{M} \\ 
\left(s_{A2}(\theta, \Theta, s_{A1}(\theta, \Theta), M) \in \arg \max_{\bm{c} \in M \cup \{\bm{o}\}}  u_{A}(\bm{c}, \theta)\right); \\ 
(ii) \ (s_{A1}, s_{A2}) \in \arg \max_{s'_A \in S_A} U_A(s'_A \mid \beta_A, \theta, \Theta) 
\end{array} \right\}.$$

According to (i) of the definition, strategy $(s_{A1}, s_{A2}) \in S_A$ is a best response at the second-round information set $(\theta, \Theta, \Theta', M)$ if $s_{A1}(\theta, \Theta) = \Theta'$ implies $u_{A}(s_{A2}(\theta, \Theta, \Theta', M), \theta) \geq u_{A}(\bm{c}, \theta)$ for all $\bm{c} \in M \cup \{\bm{o}\}$. Note that this does not depend on the belief system or the game tree $\Theta$. It does depend, indirectly, on the message disclosed by the agent in the first round because, if his first-round strategy is such that the second-round information set is not reached (e.g., by disclosing any other message than $\Theta'$), then the strategy would be automatically a best response at the unreached second-round information set.

According to (ii) of the definition of $BR_a$, strategy $(s_{A1}, s_{A2}) \in S_A$ is a best response at the first-round information set $(\theta, \Theta)$ to the belief system $\beta_A$ if: 
\begin{eqnarray*} \lefteqn{\sum_{s_P^{\Theta} \in S_P^{\Theta}} \beta_A(\theta, \Theta)(\{s_P^{\Theta}\}) \cdot u_A\left(s_{A2}(\theta, \Theta, s_{A1}(\theta, \Theta), s_P^{\Theta}(s_{A1}(\theta, \Theta))), \theta\right) \geq } \\ & &  \sum_{s_P^{\Theta} \in S_P^{\Theta}} \beta_A(\theta, \Theta)(\{s_P^{\Theta}\}) \cdot u_A\left(s'_{A2}(\theta, \Theta, s'_{A1}(\theta, \Theta), s_P^{\Theta}(s'_{A1}(\theta, \Theta))), \theta\right) 
\end{eqnarray*} for all $(s'_{A1}, s'_{A2}) \in S_A$. 

Again, since the problem is finite, the set of best responses is non-empty. 

We have all ingredients to state our solution concept, a notion of robust rationalizability for dynamic games with unawareness that incorporates restrictions on first-order beliefs. 
Define for each $\beta_P \in B_P$, $$B(\beta_P) := \left\{\beta'_P \in B_P : \forall \Theta \in \mathcal{T} \ \left(\begin{array}{l} \left(\text{supp}(\text{marg}_{\Theta} \beta_P'(\Theta)) = \text{supp}(\text{marg}_{\Theta} \beta_P(\Theta))\right) \mbox{ and } \\ \ \text{marg}_{S_A^{\Theta}} \beta'_P(\Theta) = \text{marg}_{S_A^{\Theta}} \beta_P(\Theta)\end{array} \right) \right\}.$$ The set $B(\beta_P)$ is the subset of the principal's belief systems such that, for each information set, the marginal belief of the principal on the agent's types has the same support as under $\beta_P$, while their marginal belief on the agent's strategies coincide. 

Our solution concepts features a notion of robustness based on open sets of belief systems. For each $\Theta \in \mathcal{T}$, the set $\Theta \times S^{\Theta}_{A}$ is finite, so we can identify $\Delta(\Theta \times S^{\Theta}_{A})$ with the corresponding simplex. We endow both $\Delta(\Theta \times S^{\Theta}_{A})$ with the weak topology (which coincides with the usual relative topology here). Since the projection map is continuous, this implies that for any open set of probability measures $X \subseteq \Delta(\Theta \times S^{\Theta}_{A})$, the set $\{\text{marg}_{\Theta} \ x : x \in X\}$ is open in $\Delta(\Theta)$ w.r.t. the weak topology on $\Delta(\Theta)$. The set $\prod_{\Theta \in \mathcal{T}} \Delta(\Theta \times S_A^{\Theta})$ is endowed with the corresponding product topology, while $B_{P}$ inherits the relative topology. 

\begin{defin}[$\Delta$-O Prudent Rationalizability]\label{EFR} Define recursively the following sets of strategies and belief systems: $R^{0}_{A} := S_{A}$, $R^{0}_{P} := S_{P}$, and for $k = 1, 2, \ldots$,
\begin{align*}
B^{k}_{A} & := \left\{\beta_{A} \in B_{A} : \forall \Theta \in \mathcal{T} \ \forall \theta \in \Theta \  \left(\text{supp}(\beta_{A}(\theta, \Theta)) = R^{k-1, \Theta}_{P}\right) \right\}, \\
R^{k}_{A} & := \left\{(s_{A1},s_{A2}) \in R^{k-1}_{A}: \ \exists \beta_A \in B^{k}_{A} \ \forall \Theta \in \mathcal{T} \ \forall \theta \in \Theta  \  ((s_{A1}, s_{A2}) \in BR_A(\beta_A, \theta, \Theta)) \right\};
\\ 
& \forall \Theta \in \mathcal{T} \ \left( \begin{array}{l} R^{k}_{A}(\Theta) := \left\{(s^{\Theta}_{A1}, s^{\Theta}_{A2}) \in R^{k, \Theta}_{A}: \exists \theta \in \Theta \  (s_{A1}^{\Theta}(\theta, \Theta) = \Theta) \right\} \\ 
\Theta^k := \left\{\theta \in \Theta : \forall (s^\Theta_{A1}, s^\Theta_{A2}) \in R_A^{k, \Theta} \left(s_{A1}^{\Theta}(\theta, \Theta) = \Theta \right)\right\} \end{array}\right);
\\
B^{k}_{P} & := \left\{\beta_{P} \in B_P : \begin{array}{l} \forall \Theta \in \mathcal{T} \left(\left(\Theta^k  \subseteq \text{supp}\left(p_{\Theta}\right)\right) \mbox{ and} \right. \\
\left. \left((R^{k-1}_{A}(\Theta) \neq \emptyset) \ \Rightarrow  \ \left( \beta_{P}(\Theta)\left(\bigcup_{s^{\Theta}_{A} \in R^{k-1}_{A}(\Theta)} \left((s^{\Theta}_{A1}(\Theta))^{-1}(\Theta) \times \left\{s^{\Theta}_{A}\right\}\right) \right) = 1 \right) \right) \right)
\end{array} \right\},
\\
R^{k}_{P} & := \left\{s_{P} \in R^{k-1}_{P} : \exists \beta_P \in B_P^k \ \exists \mbox{nonempty open }  O \subseteq B(\beta_P) \cap B^{k}_{P} \ \forall \Theta \in \mathcal{T} \ \left(s_{P}(\Theta) \in \bigcap_{\beta'_{P} \in O} BR_{P}(\beta'_{P}, \Theta)\right) \right\}.
\end{align*}
The set of $\Delta$-O prudent rationalizable strategies for the principal and the agent, respectively, are given by:
$$R^{\infty}_{P} := \bigcap^{\infty}_{k=1} R^{k}_{P} \ \mbox{ and } \ R^{\infty}_{A} := \bigcap^{\infty}_{k=1} R^{k}_{A}.$$
\end{defin}

The set $B^{k}_{A}$ comprises the agent's level-$k$ (first-stage) belief systems that have full support over level-$(k-1)$ $\Delta$-O prudent rationalizable strategies of the principal. The set $R^{k}_{A}$ collects all level-($k-1$) $\Delta$-O prudent rationalizable strategies of the agent for which there exists a full support belief over level-$(k-1)$ $\Delta$-O prudent rationalizable strategies of the principal that makes them a weak sequential best response. Clearly, requiring full support beliefs incorporates an admissibility criterion or cautiousness for the agent. 

For any $\Theta \in \mathcal{T}$, $R^{k}_{A}(\Theta)$ is the set of level-$k$ $\Delta$-O prudent rationalizable strategies of the agent under which some type sends message $\Theta$, while $\Theta^k$ is the set of the agent's types who disclose $\Theta$ under \emph{any} level-$k$ $\Delta$-O prudent rationalizable strategy. The latter set can be empty, as every type may have some other message which is $\Delta$-O prudent rationalizable at level $k$. 

When $\Theta$ can be disclosed with a level-$k$ $\Delta$-O prudent rationalizable strategy of the agent, the level-$(k + 1)$ principal---upon observing $\Theta$---is certain that the agent plays level-$k$ $\Delta$-O prudent rationalizable strategies. To parse the notation of $B^{k}_P$, in particular $(s^{\Theta}_{A1}(\Theta))^{-1}(\Theta)$, recall that the agent's first-round strategy is a function of both his type $\theta$ and the game tree associated with $\Theta$. Thus, $(s^{\Theta}_{A1}(\Theta))^{-1}(\Theta)$ is the subset of types in $\Theta$ in the game tree associated with $\Theta$ who disclose $\Theta$ to the principal under strategy $s_{A1}$. Our principal is allowed to make inferences about the types she is enlightened about and could rule out some types if she reasons that they would not have wanted to raise her awareness in the first place. We require her to be prudent in the sense of not excluding any type whose \emph{only} level-$(k-1)$ $\Delta$-O prudent rationalizable action is to disclose $\Theta$. On top of it, she cannot rule out all newly discovered types either (by property (iv), wariness). 

Finally, $R^{k}_{P}$ is the principal's set of level-$k$ $\Delta$-O prudent rationalizable strategies. To understand the definition of $R^{k}_P$, consider first a simplified version: $$\left\{s_{P} \in R^{k-1}_{P} : \exists \beta_P \in B_P^k \ \forall \Theta \in \mathcal{T} \ \left(s_{P}(\Theta) \in BR_{P}(\beta_{P}, \Theta)\right) \right\}.$$ This is the set of level-$(k-1)$ $\Delta$-O prudent rationalizable strategies of the principal for which there exist some belief system in $B_P^k$ with which the principal's action is a best reply at each of her information sets. The only difference between this definition and the definition of $R^{k}_P$ is that we do not just require this for some belief system, but rather for an open set of belief systems in $B^k_P$ whose marginal on types have the same support. That is, we do not just require optimality with respect to some marginal belief over types, but also regarding slight perturbations of the probabilities on types. We view this as a robustness criterion: Robustness to slight perturbations of the marginal beliefs, without perturbing the support of those beliefs---what is ruled out stays ruled out. 

Replacing properties (iv) and (v) with the stronger requirement that $\text{supp}\left(p_{\Theta}\right) \supseteq \Theta \setminus \Theta_P$ would effectively prevent our principal from ruling out any types. Imagine that the principal is aware of types $\{0.99, 1.99, 2.99\}$ and, for a given message, at some level $k$ she can rule out (for the first time) types $\{1.99, 2.99\}$. Then, she would like to offer the first-best menu for type $\theta_{1} = 0.99$. However, under no rationalizable strategy at level $(k - 1)$ is the principal allowed to offer said contract. As strategy sets are nested across rationalizability levels, the best she can do is offer the menu for all three types but with her beliefs as concentrated as possible on type $\theta_{1} = 0.99$. Even if she is certain that she is facing the latter type, she is forced to award him information rents. We find such a condition too strong and unnatural. 
  
Note that since our solution concept is an iterated elimination procedure on strategies, it yields predictions for every finite level $k$ of mutual (strong) beliefs. This is akin to level-k reasoning in experimental game theory and should prove useful for future experimental tests of our theory.\footnote{See Li and Schipper (2025, 2020) for experimental tests of prudent rationalizability without restrictions in disclosure games.} 


\section{Unawareness of High Cost Types Only\label{high}}

We first consider the case in which the principal is unaware of high cost types only. We begin with an example. 

\subsection{Three Marginal Cost Types\label{3_high}}

We characterize $\Delta$-O prudent rationalizable strategies level-by-level in the example from the previous section depicted in Figure~\ref{gen_game}, where $\bar{\Theta} = \{\theta_{1}, \theta_{2}, \theta_{3}\}$ for $\theta_{1} = 0.99$, $\theta_{2} = 1.99$, and $\theta_{3} = 2.99$, with the principal being unaware of the highest-possible cost: $\Theta_{P} = \{\theta_{1}, \theta_{2}\}$. To satisfy the assumption $\gamma > b$, we must have $100 > b$.

\bigskip 

\noindent {\bf Level 1, principal:} At every information set, any menu of contracts is a best response to the belief that the agent takes the outside option anyway. Thus, $R_{P}^{1} = S_P$.

\noindent {\bf Level 1, agent:} When faced with a menu of contracts in his second stage, every type of the agent selects a contract that maximizes his payoff unless they all yield negative profit for him, in which case he selects the outside option. Notice that this implies that the agent behaves in accordance with his incentive and participation constraints. 

At his first stage, both actions of raising the principal's awareness or keeping her in the dark about $\theta_{3}$ are a weak sequential best response, depending on what the agent believes about the menus offered by the principal thereafter. At this first level of the rationalizability procedure, the beliefs of the agent are unconstrained. This holds for any type of the agent in both trees. Thus, 
$$R_A^1 = \left\{ (s_{A1}, s_{A2}) \in S_A : \begin{array}{l} (i) \quad \forall \theta \in \{\theta_1, \theta_2\} \ (s_{A1}(\theta, \{\theta_1, \theta_2\}) = \{\theta_1, \theta_2\}) \\
(ii) \  \  \forall \theta \in \{\theta_1, \theta_2, \theta_3\} \ (s_{A1}(\theta, \{\theta_1, \theta_2, \theta_3\}) \in \{\{\theta_1, \theta_2\}, \{\theta_1, \theta_2, \theta_3\}\}) \\
(iii) \ \forall \Theta \in \{\{\theta_1, \theta_2\}, \{\theta_1, \theta_2, \theta_3\}\} \ \forall \theta \in \Theta \ \forall M \in \mathcal{M} \\
\quad \quad (s_{A2}(\theta, \Theta, s_{A1}(\theta, \Theta), M) \in \arg \max_{\bm{c} \in M \cup \{\bm{o}\}} u_A(\bm{c}, \theta))
\end{array}\right\}.$$  

\bigskip 

\noindent {\bf Level 2, principal:} At level 2, the principal is certain of first-level $\Delta$-O prudent rationalizable strategies of the agent, $R_A^1$. Thus, she knows from $(iii)$ in $R_A^1$ that the agent observes the standard incentive and participation constraints when choosing a contract from a menu. Moreover, according to $(i)$ and $(ii)$ of the definition $R_A^1$, all types of the agent may or may not disclose. As shown in Figure~\ref{gen_game}, the principal has two information sets, one in which she remains unaware of $\theta_3$ and one at which she is aware of $\theta_3$. We consider both cases:

{\bf Information set $\Theta_P = \{\theta_1, \theta_2\}$:} Pick any $\beta_{P} \in B^{2}_{P}$ and recall the notation $p_{\Theta_{P}}=\text{marg}_{\Theta_{P}}\beta_{P}(\Theta_{P})$; since all types in $\Theta_P$ must disclose $\Theta_P$, this marginal belief must have full support on $\Theta_P$. With any such $\beta_{P}(\Theta_{P})$, the principal is certain that any type of the agent she can conceive of picks a best response from any menu of contracts offered and thus will observe incentive compatibility and participation constraints. Assume for the moment that the principal considers only menus of at most two contracts---we will consider larger menus afterward. The principal's optimal menu of contracts solves the constrained optimization problem: 
$$\max_{\left(q^\theta_{\Theta_{P}}, t^\theta_{\Theta_{P}}\right)_{\theta \in \{\theta_{1},\theta_{2}\}} \in \{0, 1, ..., b\}^{4}} p_{\Theta_{P}}(\theta_{1})\left(v\left(q_{\Theta_{P}}^{ \theta_{1}}\right) - t^{ \theta_{1}}_{\Theta_{P}}\right) + p_{\Theta_{P}}(\theta_{2})\left(v\left(q_{\Theta_{P}}^{ \theta_{2}}\right) - t^{ \theta_{2}}_{\Theta_{P}}\right)$$ subject to the incentive compatibility constraints: 
\begin{itemize} 
\item[] IC$^{\Theta_{P}}_{1, 2}$: $$t^{ \theta_{1}}_{\Theta_{P}} - \theta_{1}q^{ \theta_{1}}_{\Theta_{P}} \geq t^{ \theta_{2}}_{\Theta_{P}} - \theta_{1}q^{ \theta_{2}}_{\Theta_{P}},$$
\item[] IC$^{\Theta_{P}}_{2, 1}$: $$t^{ \theta_{2}}_{\Theta_{P}} - \theta_{2}q^{ \theta_{2}}_{\Theta_{P}} \geq t^{ \theta_{1}}_{\Theta_{P}} - \theta_{2}q^{ \theta_{1}}_{\Theta_{P}};$$
\end{itemize} and the participation constraints:
\begin{itemize} 
\item[] PC$^{\Theta_{P}}_{1}$: $$t^{ \theta_{1}}_{\Theta_{P}} - \theta_{1}q^{ \theta_{1}}_{\Theta_{P}} \geq 0,$$ 
\item[] PC$^{\Theta_{P}}_{2}$: $$t^{ \theta_{2}}_{\Theta_{P}} - \theta_{2}q^{ \theta_{2}}_{\Theta_{P}} \geq 0.$$ 
\end{itemize} 
As usual, the two IC constraints yield the following weak monotonicity condition: $q^{ \theta_{1}}_{\Theta_{P}}\geq q^{ \theta_{2}}_{\Theta_{P}}$.

This screening problem looks deceptively standard, but it is far from it. It is an integer programming problem, since not only types but also quantities and transfers are discrete. It is not even immediately clear that we can use the standard tricks of reducing the number of constraints, and we can certainly not rely on the familiar continuous calculus-based first-order approach. 

Francetich and Schipper (2025) show how the structure of our type space allows us to replicate many of the usual constraint simplification results. The participation constraint PC$^{\Theta_{P}}_{2}$ ``binds'' in the sense that: 
$$t^{ \theta_{2}}_{\Theta_{P}} = \left\lceil \theta_{2} q^{ \theta_{2}}_{\Theta_{P}} \right\rceil = \left\lceil \theta_{2}\right\rceil q^{ \theta_{2}}_{\Theta_{P}} = 2 q^{ \theta_{2}}_{\Theta_{P}},$$
 where $\lceil \cdot \rceil$ is the usual ceiling function defined by $\lceil x \rceil := \min(\{n \in \mathbb{Z} : x \leq n \})$, and the incentive compatibility constraint IC$^{\Theta_{P}}_{1, 2}$ also ``binds'' in the following sense:
\begin{align*}
t^{ \theta_{1}}_{\Theta_{P}} = t^{ \theta_{2}}_{\Theta_{P}} + \left\lceil \theta_{1} (q^{ \theta_{1}}_{\Theta_{P}} - q^{ \theta_{2}}_{\Theta_{P}}) \right\rceil = t^{ \theta_{2}}_{\Theta_{P}} + \left\lceil \theta_{1} \right\rceil (q^{ \theta_{1}}_{\Theta_{P}} - q^{ \theta_{2}}_{\Theta_{P}}) = t^{ \theta_{2}}_{\Theta_{P}} + q^{ \theta_{1}}_{\Theta_{P}} - q^{ \theta_{2}}_{\Theta_{P}} = q^{ \theta_{1}}_{\Theta_{P}} + q^{ \theta_{2}}_{\Theta_{P}}.
\end{align*}
Notice that it is generally not true that $\lceil\alpha n\rceil = \lceil\alpha\rceil n$ for an integer $n$ and a positive number $\alpha$. For example, $\lceil 2.5 \cdot 2 \rceil = 5 < 6 = \lceil 2.5 \rceil 2$. It is true, however, under our type structure (Francetich and Schipper, 2025, Lemma 4).

Notice that PC$^{\Theta_{P}}_{1}$ is implied by PC$^{\Theta_{P}}_{2}$ and IC$^{\Theta_{P}}_{1, 2}$. Moreover, under weak monotonicity, the binding IC$^{\Theta_{P}}_{1, 2}$ implies IC$^{\Theta_{P}}_{2, 1}$. Thus, like in the standard problem, the structure of our problem allows us to reduce the number of constraints to only the local upward incentive constraints and the participation constraint for the least-efficient type.  

Substituting the remaining constraints into the principal's expected payoff function yields the unconstrained discrete maximization problem: 
$$\max_{\left(q^\theta_{\Theta_{P}}, t^\theta_{\Theta_{P}}\right)_{\theta \in \{\theta_{1},\theta_{2}\}} \in \{0, 1, ..., b\}^{4}} p_{\Theta_{P}}(\theta_{1})\left(v\left(q_{\Theta_{P}}^{ \theta_{1}}\right) - q^{ \theta_{1}}_{\Theta_{P}} - q^{ \theta_{2}}_{\Theta_{P}}\right) + p_{\Theta_{P}}(\theta_{2})\left(v\left(q_{\Theta_{P}}^{ \theta_{2}}\right) - 2 q^{ \theta_{2}}_{\Theta_{P}}\right).$$ 
The discrete derivatives allow us to mimic the usual first-order approach. Yet, instead of two conditions, we now get four first-order conditions, involving backward and forward (discrete) derivatives. Denoting optimal quantities by $\hat{q}^{ \theta_{1}}_{\Theta_{P}}$ and $\hat{q}^{ \theta_{2}}_{\Theta_{P}}$, respectively, we can write the combined first-order conditions as:
$$\Delta^{+} v\left(\hat{q}^{ \theta_{1}}_{\Theta_{P}}\right) \leq 1 \leq \Delta^{-} v\left(\hat{q}^{\theta_{1}}_{\Theta_{P}}\right) ;  \quad \Delta^{+} v\left(\hat{q}^{\theta_{2}}_{\Theta_{P}}\right) \leq 2 + \frac{p_{\Theta_{P}}(\theta_{1})}{p_{\Theta_{P}}(\theta_{2})} \leq \Delta^{-} v\left(\hat{q}^{\theta_{2}}_{\Theta_{P}} \right).$$

Showing that the first-order conditions imply (strict) monotonicity in the discrete screening problem is a bit more involved than in standard screening problems. It requires not only discrete strict concavity (Assumption~\ref{properties_v}.\ref{sdc}), but also Assumption~\ref{properties_v}.\ref{not_too_sdc}. From discrete strict concavity of $v$, we know from Francetich and Schipper (2025, Lemma A7 (iii)) that: $$\hat{q}^{ \theta_{1}}_{\Theta_{P}} > \hat{q}^{ \theta_{2}}_{\Theta_{P}} \quad \Leftrightarrow \quad \left(\Delta^+ v\left(\hat{q}^{ \theta_{1}}_{\Theta_{P}}\right) < \Delta^- v\left(\hat{q}^{ \theta_{2}}_{\Theta_{P}}\right) \quad \mbox{ and } \quad  \hat{q}^{ \theta_{1}}_{\Theta_{P}} \neq \hat{q}^{ \theta_{2}}_{\Theta_{P}}\right).$$ Combined with the first-order conditions, we get: $$\Delta^+ v\left(\hat{q}^{ \theta_{1}}_{\Theta_{P}}\right) - \Delta^- v\left(\hat{q}^{ \theta_{2}}_{\Theta_{P}}\right) < -1 - \frac{p_{\Theta_{P}}(\theta_{1})}{p_{\Theta_{P}}(\theta_{2})}.$$ 
Assumption~\ref{properties_v}.\ref{not_too_sdc} is equivalent to $\Delta^+ v(q) - \Delta^- v(q) \geq -1$. Suppose now to the contrary that $\hat{q}^{ \theta_{1}}_{\Theta_{P}} = \hat{q}^{ \theta_{2}}_{\Theta_{P}}$. Then, $\Delta^+ v(\hat{q}^{ \theta_{1}}_{\Theta_{P}}) - \Delta^- v(\hat{q}^{ \theta_{2}}_{\Theta_{P}}) \geq -1$, a contradiction. We conclude that $\hat{q}^{ \theta_{1}}_{\Theta_{P}} > \hat{q}^{ \theta_{2}}_{\Theta_{P}}$. 

So far, we glossed over the fact that, due to discreteness, the optimizers may not be unique \textit{even under discrete strict concavity} (Francetich and Schipper, 2025). However, there can be at most two optimal quantities for each type, and they must be adjacent to each other on the grid of quantities. Our assumptions and solution concept rule out non-uniqueness. 

Imagine that both $q$ and $q+1$ are optimizers for type $\theta_{1}$; then we would have: 
\begin{eqnarray*}  \Delta^{+} v\left(q \right) \leq & 1 & \leq \Delta^{-} v\left(q\right) \\
\Delta^{+} v\left(q + 1 \right) \leq & 1 & \leq \Delta^{-} v\left(q + 1\right) ,
\end{eqnarray*} which implies: 
\begin{eqnarray*} \Delta^{+} v\left(q \right) \leq & 1 & \leq \Delta^{-} v\left(q + 1\right).  
\end{eqnarray*} However, by the definition of discrete derivatives, $$\Delta^{+} v\left(q \right)  = \Delta^{-} v\left(q + 1\right).$$ Thus, we must have $\Delta^{+} v\left(q \right) = 1$, which is ruled out by Assumption~\ref{properties_v}.\ref{efficient_unique}.

For $\theta_{2}$, non-uniqueness occurs when:  $$\Delta^{+} v\left(\hat{q}^{\theta_{2}}_{\Theta_{P}}\right) = 2 + \frac{p_{\Theta_{P}}(\theta_{1})}{p_{\Theta_{P}}(\theta_{2})}.$$ However, for every $s_P \in R_P^2$, there must exist a nonempty open set of belief systems $O \subseteq B_P(\beta_P) \cap B_P^2$ such that $s_P(\Theta_P) \in \bigcap_{\beta_P \in O} BR_P(\beta_P, \Theta_P)$. A menu with non-unique optimal contracts for $\theta_{2}$ cannot be a best response for beliefs whose marginals are nearby $p_{\Theta_P}$, because perturbing $p_{\Theta_P}$ slightly would break the equality above. Here we see the power of $\Delta$-O prudent rationalizability involving robustness of best responses to open sets of beliefs. 

This does not imply that the principal has a unique best reply even to beliefs whose marginal on types yields a unique optimal menu of contracts, however. The reason is that at level-2, the principal is certain that the agent only selects his best responses. Thus, it does not hurt the principal to offer additional ``irrelevant'' contracts, namely contracts that no type would choose. Denoting by $M$ the menu of uniquely optimal contracts for types in $\Theta_P$ for some $\beta_P \in B_P^2$ with $p_{\Theta_{P}}$ being log-concave and full support, we have $s_P(\Theta_P) \supseteq M$ for any $s_P \in R_P^2$. 

{\bf Information set $\bar{\Theta} = \{\theta_{1}, \theta_{2}, \theta_{3}\}$:} Pick any $\beta_{P}(\Theta_{P})\in B^{2}_{P}$ and take the corresponding $p_{\bar{\Theta}} \in \Delta(\bar{\Theta})$. Since $R^1_A(\bar{\Theta}) \neq \emptyset$, this marginal belief $p_{\bar{\Theta}}$ can have the following supports: $\{\theta_1, \theta_2, \theta_3\} = \bar{\Theta}$, $\{\theta_2, \theta_3\}$, and $\{\theta_3\}$. Support $\{\theta_1, \theta_3\}$ is ruled out by log-concavity, while wariness (iv) rules out $\{\theta_1, \theta_2\} = \Theta_P$, $\{\theta_1\}$, and $\{\theta_2\}$. 

{\bf Support $\{\theta_1, \theta_2, \theta_3\}$:} In Francetich and Schipper (2025), we show how to generalize the discrete screening problem in the two-type example to any finite number of types. In particular, using our assumption on the grids of types and quantities, we can reduce the number of constraints to just the participation constraint for the highest type and the local upward incentive compatibility constraints. The discrete first-order conditions are now: 
$$\Delta^{+} v\left(\hat{q}^{\theta_{1}}_{\bar{\Theta}}\right) \leq 1 \leq \Delta^{-} v\left(\hat{q}^{\theta_{1}}_{\bar{\Theta}}\right); \quad \Delta^{+} v\left(\hat{q}^{\theta_{2}}_{\bar{\Theta}}\right) \leq 2 + \frac{p_{\bar{\Theta}}(\theta_{1})}{p_{\bar{\Theta}}(\theta_{2})} \leq \Delta^{-} v\left(\hat{q}^{\theta_{2}}_{\bar{\Theta}}\right);$$
$$\Delta^{+} v\left(\hat{q}^{\theta_{3}}_{\bar{\Theta}}\right) \leq 3 + \frac{p_{\bar{\Theta}}(\theta_{1}) + p_{\bar{\Theta}}(\theta_{2})}{p_{\bar{\Theta}}(\theta_{3})}\leq \Delta^{-} v\left(\hat{q}^{\theta_{3}}_{\bar{\Theta}}\right).$$

Showing monotonicity makes now use of log-concavity of the marginal beliefs $p_{\bar{\Theta}}$ (see Francetich and Schipper, 2025, Proposition 2) in addition to Assumptions~\ref{properties_v}.\ref{sdc} and~\ref{properties_v}.\ref{not_too_sdc}. Non-uniqueness of contracts is ruled out by the requirement of the menu of optimal contracts being a best response to an open set of marginal beliefs with full support on $\bar{\Theta}$. Besides the menu of optimal contracts, any best response of the principal may still include ``redundant'' contracts that are not a best response for any type in $\bar{\Theta}$. 

We now compare the quantities and thus menus of contracts across information sets. Consider any $s_P \in R_P^2$ for which there exists $\beta_P \in B_P^2$ and an open set $O \subseteq B_P(\beta_P) \cap B_P^2$ such that $s_P(\bar{\Theta}) \in \bigcap_{\beta'_P \in O} BR_P(\beta'_P, \bar{\Theta})$ and $s_P(\Theta_P) \in \bigcap_{\beta'_P \in O} BR_P(\beta'_P, \Theta_P)$; recall that $s_P(\bar{\Theta}) \supseteq M$ and $s_P(\Theta_P) \supseteq M'$, where $M$ and $M'$ are the menus of uniquely optimal contracts for $p'_{\bar{\Theta}}$ and $p'_{\Theta_P}$, respectively. Reverse Bayesianism allows us to relate the menus $M$ and $M'$. Both must include contracts for types $\theta_1$ and $\theta_2$ that satisfy the same discrete first-order conditions. Type $\theta_{1}$ is awarded the first-best quantity, which is independent of beliefs. For type $\theta_2$, note that: $$\Delta^{+} v\left(\hat{q}^{\theta_{2}}_{\Theta_{P}}\right) \leq 2 + \frac{p_{\Theta_{P}}(\theta_{1})}{p_{\Theta_{P}}(\theta_{2})} \leq \Delta^{-} v\left(\hat{q}^{\theta_{2}}_{\Theta_{P}} \right); \quad \Delta^{+} v\left(\hat{q}^{\theta_{2}}_{\bar{\Theta}}\right) \leq 2 + \frac{p_{\bar{\Theta}}(\theta_{1})}{p_{\bar{\Theta}}(\theta_{2})} \leq \Delta^{-} v\left(\hat{q}^{\theta_{2}}_{\bar{\Theta}}\right)$$ must be equivalent because reverse Bayesianism implies: $$\frac{p_{\Theta_{P}}(\theta_{1})}{p_{\Theta_{P}}(\theta_{2})} = \frac{p_{\bar{\Theta}}(\theta_{1})}{p_{\bar{\Theta}}(\theta_{2})}.$$ Since the menus of optimal contracts $M$ and $M'$ for $\bar{\Theta}$ and $\Theta_P$, respectively, are unique, the discrete first-order conditions imply now for the optimal contract quantities $\hat{q}_{\bar{\Theta}}^{\theta_1} = \hat{q}_{\Theta_P}^{\theta_1}$ and $\hat{q}_{\bar{\Theta}}^{\theta_2} = \hat{q}_{\Theta_P}^{\theta_2}$. 


{\bf Support $\{\theta_2, \theta_3\}$:} The discrete first-order conditions are now: 
$$\Delta^{+} v\left(\hat{q}^{ \theta_{2}}_{\bar{\Theta}}\right) \leq 2 \leq \Delta^{-} v\left(\hat{q}^{\theta_{2}}_{\bar{\Theta}}\right) ;  \quad \Delta^{+} v\left(\hat{q}^{\theta_{3}}_{\bar{\Theta}}\right) \leq 3 + \frac{p_{\bar{\Theta}}(\theta_{2})}{p_{\bar{\Theta}}(\theta_{3})} \leq \Delta^{-} v\left(\hat{q}^{\theta_{3}}_{\bar{\Theta}} \right).$$ 

To compare the quantities of type $\theta_2$ across awareness levels, we use the discrete first-order conditions to obtain: 
$$\Delta^{+} v\left(\hat{q}^{\theta_{2}}_{\bar{\Theta}}\right) \leq 2 < 2 + \frac{p_{\Theta_P}(\theta_{1})}{p_{\Theta_P}(\theta_{2})} \leq \Delta^{-} v\left(\hat{q}^{\theta_{2}}_{\Theta_P}\right).$$ By discrete strict concavity of $v$, $\Delta^{+} v\left(\hat{q}^{\theta_{2}}_{\bar{\Theta}}\right) < \Delta^{-} v\left(\hat{q}^{\theta_{2}}_{\Theta_P}\right)$ implies $\hat{q}^{\theta_{2}}_{\bar{\Theta}} \geq \hat{q}^{\theta_{2}}_{\Theta_P}$ (see Francetich and Schipper, 2025, Lemma A7 (iii)). So, the quantity offered by the principal to type $\theta_2$ upon becoming aware of $\theta_3$ cannot decrease. Note that this is unrelated to reverse Bayesianism, which is moot in this case. When would the principal offer a strictly larger quantity to type $\theta_2$ upon becoming aware of $\theta_3$ in this case? Note that the combined discrete first-order conditions above imply: 
\begin{eqnarray} \Delta^{+} v\left(\hat{q}^{\theta_{2}}_{\bar{\Theta}}\right)  - \Delta^{-} v\left(\hat{q}^{\theta_{2}}_{\Theta_P}\right) & \leq & - \frac{p_{\Theta_P}(\theta_{1})}{p_{\Theta_P}(\theta_{2})}. \label{example_high_2} 
\end{eqnarray} Suppose to the contrary that $\hat{q}^{\theta_{2}}_{\Theta_P} = \hat{q}^{\theta_{2}}_{\bar{\Theta}}$. Then $\Delta^{+} v\left(\hat{q}^{\theta_{2}}_{\bar{\Theta}}\right)  - \Delta^{-} v\left(\hat{q}^{\theta_{2}}_{\Theta_P}\right) = \Delta^+ \Delta^- v\left(\hat{q}^{\theta_{2}}_{\bar{\Theta}}\right) \geq -1$, where the last inequality comes from Assumption~\ref{properties_v}.\ref{not_too_sdc}. We see now from (\ref{example_high_2}) that the last inequality would be violated if $p_{\Theta_P}(\theta_{1}) > p_{\Theta_P}(\theta_{2})$. That is, if the principal assigns a larger probability to $\theta_1$ than to $\theta_2$, then she will offer a strictly larger quantity to $\theta_2$ upon becoming aware of $\theta_3$.

{\bf Support $\{\theta_3\}$:} In such a case, the principal believes that she has nothing to screen for and would simply offer the (by Assumption~\ref{properties_v}.\ref{efficient_unique} unique) first-best contract $$\Delta^{+} v\left(\hat{q}^{\theta_{3}}_{\theta_3} \right) \leq 3 \leq \Delta^{-} v\left(\hat{q}^{\theta_{3}}_{\theta_3} \right).$$ Such a contract would satisfy the participation constraint of type $\theta_3$ because of the ``rounding'' rent, $\frac{1}{100} \hat{q}^{\theta_3}_{\theta_3}$. The principal may include further ``redundant'' contracts into the menu of contracts offered as long as these contracts are not a best response for type $\theta_3$.



\noindent {\bf Level 2, agent:} At level 2 of the procedure, the agent is certain that the principal follows a strategy in $R_P^1$ at every one of her information sets. This imposes no restrictions on his belief of whether or not he is to be presented with a better menu of contracts upon raising the principal's awareness of $\theta_3$. Moreover, no matter what menu he is presented with by the principal, he chooses a best response. Thus, $R_A^2 = R_A^1$. 

\bigskip 

\noindent {\bf Level 3, principal:} Since $R_A^2 = R_A^1$ we have $B_P^3 = B_P^2$. Thus, $R_P^3 = R_P^2$. 

\noindent {\bf Level 3, agent:} At level 3, for both game trees $\Theta \in \{\Theta_P, \bar{\Theta}\}$, any type $\theta \in \Theta$ is certain of the principal's strategies in $R_P^{2, \Theta}$. In game tree $\Theta_P$, the only disclosure action of any type $\theta \in \Theta_P$ is $s_{A1}(\theta, \Theta_P) = \Theta_P$. Thus, we focus on the disclosure actions of types in game tree $\bar{\Theta}$. 

{\bf Type $\theta_3$:}  We claim that the only weak sequential best response for type $\theta_3$ at game tree $\bar{\Theta}$ is to raise awareness of himself, $s_{A1}(\theta_3, \bar{\Theta}) = \bar{\Theta}$. Without disclosure, his weak sequential best response is to take the outside option, $\bm{o}$, earning him zero profit. This is because he would make strictly negative profit with any contract offered by the principal. Notice that this applies not just to any the contracts optimal for $\theta_1$ and $\theta_2$ offered by the principal, but also to any other possible ``redundant'' contracts. Since those redundant contracts are not a best response by $\theta_2$, they cannot be a best response for the even higher cost type $\theta_3$ either. With disclosure, type $\theta_3$ is ``held to the profit of his outside option'' due to his binding participation constraint but earns a \textit{round-up rent} given by $\lceil \theta_3 \rceil \hat{q}^{\theta_3}_{\bar{\Theta}} - \theta_3 \hat{q}^{\theta_3}_{\bar{\Theta}}  = \frac{1}{100} \hat{q}^{\theta_3}_{\bar{\Theta}}$. These arguments hold for any belief system of the agent over level-3 $\Delta$-O prudent rationalizable strategies of the principal. 

{\bf Type $\theta_2$:} The agent forms full support beliefs over the principal's strategies in $R_P^2$. These include strategies that are best responses for the principal, at his information set $\bar{\Theta}$, to beliefs with various supports. We discuss the cases in turn: 

{\bf Support $\bar{\Theta}$:} If the principal plays $s_P(\bar{\Theta}) \in R_P^2$ that is optimal under a marginal belief with support $\bar{\Theta}$, then the contract offered to $\theta_2$ satisfies $\hat{q}_{\bar{\Theta}}^{\theta_2} = \hat{q}_{\Theta_P}^{\theta_2}$. Yet, $\hat{t}_{\bar{\Theta}}^{\theta_2} > \hat{t}_{\Theta_P}^{\theta_2}$ because $\theta_2$ just earns his round-up rent from being held to his outside option when not disclosing, while earning additional information rents from the presence of $\theta_3$ when disclosing $\theta_3$: $$u_A(\hat{q}_{\bar{\Theta}}^{\theta_2}, \hat{t}_{\bar{\Theta}}^{\theta_2}, \theta_2) = \hat{q}_{\bar{\Theta}}^{\theta_3} + \frac{1}{100} \hat{q}_{\bar{\Theta}}^{\theta_2} > \frac{1}{100} \hat{q}_{\Theta_P}^{\theta_2} = u_A(\hat{q}_{\Theta_P}^{\theta_2}, \hat{t}_{\Theta_P}^{\theta_2}, \theta_2).$$ 


{\bf Support $\{\theta_2, \theta_3\}$:} If the principal plays $s_P(\bar{\Theta}) \in R_P^2$ that is optimal under a marginal belief with support $\{\theta_2, \theta_3\}$, then the contract offered to $\theta_2$ satisfies $\hat{q}_{\bar{\Theta}}^{\theta_2} \geq \hat{q}_{\Theta_P}^{\theta_2}$, with strict inequality for some of the principal's marginal beliefs. For transfers,  $\hat{t}_{\bar{\Theta}}^{\theta_2} > \hat{t}_{\Theta_P}^{\theta_2}$ because $\theta_2$ just earns his round-up rent from being held to his outside option when not disclosing, while earning additional information rents from the presence of $\theta_3$ when disclosing $\theta_3$: $$u_A(\hat{q}_{\bar{\Theta}}^{\theta_2}, \hat{t}_{\bar{\Theta}}^{\theta_2}, \theta_2) = \hat{q}_{\bar{\Theta}}^{\theta_3} + \frac{1}{100} \hat{q}_{\bar{\Theta}}^{\theta_2} > \frac{1}{100} \hat{q}_{\Theta_P}^{\theta_2} = u_A(\hat{q}_{\Theta_P}^{\theta_2}, \hat{t}_{\Theta_P}^{\theta_2}, \theta_2).$$ 

{\bf Support $\{\theta_3\}$:} If the principal plays $s_P(\bar{\Theta}) \in R_P^2$ that is optimal given a marginal belief with support $\{\theta_3\}$, then $\theta_2$ when taking the contract for $\theta_3$ after disclosure earns a larger payoff than from the round-up rent from not disclosing. We have: $$u_A(\hat{q}_{\bar{\Theta}}^{\theta_3}, \hat{t}_{\bar{\Theta}}^{\theta_3}, \theta_2) = \hat{q}_{\bar{\Theta}}^{\theta_3} + \frac{1}{100} \hat{q}_{\bar{\Theta}}^{\theta_3} > \frac{1}{100} \hat{q}_{\Theta_P}^{\theta_2} = u_A(\hat{q}_{\Theta_P}^{\theta_2}, \hat{t}_{\Theta_P}^{\theta_2}, \theta_2).$$ Although we would expect $\hat{q}_{\bar{\Theta}}^{\theta_3} \leq \hat{q}_{\Theta_P}^{\theta_2}$ in this case, the assumption that $100 > b$ implies the inequality for any non-zero $\hat{q}_{\bar{\Theta}}^{\theta_3}$. 



We conclude that in all cases it is a weak sequential best response for type $\theta_2$ to raise the principal's awareness of $\theta_3$. Since the agent has a full support belief over the principal's level-2 $\Delta$-O prudent rationalizable strategies, only disclosure at level-3 is $\Delta$-O prudent rationalizable for the $\theta_2$ in game tree $\bar{\Theta}$. 

{\bf Type $\theta_1$:} Again, we discuss each support case in turn: 

{\bf Support $\bar{\Theta}$:} If the principal plays $s_P(\bar{\Theta}) \in R_P^2$ that is optimal given a marginal belief with support $\bar{\Theta}$, then the contract offered to $\theta_1$ satisfies $\hat{q}_{\bar{\Theta}}^{\theta_1} = \hat{q}_{\Theta_P}^{\theta_1}$. Yet, $\hat{t}_{\bar{\Theta}}^{\theta_1} > \hat{t}_{\Theta_P}^{\theta_2}$ because $\theta_1$ just earns additional information rents from the presence of $\theta_3$ when disclosing $\theta_3$: $$u_A(\hat{q}_{\bar{\Theta}}^{\theta_1}, \hat{t}_{\bar{\Theta}}^{\theta_1}, \theta_1) = \hat{q}_{\bar{\Theta}}^{\theta_3} + \hat{q}_{\bar{\Theta}}^{\theta_2} + \frac{1}{100} \hat{q}_{\bar{\Theta}}^{\theta_1} > \hat{q}_{\bar{\Theta}}^{\theta_2} + \frac{1}{100} \hat{q}_{\Theta_P}^{\theta_1} = u_A(\hat{q}_{\Theta_P}^{\theta_1}, \hat{t}_{\Theta_P}^{\theta_1}, \theta_1).$$ 


{\bf Support $\{\theta_2, \theta_3\}$:} If the principal plays $s_P(\bar{\Theta}) \in R_P^2$ that is optimal given a marginal belief with support $\{\theta_2, \theta_3\}$, then $\theta_1$ when taking the contract for $\theta_2$ after disclosure earns a larger payoff than from not disclosing. We have: $$u_A(\hat{q}_{\bar{\Theta}}^{\theta_2}, \hat{t}_{\bar{\Theta}}^{\theta_2}, \theta_1) = \hat{q}_{\bar{\Theta}}^{\theta_3} + \hat{q}_{\bar{\Theta}}^{\theta_2} + \frac{1}{100} \hat{q}_{\bar{\Theta}}^{\theta_2} > \hat{q}_{\Theta_P}^{\theta_2} + \frac{1}{100} \hat{q}_{\Theta_P}^{\theta_1} = u_A(\hat{q}_{\Theta_P}^{\theta_1}, \hat{t}_{\Theta_P}^{\theta_1}, \theta_1).$$ Again, although we would expect $\hat{q}_{\bar{\Theta}}^{\theta_2} \leq \hat{q}_{\Theta_P}^{\theta_1}$ in this case, the assumption that $100 > b$ and our earlier observation of $\hat{q}_{\bar{\Theta}}^{\theta_2} \geq \hat{q}_{\Theta_P}^{\theta_2}$ (with strict inequality for some marginal beliefs of the principal) imply the inequality for any non-zero $\hat{q}_{\bar{\Theta}}^{\theta_3}$. 

{\bf Support $\{\theta_3\}$:} If the principal plays $s_P(\bar{\Theta}) \in R_P^2$ that is optimal given a marginal belief with support $\{\theta_3\}$, then $\theta_1$ taking the contract for $\theta_3$ after disclosure earns a larger payoff than from non-disclosure if $\hat{q}_{\bar{\Theta}}^{\theta_3} > \frac{1}{2} \hat{q}_{\Theta_P}^{\theta_2}$. This condition, together with $100 > b$ and $\hat{q}_{\bar{\Theta}}^{\theta_3}$ being non-zero, implies: 
$$u_A(\hat{q}_{\bar{\Theta}}^{\theta_3}, \hat{t}_{\bar{\Theta}}^{\theta_3}, \theta_1) = 2 \hat{q}_{\bar{\Theta}}^{\theta_3} + \frac{1}{100} \hat{q}_{\bar{\Theta}}^{\theta_3} > \hat{q}_{\Theta_P}^{\theta_2} + \frac{1}{100} \hat{q}_{\Theta_P}^{\theta_1} = u_A(\hat{q}_{\Theta_P}^{\theta_1}, \hat{t}_{\Theta_P}^{\theta_1}, \theta_1).$$ Moreover, there might be ``redundant'' contracts in the menu offered by the principal in this case that afford type $\theta_1$ strict positive payoffs. Yet, disclosure may be worse than non-disclosure for type $\theta_1$ if $\hat{q}_{\bar{\Theta}}^{\theta_3} < \frac{1}{2} \hat{q}_{\Theta_P}^{\theta_2}$ and none of the ``redundant'' contracts are attractive to him.

We conclude that both disclosure of $\theta_3$ and non-disclosure of $\theta_3$ are weak sequential best responses for type $\theta_1$ in game tree $\bar{\Theta}$ depending on his full support belief over $R_P^2$. 

Overall, we conclude that $R_A^3 := \{(s_{A1}, s_{A2}) \in R_A^2 : s_{A1}(\theta_2, \bar{\Theta}) = s_{A1}(\theta_3, \bar{\Theta}) = \bar{\Theta}\}$. 

\bigskip 

\noindent {\bf Level 4, principal:} At level 4, the principal is certain of level-3 $\Delta$-O prudent rationalizable strategies of the agent. Any type may disclose, although disclosure is the only action in level-3 $\Delta$-O prudent rationalizable strategies for types $\theta_3$ and $\theta_2$. Thus, both types must be in the support of her marginal belief upon disclosure of $\bar{\Theta}$. With such a belief, the case of marginal beliefs with support $\{\theta_3\}$ is ruled out in the analysis of her $\Delta$-O prudent rationalizable strategies at level 2. Thus, $R^4_P \subset R^3_P$.

\noindent {\bf Level 4, agent:} Since $R_P^3 = R_P^2$ we have $B_A^4 = B_A^3$. Thus, $R_A^4 = R_A^3$. 

\bigskip 

\noindent {\bf Level 5, principal:} Since $R_A^4 = R_A^3$ we have $B_P^5 = B_P^4$. Thus, $R_P^5 = R_P^4$. 

\noindent {\bf Level 5, agent:} At level 5, for both game trees $\Theta \in \{\Theta_P, \bar{\Theta}\}$, any type $\theta \in \Theta$ is certain of the principal's strategies in $R_P^{4, \Theta}$. In game tree $\Theta_P$, $s_{A1}(\theta, \Theta_P) = \Theta_P$ for all $\theta \in \Theta_P$. Thus, we focus on the disclosure actions of types in game tree $\bar{\Theta}$. The analysis follows similar arguments as for level 3:

Both types $\theta_2$ and $\theta_3$ find it weak sequentially rational to disclose at their information set, by the same arguments as at level 3. 

Type $\theta_1$ is now certain at his information $(\theta_1, \bar{\Theta})$ that the principal best responds at her information set $\bar{\Theta}$ only to marginal beliefs with supports $\bar{\Theta}$ or $\{\theta_2, \theta_3\}$. For all such strategies of the principal, the only weak sequential best response for type $\theta_1$ is to raise the principal's awareness of $\theta_3$ by disclosing $\bar{\Theta}$. This follows from the arguments for level 3. 

That is, at level 5, the only weak sequential best response for \emph{all} types of the agent is to disclose $\bar{\Theta}$ in the upmost tree $\bar{\Theta}$. 

\bigskip 

\noindent {\bf Level 6, principal:} The principal is now certain of $R^5_A$. Thus, at information set $\bar{\Theta}$, the support of her marginal beliefs over marginal cost types must include all types $\bar{\Theta}$. This further refines strategies, $R_P^6 \subseteq R_P^5$. 

\noindent {\bf Level 6, agent:} Since $R_P^5 = R_P^4$ we have $B_A^6 = B_A^5$. Thus, $R_A^6 = R_A^5$. 

\bigskip

\noindent {\bf Level 7, agent:} Every type $\theta$ is now certain at his information set $(\theta, \bar{\Theta})$ that the principal best responds at her information set $\bar{\Theta}$ only to marginal beliefs with support $\bar{\Theta}$. Again, it is weakly sequentially optimal for any type $\theta$ to raise the principal's awareness of $\theta_3$ by disclosing $\bar{\Theta}$.

\noindent {\bf Level 7, principal:} Since $R_A^6 = R_A^5$ we have $B_P^7 = B_P^6$. Thus, $R_P^7 = R_P^6$. 

\bigskip 

\noindent No further reduction is induced by $\Delta$-O prudent rationalizability at higher levels. We conclude $R^{\infty}_A = R^5_A$ and $R^{\infty}_P = R_P^6$. 

\bigskip 

In any $\Delta$-O prudent rationalizable outcome, in the upmost tree $\bar{\Theta}$, we have that all raise the principal's awareness, and she offers a menu of contracts that includes optimal contracts for all types plus, possibly, some redundant ones that no type would select. 

Note the necessity of the requirement that the support of the principal's marginal belief over types of the agent must include all agents whose only prior level rationalizable action is to disclose. Without such an assumption, upon disclosure of $\bar{\Theta}$, the principal may assign probability 1 to the newly-discovered type $\theta_3$. When type $\theta_1$ believes with sufficiently large probability in a strategy of the principal that is a best response to such a belief of the principal, he might find it optimal not to disclose, upsetting the entire argument.

\subsection{Any Finite Number of Marginal Cost Types\label{m_high}}

The analysis of the example with three types illustrates the main arguments. We now extend it to a setting with an arbitrary number of finite types, summarized in Theorem \ref{m_high_theorem} below. 

\begin{theo}\label{m_high_theorem}  Consider the case when the principal is unaware only of high-cost types: $\min(\Theta_{P})=\min(\bar{\Theta})$ and $\max(\Theta_{P})<\max(\bar{\Theta})$. 
\begin{itemize}
\item[(i)] 
Fix $\Theta\in\mathcal{T}$ and let $M$ be the menu of uniquely-optimal contracts for some log-concave $p \in \Delta(\Theta)$. For any $s_P \in R_P^{\infty}$, $s_P(\Theta) \supseteq M$.
  
\item[(ii)] 
For any $s_P \in R_P^{\infty}$ and $\Theta \in \mathcal{T}(\bar{\Theta})$, consider any $\theta \in \Theta_{P}$ for which there is a contract in menu $s_P(\Theta)$ with quantity $q_{\Theta}^\theta$ and a contract in menu $s_P(\Theta_{P})$ with quantity $q^{\theta}_{\Theta_{P}}$. For any $\beta_{P}\in B_{P}$ such that $\min(\text{supp}(p_{\Theta})) = \min(\text{supp}(p_{\Theta_{P}}))$,\footnote{Recall our notation $p_\Theta := marg_{\Theta} \ \beta_P(\Theta)$.}  we have $q^{\theta}_{\Theta} = q^{\theta}_{\Theta_{P}}$; otherwise,  $q^{\theta}_{\Theta} \geq q^{\theta}_{\Theta_{P}}$, with strict inequality for some $\beta'_{P}\in B_{P}$.

\item[(iii)] 
For any $s_A \in R_A^{\infty}$ and $\Theta \in \mathcal{T}(\bar{\Theta})$, $s_{A1}(\theta, \Theta) = \Theta$ for all $\theta \in \Theta \setminus \Theta_P$. In particular,  $s_{A1}(\theta, \bar{\Theta}) = \bar{\Theta}$ for all $\theta \in \bar{\Theta}\setminus \Theta_P$. 

\end{itemize}
\end{theo}

\noindent \textsc{Proof. } The proof is by induction of the levels of the rationalizability procedure. It relies heavily on discrete concave screening, developed in Francetich and Schipper (2025). 

\bigskip 

\noindent {\bf Level 1, principal:} At every information set, any menu of contracts is a best response to the belief that the agent takes the outside option anyway. Thus, $R_{P}^{1} = S_P$.

\noindent {\bf Level 1, agent:} At any second-stage information set of the agent allowed by his own strategy, $(\theta, \Theta, s_{A1}(\theta, \Theta), M)$ for $\theta \in \Theta, \Theta \in \mathcal{T}, s_{A1}(\theta, \Theta) \in \mathcal{T}(\Theta)$, and $M \in \mathcal{M}$, type $\theta$ in game tree $\Theta$ is certain of the menu $M$ he received from the principal, and his continuation strategy selects a best response in $\arg \max_{\bm{c} \in M \cup \{\bm{o}\}} u_A(\bm{c}, \theta)$. Notice that this implies that the agent behaves in accordance with his incentive compatibility and participation constraints. No restrictions are imposed on second-stage information sets not allowed by his strategy. 

At any of his first-stage information sets, $(\theta, \Theta)$ for $\theta \in \Theta, \Theta \in \mathcal{T}$, any disclosure action $\Theta' \in \mathcal{T}(\Theta)$ is a weak sequential best response to a belief $\beta_A(\theta, \Theta)$ that puts sufficiently large probability on a strategy of the principal that she presents (i) a menu $M'$ with a contract yielding a large profit to the agent when he discloses $\Theta'$, and (ii) a menu $M''$ in which all contracts yield a loss to the agent upon disclosing any other $\Theta'' \in \mathcal{T}(\Theta)$, $\Theta'' \neq \Theta'$. Thus, level-1 imposes no restriction on first-stage actions of the agent. Thus, 
$$R_A^1 = \left\{ (s_{A1}, s_{A2}) \in S_A : \begin{array}{l} 
\ \forall \Theta \in \mathcal{T} \ \forall \theta \in \Theta \ \forall M \in \mathcal{M}  \\
\ (s_{A2}(\theta, \Theta, s_{A1}(\theta, \Theta), M) \in \arg \max_{\bm{c} \in M \cup \{\bm{o}\}} u_A(\bm{c}, \theta))
\end{array}\right\}.$$ 

\bigskip 

\noindent {\bf Level 2, principal:} With any $\beta_P \in B_P^2$ and at any information set $\Theta \in \mathcal{T}$, the principal is certain of $R_A^{1, \Theta}$. Since $R_A^{1, \Theta}$ imposes no restrictions on first-stage disclosure by the agent, she can infer nothing about his types. Yet, she is certain that any type in $\Theta$ observes incentive compatibility and participation constraints when choosing $\bm{c} \in M \cup \{\bm{o}\}$ for any $M \in \mathcal{M}$.  

For any information set $\Theta \in \mathcal{T}$, let $\text{Supp}(\Theta) := \{\text{supp}(\text{marg}_{\Theta} \beta_P(\Theta)):\beta_P \in B_P^2\}$ denote the set of all supports of all marginal distributions for beliefs $\beta_P \in B_P^2$. Note that for any $\theta \in \bar{\Theta}$ with $\min \text{supp}\left(p_{\Theta}\right) \leq \theta \leq \max \text{supp}\left(p_{\Theta}\right)$, we have $\theta \in \text{supp}\left(p_{\Theta}\right)$.  This follows from log-concavity.

For any strategy $s_P \in R_P^2$, there must exist $\beta_P \in B_P^2$ and $O \subseteq B(\beta_P) \cap B_P^2$ such that $s_P(\Theta) \in \bigcap_{\beta'_P \in O} BR_P(\beta_P', \Theta)$ for all $\Theta \in \mathcal{T}$. Since $O \subseteq B(\beta_P)$, we must have $\text{supp}\left(p_{\Theta}\right) = \text{supp}\left(p'_{\Theta}\right) \in \text{Supp}(\Theta)$ for all $\Theta \in \mathcal{T}$. 

Given any $\beta_{P}$, $\Theta\in\mathcal{T}$, and the corresponding marginal $p_{\Theta}$ such that the optimal contract menu is unique, let $\mathcal{M}(p_{\Theta})$ be a menu of contracts consisting of the unique optimal menu and additional contracts that no type in $\Theta$ finds to be a best response. By Francetich and Schipper (2025, Lemma 9), for any $\beta_P \in O$ and $\Theta \in \mathcal{T}$, $\{s_P(\Theta)\} = \mathcal{M}(p_{\Theta})$. In particular, no type in $\text{supp}\left(p_{\Theta}\right)$ is indifferent over contracts in $s_P(\Theta) \cup \{\bm{o}\}$ because the agent's incentive compatibility and participation constraints hold strictly for all types. This proves (i) of the theorem. 

For any two $\Theta, \Theta' \in \mathcal{T}$ with $\Theta' \in \mathcal{T}(\Theta)$ and any type $\theta \in \text{supp}\left(p_{\Theta'}\right) \cap \text{supp}\left(p_{\Theta}\right)$, let $(q^\theta_{\Theta}, t^\theta_{\Theta})$ be the contract picked by type $\theta$ in game tree $\Theta$ from $s_P(\Theta)$ and $(q^\theta_{\Theta'}, t^\theta_{\Theta'})$ be the contract picked by type $\theta$ in game tree $\Theta'$ from $s_P(\Theta')$. 

If $\min \left(\text{supp}(p_{\Theta'})\right) = \min\left(\text{supp}(p_{\Theta})\right)$, then $q^\theta_{\Theta'} = q^\theta_{\Theta}$. This follows from Lemma~\ref{Karni_Viero} (i) in the appendix and Francetich and Schipper (2025, Lemma 9), because $p_{\Theta}$ and $p_{\Theta'}$ satisfy reverse Bayesianism and because optimal contracts in $s_P(\Theta)$ and $s_P(\Theta')$ are unique for each type of the agent, being the principal's best responses to open sets of beliefs. 

If $\min\left(\text{supp}(p_{\Theta'})\right) \neq \min\left(\text{supp}(p_{\Theta})\right)$, we must have $\min\left(\text{supp}(p_{\Theta'})\right) < \min \left(\text{supp}(p_{\Theta})\right)$ by monotone supports (v). Then, by Lemma~\ref{Karni_Viero} (ii) in the appendix and Francetich and Schipper (2025, Lemma 9), we must have $q^\theta_{\Theta} \geq q^\theta_{\Theta'}$ since: (1) all terms included in the sum of the numerator of the inverse hazard rate for $\Theta$ are also included in the sum making up the numerator of the inverse hazard rate for $\Theta'$; (2) $p_{\Theta}$ and $p_{\Theta'}$ satisfy reverse Bayesianism; and (3) the optimal contracts in $s_P(\Theta)$ and $s_P(\Theta')$ are unique. Also by Lemma~\ref{Karni_Viero} (ii) in the appendix, we must have that $q^\theta_{\Theta} > q^\theta_{\Theta'}$ for some belief systems in $B_P^2$ and thus for strategies in $R_P^2$.  

This proves part (ii) of the theorem. 

\noindent {\bf Level 2, agent:} At level 2 of the procedure, the agent with any belief system is certain at every one of her information sets $(\theta, \Theta)$ for $\Theta \in \mathcal{T}$ and $\theta \in \Theta$ that the principal follows a strategy in $R_P^{1, \Theta}$. This imposes no restrictions on his belief of whether or not he is presented with a better menu of contracts upon raising the principal's awareness from $\Theta_P$ to $\Theta' \in \mathcal{T}(\Theta)$. Moreover, no matter what menu he is presented with by the principal at his second-stage information sets allowed by his strategy, he chooses a best response. Thus, $R_A^2 = R_A^1$.

\bigskip 

\noindent {\bf Level 3, principal:} Since $R_A^2 = R_A^1$ we have $B_P^3 = B_P^2$. Thus, $R_P^3 = R_P^2$. 

\noindent {\bf Level 3, agent:} We claim that for all $s_A \in R_A^3$, $s_A(\theta, \Theta) = \Theta$ for all $\Theta \in \mathcal{T}$ and $\theta \in \Theta \setminus \Theta_P$. 

Consider $\beta_A \in B_A^3$. Let $s_P \in \text{supp}(\beta_A)$. Let $s_P$ be sequential rational under $\beta_P$. Further, let $\Theta \in \mathcal{T}$, $\Theta' \in \mathcal{T}(\Theta)$, $\Theta' \subsetneq \Theta$, $\theta \in \Theta$. 

We distinguish five cases: 

\textbf{Case I:} $\theta \in \text{supp}(p_{\Theta}) \cap \text{supp}(p_{\Theta'})$. We show: 
\begin{align} 
u_A(\hat{q}_{\Theta}^{\theta}, \hat{t}_{\Theta}^{\theta}, \theta) & > u_A(\hat{q}_{\Theta'}^{\theta}, \hat{t}_{\Theta'}^{\theta}, \theta)  \nonumber
\\
\hat{t}_{\Theta}^{\theta} - \theta \hat{q}_{\Theta}^{\theta} & > \hat{t}_{\Theta'}^{\theta} - \theta \hat{q}_{\Theta'}^{\theta}  \nonumber
\\
\lceil \theta \rceil \hat{q}_{\Theta}^{\theta} + \sum_{\theta' > \theta,\, \theta' \in \text{supp}(p_{\Theta})} \hat{q}_{\Theta}^{\theta'} - \theta \hat{q}_{\Theta}^{\theta} & > \lceil \theta \rceil \hat{q}_{\Theta'}^{\theta} + \sum_{\theta' > \theta, \,\theta' \in \text{supp}(p_{\Theta'})} \hat{q}_{\Theta'}^{\theta'} - \theta \hat{q}_{\Theta'}^{\theta}  \nonumber
\\ 
\frac{1}{\gamma} (\hat{q}_{\Theta}^{\theta} - \hat{q}_{\Theta'}^{\theta}) & > \sum_{\theta' > \theta, \,\theta' \in \text{supp}(p_{\Theta'})} \hat{q}_{\Theta'}^{\theta'} -  \sum_{\theta' > \theta, \,\theta' \in \text{supp}(p_{\Theta})} \hat{q}_{\Theta}^{\theta'}.\label{summies}
\end{align} 
The l.h.s. of \eqref{summies} is non-negative by Lemma~\ref{Karni_Viero} (ii) in the appendix because any strategy $s_P \in R_P^2$ dictates, at $\Theta$, a sequential best reply to a belief system with $\min\left(\text{supp} (p_{\Theta'})\right) < \min\left(\text{supp} (p_{\Theta})\right)$.  The r.h.s. is strictly smaller zero because: (1) all terms in the first sum have corresponding terms in the second sum, and (2) the terms in the first sum are weakly smaller than their counterparts in the second sum by Lemma~\ref{Karni_Viero} (ii) in the appendix, due to monotone supports.  

\textbf{Case II:} $\theta \in \text{supp}(p_{\Theta})$ and $\theta \notin \text{supp}(p_{\Theta'})$. We show: 
\begin{align*} 
u_A(\hat{q}_{\Theta}^{\theta}, \hat{t}_{\Theta}^{\theta}, \theta) & > u_A(\hat{q}_{\Theta'}^{\max \Theta'}, \hat{t}_{\Theta'}^{\max \Theta'}, \theta)  
\\
\lceil \theta \rceil \hat{q}_{\Theta}^{\theta} + \sum_{\theta' > \theta, \theta' \in \text{supp}(p_{\Theta})} \hat{q}_{\Theta}^{\theta'} - \theta \hat{q}_{\Theta}^{\theta} & > \lceil \max \Theta' \rceil \hat{q}_{\Theta'}^{\max \Theta'} - \theta \hat{q}_{\Theta'}^{\max \Theta'} 
\\
\frac{1}{\gamma} \hat{q}_{\Theta}^{\theta} + \sum_{\theta' > \theta, \theta' \in \text{supp}(p_{\Theta})} \hat{q}_{\Theta}^{\theta'} & > (\lceil \max \Theta' \rceil - \theta)\hat{q}_{\Theta'}^{\max \Theta'}
\end{align*} 
By wariness, $\max \Theta' \in \text{supp}(p_{\Theta'})$. In the last line, the r.h.s. is strictly negative because in this case we must have $\theta > \lceil \max \Theta' \rceil$, while the l.h.s. is strictly positive. 

\textbf{Case III:} $\theta \notin \text{supp}(p_{\Theta})$ and $\theta \notin \text{supp}(p_{\Theta'})$ with $\min\left(\text{supp}(p_{\Theta})\right) > \theta > \max(\Theta')$. (By wariness, we have $\max(\Theta')\in \text{supp}(p_{\Theta'})$.) Note that when $\Theta'$ is the first-stage action, the best second-stage action of type $\theta$ is to take the outside option $\bm{o}$, yielding a payoff of zero. Thus, denoting $\underline{\theta} := \min\left(\text{supp}(p_{\Theta})\right)$, we show: 
\begin{align*} u_A(\hat{q}_{\Theta}^{\underline{\theta}}, \hat{t}_{\Theta}^{\underline{\theta}}, \theta) & > 0 \\
\lceil \underline{\theta} \rceil \hat{q}_{\Theta}^{\underline{\theta}} + \sum_{\theta' > \underline{\theta}} \hat{q}_{\Theta}^{\theta'} - \theta \hat{q}_{\Theta}^{\underline{\theta}} & > 0 \\
(\lceil \underline{\theta} \rceil - \theta) \hat{q}_{\Theta}^{\underline{\theta}} + \sum_{\theta' > \underline{\theta}} \hat{q}_{\Theta}^{\theta'} & > 0. 
\end{align*} 
The l.h.s. in the last line is clearly positive, as desired. 

We conclude from these three cases that any $\theta \in \Theta \setminus \Theta_P$ prefers to disclose $\Theta$ rather than any other $\Theta'$ in $\mathcal{T}(\Theta)$. 

In other cases, both disclosure and non-disclosure may be level-3 $\Delta$-O prudent rationalizable. 

\textbf{Case IV:} $\theta \notin \text{supp}(p_{\Theta})$ and $\theta \notin \text{supp}(p_{\Theta'})$, with $\min\left(\text{supp}(p_{\Theta})\right) > \max (\Theta') > \theta$. As in Case III, let $\underline{\theta} := \min\left(\text{supp}(p_{\Theta})\right)$. Analogously, denote by $\underline{\theta}' := \min \left(\text{supp}(p_{\Theta'})\right)$. Observe that: 
\begin{align} 
u_A(\hat{q}_{\Theta}^{\underline{\theta}}, \hat{t}_{\Theta}^{\underline{\theta}}, \theta) & > u_A(\hat{q}_{\Theta'}^{\underline{\theta}'}, \hat{t}_{\Theta'}^{\underline{\theta}'}, \theta) \nonumber \\
\lceil \underline{\theta} \rceil \hat{q}_{\Theta}^{\underline{\theta}} + \sum_{\theta' > \underline{\theta}} \hat{q}_{\Theta}^{\theta'} - \theta \hat{q}_{\Theta}^{\underline{\theta}} & >\lceil \underline{\theta}' \rceil \hat{q}_{\Theta'}^{\underline{\theta}'} + \sum_{\theta' > \underline{\theta}', \,\theta' \in \Theta'} \hat{q}_{\Theta'}^{\theta'} - \theta \hat{q}_{\Theta'}^{\underline{\theta}'} \nonumber \\
(\lceil \underline{\theta} \rceil - \theta) \hat{q}_{\Theta}^{\underline{\theta}} - (\lceil \underline{\theta}' \rceil - \theta) \hat{q}_{\Theta'}^{\underline{\theta}'}  & > \sum_{\theta' > \underline{\theta}'} \hat{q}_{\Theta'}^{\theta'} - \sum_{\theta' > \underline{\theta}} \hat{q}_{\Theta}^{\theta'}. \label{summies2}
\end{align} 
The l.h.s. of \eqref{summies2} is non-negative since $(\lceil \underline{\theta} \rceil - \theta) > (\lceil \underline{\theta}' \rceil - \theta)$ and $\hat{q}_{\Theta}^{\underline{\theta}} \geq \hat{q}_{\Theta'}^{\underline{\theta}'}$ by Lemma~\ref{Karni_Viero} in the appendix because of monotone supports. If $\sum_{\theta' > \underline{\theta}} \hat{q}_{\Theta}^{\theta'} > \sum_{\theta' > \underline{\theta}'} \hat{q}_{\Theta'}^{\theta'}$, then at information set $(\Theta, \theta)$, disclosure of $\Theta$ is the only level-3 $\Delta$-O prudent rationalizable action for agent $\theta$ in tree $\Theta$. Otherwise, disclosure of some $\Theta'$ may be a better reply in this subcase. 

\textbf{Case V:} $\theta \notin \text{supp}(p_{\Theta})$ and $\theta \in \text{supp}(p_{\Theta'})$. As in the prior case, we define $\underline{\theta} := \min\left(\text{supp}(p_{\Theta})\right)$. We observe: 
\begin{align*}  
u_A(\hat{q}_{\Theta}^{\underline{\theta}}, \hat{t}_{\Theta}^{\underline{\theta}}, \theta) & > u_A(\hat{q}_{\Theta'}^{\theta}, \hat{t}_{\Theta}^{\theta}, \theta) 
\\
\lceil \underline{\theta} \rceil \hat{q}_{\Theta}^{\underline{\theta}} + \sum_{\theta' > \underline{\theta}} \hat{q}_{\Theta}^{\theta'} - \theta \hat{q}_{\Theta}^{\underline{\theta}} & > \lceil \theta \rceil \hat{q}_{\Theta'}^{\theta} + \sum_{\theta' > \theta, \theta' \in \Theta'} \hat{q}_{\Theta'}^{\theta'} - \theta \hat{q}_{\Theta'}^{\theta} 
\\
(\lceil \underline{\theta} \rceil - \theta) \hat{q}_{\Theta}^{\underline{\theta}} & > \frac{1}{\gamma} \hat{q}_{\Theta'}^\theta +  \sum_{\theta' > \theta, \theta' \in \Theta'} \hat{q}_{\Theta'}^{\theta'} - \sum_{\theta' > \underline{\theta}} \hat{q}_{\Theta}^{\theta'}
\end{align*} 
In the last line, the l.h.s. is strictly larger than zero, while the r.h.s. may be positive or negative. Thus, both disclosure of $\Theta$ or $\Theta'$ may be level-3 $\Delta$-O prudent rationalizable for type $\theta$ in tree $\Theta$ in this case. 

\bigskip 
 
\noindent {\bf Level 4, principal:} For any $\Theta \in \mathcal{T}$, the principal at $\Theta$ is certain that $s_{A1}(\theta,\Theta)=\Theta$ for all $\theta\in\Theta\setminus\Theta_{P}$. By log-concavity, $\Theta \setminus \Theta_P \subseteq \text{supp} (\text{marg}_{\Theta} \beta_P(\Theta))$ for any $\beta_P \in B_P^4$. Thus, for any $s_P \in R_P^4$ and $\Theta \in \mathcal{T}$, $s_P(\Theta)$ contains at least $|\Theta \setminus \Theta_P|$ contracts optimal for types in $\Theta \setminus \Theta_P$ and possibly more contracts for types in $\Theta_P$, as well as possibly redundant contracts.  

\noindent {\bf Level 4, agent:} Since $R_P^3 = R_P^2$ we have $B_A^3 = B_A^2$. Thus, $R_A^4 = R_A^3$. 

\bigskip 

\noindent {\bf Level 5, principal:} Since $R_A^4 = R_A^3$ we have $B_P^5 = B_P^4$. Thus, $R_P^5 = R_P^4$.

\noindent {\bf Level 5, agent:} Note that Case III of the analysis of the agent's level 3 is now ruled out. Thus, no further refinement occurs in the general case. We conclude $R^{\infty}_A \subseteq R^5_A$ and $R^{\infty}_P \subseteq R_P^4$.

\bigskip 

\noindent This completes the proof of (iii), and thus of the Theorem. \hfill $\Box$

\bigskip 

Part (i) in Theorem \ref{m_high_theorem} states that, in any $\Delta$-O prudent rationalizable outcome, the principal offers the (unique) optimal contract menu designed to however many types is made aware of under some log-concave beliefs---possibly expanded to include redundant contracts.

Part (ii) states that the quantities awarded (in a $\Delta$-O prudent rationalizable outcome) to types of which the principal is initially aware of remain the same upon her being made aware of other types---as long as the former are not ruled out \textit{and} their position in support of the principal's belief relative to lower cost types does not change. If they ``move down'' in the ranking of high costs, they will suffer a smaller distortion and be awarded a higher quantity. 

Part (iii) characterizes disclosure in $\Delta$-O prudent rationalizable outcomes. It states that, for any game tree $\Theta \in \mathcal{T}(\bar{\Theta})$, all types of the agent that the principal is initially unaware of fully raise her awareness of $\Theta$. In particular, all types of the agent in $\bar{\Theta} \setminus \Theta_P$ raise the principal's awareness of all types $\bar{\Theta}$; under the uninformative message $\Theta_{P}$, these types would opt for their outside option, so they have nothing to lose. 

In the three-type example, we showed something stronger: $s_{A1}(\theta,\bar{\Theta}) = \bar{\Theta}$ for all $\theta\in\bar{\Theta}$.\footnote{This accounts for the rationalizability process taking more steps in the 3-type example than in the proof of Theorem \ref{m_high_theorem}.} Intuitively, $\theta_{1}$ and $\theta_{2}$ alerting the principal about $\theta_{3}$ make them look even better. With more than three types, however, this simple intuition can fail. At level 3 of the rationalizability process, upon receiving the message $\bar{\Theta}$, the principal may infer that only the highest types in $\bar{\Theta}$ would have raised her awareness. (See Case V in the proof of Theorem \ref{m_high_theorem}.) In the ``default'' menu following message $\Theta_{P}$, a type $\theta\leq\max(\Theta_{P})$ is ``included'' and earns some information rents; in the menu following message $\bar{\Theta}$, such a type would be ``excluded'' and would choose the contract meant for the lowest-cost type on the support of $p_{\bar{\Theta}}$, the marginal of the principal's beliefs. The latter contract yields a higher information rent per unit but awards a lower quantity, making it possible for the message $\bar{\Theta}$ to become unappealing. This is demonstrated in the following example.      

\bigskip 

\begin{ex}\label{counterex}
Take $D=\{0,1,\ldots,99\}$, $\gamma=100$, and $v(q)=50q-0.25q^{2}$; we have $\Delta^{+}v(q)=49.75-0.5q$ and $\Delta^{-}v(q)=50.25-0.5q$. The marginal cost type space is $\bar{\Theta}=\{0.99,\ldots,4.99\}$ but the principal is initially aware only of the types in $\Theta_{P}=\{0.99,\ldots,3.99\}$. At level 3, type $0.99$ considers reporting $\bar{\Theta}$. Take $\beta_{P}$ such that:
\[
\text{marg}_{\Theta_{P}}\beta_{P}(\Theta_{P})=\left(0.5,0.15,0.3,0.5\right)
\]
and:
\[
\text{marg}_{\bar{\Theta}}\beta_{P}\left(\bar{\Theta}\right)=\left(0,0,0,\frac{89}{91},\frac{2}{91}\right),
\]
so that $\text{supp}\left(\text{marg}_{\Theta_{P}}\beta_{P}(\Theta_{P})\right)=\Theta_{P}$ and $\text{supp}\left(\text{marg}_{\bar{\Theta}}\beta_{P}\left(\bar{\Theta}\right)\right)=\{3.99,4.99\}$. Notice that these beliefs satisfy logconcavity, wariness, and monotone supports, with reverse Bayesianism being trivially satisfied. If type $0.99$ reports $\Theta=\Theta_{P}$, his payoff is:
\[
u_{A}\left(\hat{q}^{0.99}_{\Theta_{P}},\hat{t}^{0.99}_{\Theta_{P}},0.99\right)=0.01\hat{q}^{0.99}_{\Theta_{P}}+\hat{q}^{1.99}_{\Theta_{P}}+\hat{q}^{2.99}_{\Theta_{P}}+\hat{q}^{3.99}_{\Theta_{P}}.
\] 
The quantities offered are as follows:
\begin{align*}
&49.75-0.5\hat{q}^{0.99}_{\Theta_{P}}\leq 1\leq 49.75-0.5\hat{q}^{0.99}_{\Theta_{P}} \ \Rightarrow \ \hat{q}^{0.99}_{\Theta_{P}}=98;
\\
&49.75-0.5\hat{q}^{1.99}_{\Theta_{P}}\leq 2+\frac{0.05}{0.15}\leq 49.75-0.5\hat{q}^{1.99}_{\Theta_{P}} \ \Rightarrow \ \hat{q}^{1.99}_{\Theta_{P}}=95;
\\
&49.75-0.5\hat{q}^{2.99}_{\Theta_{P}}\leq 3+\frac{0.2}{0.3}\leq 49.75-0.5\hat{q}^{2.99}_{\Theta_{P}} \ \Rightarrow \ \hat{q}^{2.99}_{\Theta_{P}}=93;
\\
&49.75-0.5\hat{q}^{3.99}_{\Theta_{P}}\leq 4+\frac{0.5}{0.5}\leq 49.75-0.5\hat{q}^{3.99}_{\Theta_{P}} \ \Rightarrow \ \hat{q}^{3.99}_{\Theta_{P}}=90.
\end{align*}
Thus, $u_{A}\left(\hat{q}^{\theta_{0.99}}_{\Theta_{P}},\hat{t}^{\theta_{0.99}}_{\Theta_{P}},0.99\right)=0.98+95+93+90=278.98$. Reporting $\Theta=\bar{\Theta}$, type $0.99$ would choose the contract intended for type $\theta_{4}$ for a payoff of: 
\[
u_{A}\left(\hat{q}^{3.99}_{\bar{\Theta}},\hat{t}^{3.99}_{\bar{\Theta}},0.99\right)=3.01\hat{q}^{3.99}_{\bar{\Theta}}+\hat{q}^{4.99}_{\bar{\Theta}}.
\] 
Now,
\begin{align*}
&49.75-0.5\hat{q}^{3.99}_{\bar{\Theta}}\leq 4\leq 49.75-0.5\hat{q}^{3.99}_{\bar{\Theta}} \ \Rightarrow \ \hat{q}^{3.99}_{\bar{\Theta}}=92;
\\
&49.75-0.5\hat{q}^{4.99}_{\bar{\Theta}}\leq 5+\frac{89}{2}\leq 49.75-0.5\hat{q}^{4.99}_{\bar{\Theta}} \ \Rightarrow \ \hat{q}^{4.99}_{\bar{\Theta}}=1.
\end{align*}
This type's payoff is now $u_{A}\left(\hat{q}^{3.99}_{\bar{\Theta}},\hat{t}^{3.99}_{\bar{\Theta}},0.99\right)=3.01\cdot92+1=277.92$. Hence, the agent will not report $\bar{\Theta}$. \hfill $\Box$
\end{ex}


\section{Unawareness of Low Cost Types Only\label{low}}

\subsection{Three Marginal Cost Types\label{3_low}}

In this section, we return to the 3-type setting $\bar{\Theta} = \{\theta_{1},\theta_{2},\theta_{3}\}$ with $\theta_{1} = 0.99$, $\theta_{2} = 1.99$, and $\theta_{3} = 2.99$ but make the principal initially unaware of the lowest-possible marginal cost type: $\Theta_{P} = \{\theta_{2},\theta_{3}\}$. We characterize $\Delta$-O prudent rationalizable strategies level-by-level. 

\bigskip

\noindent {\bf Level 1:} This level works just like in the high-cost type example, so further details are omitted.

\bigskip 

\noindent {\bf Level 2, principal:} At level 2, the principal is certain of first-level $\Delta$-O prudent rationalizable strategies of the agent, $R_A^1$. Thus, she knows that the agent observes the standard incentive and participation constraints when choosing a contract from a menu, and that all types of the agent in tree $\bar{\Theta}$ may disclose. 

\textbf{Information set $\Theta_P = \{\theta_2, \theta_3\}$:} Pick any $\beta_{P}(\Theta_{P})\in B^{2}_{P}$ and take the marginal on $\Theta_P$ denoted as before by $p_{\Theta_{P}}$. As in Section \ref{3_high}, since all types in the tree associated with $\Theta_P$ must disclose $\Theta_P$, this marginal belief must have full support on $\Theta_P$. We first look at the optimal menu of at most two contracts: 
$$\max_{\left(q^\theta_{\Theta_{P}}, t^\theta_{\Theta_{P}}\right)_{\theta \in \{\theta_{2}, \theta_{3}\}} \in \{0, 1, ..., b\}^{4}} p_{\Theta_{P}}(\theta_{2})\left(v\left(q_{\Theta_{P}}^{ \theta_{2}}\right) - t^{ \theta_{2}}_{\Theta_{P}}\right) + p_{\Theta_{P}}(\theta_{3})\left(v\left(q_{\Theta_{P}}^{ \theta_{3}}\right) - t^{ \theta_{3}}_{\Theta_{P}}\right),$$ subject to the incentive compatibility constraints: 
\begin{itemize} 
\item[] IC$^{\Theta_{P}}_{2, 3}$: $$t^{ \theta_{2}}_{\Theta_{P}} - \theta_{2}q^{ \theta_{2}}_{\Theta_{P}} \geq t^{ \theta_{3}}_{\Theta_{P}} - \theta_{2}q^{ \theta_{3}}_{\Theta_{P}},$$
\item[] IC$^{\Theta_{P}}_{3, 2}$: $$t^{ \theta_{3}}_{\Theta_{P}} - \theta_{3}q^{ \theta_{3}}_{\Theta_{P}} \geq t^{ \theta_{2}}_{\Theta_{P}} - \theta_{3}q^{ \theta_{2}}_{\Theta_{P}};$$
\end{itemize} and the participation constraints:
\begin{itemize} 
\item[] PC$^{\Theta_{P}}_{2}$: $$t^{ \theta_{2}}_{\Theta_{P}} - \theta_{2}q^{ \theta_{2}}_{\Theta_{P}} \geq 0,$$ 
\item[] PC$^{\Theta_{P}}_{3}$: $$t^{ \theta_{3}}_{\Theta_{P}} - \theta_{3}q^{ \theta_{3}}_{\Theta_{P}} \geq 0.$$ 
\end{itemize} 

This problem is entirely analogous to its counterpart in Section \ref{3_high}. Following similar steps leads to the following unconstrained optimization problem:
$$\max_{\left(q^\theta_{\Theta_{P}}, t^\theta_{\Theta_{P}}\right)_{\theta \in \{\theta_{2},\theta_{3}\}} \in \{0, 1, ..., b\}^{4}} p_{\Theta_{P}}(\theta_{2})\left(v\left(q_{\Theta_{P}}^{ \theta_{2}}\right) - 2 q^{ \theta_{2}}_{\Theta_{P}} - q^{ \theta_{3}}_{\Theta_{P}}\right) + p_{\Theta_{P}}(\theta_{3})\left(v\left(q_{\Theta_{P}}^{ \theta_{3}}\right) - 3 q^{ \theta_{3}}_{\Theta_{P}}\right).$$ Denoting optimal quantities by $\hat{q}^{ \theta_{2}}_{\Theta_{P}}$ and $\hat{q}^{ \theta_{3}}_{\Theta_{P}}$, respectively, the combined first-order conditions are:
$$\Delta^{+} v\left(\hat{q}^{ \theta_{2}}_{\Theta_{P}}\right) \leq 2 \leq \Delta^{-} v\left(\hat{q}^{\theta_{2}}_{\Theta_{P}}\right) ;  \quad \Delta^{+} v\left(\hat{q}^{\theta_{3}}_{\Theta_{P}}\right) \leq 3 + \frac{p_{\Theta_{P}}(\theta_{2})}{p_{\Theta_{P}}(\theta_{3})} \leq \Delta^{-} v\left(\hat{q}^{\theta_{3}}_{\Theta_{P}} \right).$$

As in Section \ref{3_high}, strict monotonicity $\hat{q}^{ \theta_{2}}_{\Theta_{P}} > \hat{q}^{ \theta_{3}}_{\Theta_{P}}$ also makes use of Assumption~\ref{properties_v}.\ref{not_too_sdc}. 
%
Non-uniqueness for the first-order conditions involving $\hat{q}^{ \theta_{2}}_{\Theta_{P}}$ is ruled out by Assumption~\ref{properties_v}.\ref{efficient_unique}, while non-uniqueness regarding $\hat{q}^{ \theta_{3}}_{\Theta_{P}}$ is precluded by our requirement of best replies to an open set of beliefs as defined in our solution concept (Definition~\ref{EFR}). Nonetheless, it does not hurt the principal to include redundant contracts.

{\bf Information set $\bar{\Theta} = \{\theta_{1}, \theta_{2}, \theta_{3}\}$:} Pick any $\beta_{P}(\Theta_{P})\in B^{2}_{P}$ and its marginal $p_{\bar{\Theta}}$. Since $R^1_A(\bar{\Theta}) \neq \emptyset$, $p_{\bar{\Theta}}$ can only have one of the following supports: $\{\theta_1, \theta_2, \theta_3\} = \bar{\Theta}$, $\{\theta_1, \theta_2\}$, and $\{\theta_1\}$. Supports $\Theta_P = \{\theta_2, \theta_3\}$, $\{\theta_2\}$, and $\{\theta_3\}$ are ruled out by wariness, while $\{\theta_1, \theta_3\}$ violates log-concavity. 

{\bf Support $\{\theta_1, \theta_2, \theta_3\}$:} This case in identical to it's high-cost counterpart.

To compare menus across awareness levels, take any $s_P \in R_P^2$ for which there exists $\beta_P \in B_P^2$ and an open set $O \subseteq B_P(\beta_P) \cap B_P^2$ for which $s_P(\bar{\Theta}) \in \bigcap_{\beta'_P \in O} BR_P(\beta'_P, \bar{\Theta})$ and $s_P(\Theta_P) \in \bigcap_{\beta'_P \in O} BR_P(\beta'_P, \Theta_P)$. Denote by $M$ and $M'$ the uniquely optimal menus corresponding to the marginals $p_{\Theta_P}$ and $p'_{\Theta_P}$, respectively; $s_P(\bar{\Theta}) \supseteq M$ and $s_P(\Theta_P) \supseteq M'$.

Focus first on the quantities for type $\theta_2$. From the first-order conditions, we obtain: 
$$\Delta^{+} v\left(\hat{q}^{\theta_{2}}_{\Theta_P}\right) \leq 2 < 2 + \frac{p_{\bar{\Theta}}(\theta_{1})}{p_{\bar{\Theta}}(\theta_{2})} \leq \Delta^{-} v\left(\hat{q}^{\theta_{2}}_{\bar{\Theta}}\right).$$ By discrete strict concavity of $v(q)$, the strict inequality 
implies that $\hat{q}^{\theta_{2}}_{\Theta_P} \geq \hat{q}^{\theta_{2}}_{\bar{\Theta}}$ (see Francetich and Schipper, 2025, Lemma A7 (iii)). So, the quantity offered by the principal to type $\theta_2$ upon becoming aware of $\theta_1$ cannot increase. This does not hinge on the principal's specific belief, as they are all full support; neither does it rely on reverse Bayesianism, which is moot in the case of $\theta_1$. If $p_{\bar{\Theta}}(\theta_{1}) > p_{\bar{\Theta}}(\theta_{2})$, the principal offers a strictly smaller quantity to type $\theta_2$ upon becoming aware of $\theta_1$. To see this, notice that $p_{\bar{\Theta}}(\theta_{1}) > p_{\bar{\Theta}}(\theta_{2})$ implies: 
\begin{eqnarray} \Delta^{+} v\left(\hat{q}^{\theta_{2}}_{\Theta_P}\right)  - \Delta^{-} v\left(\hat{q}^{\theta_{2}}_{\bar{\Theta}}\right) & \leq & - \frac{p_{\bar{\Theta}}(\theta_{1})}{p_{\bar{\Theta}}(\theta_{2})}<-1; 
\end{eqnarray} but $\hat{q}^{\theta_{2}}_{\Theta_P} = \hat{q}^{\theta_{2}}_{\bar{\Theta}}$ would imply that $\Delta^{+} v\left(\hat{q}^{\theta_{2}}_{\Theta_P}\right)  - \Delta^{-} v\left(\hat{q}^{\theta_{2}}_{\bar{\Theta}}\right) = \Delta^+ \Delta^- v\left(\hat{q}^{\theta_{2}}_{\Theta_P}\right) \geq -1$, where the last inequality comes from Assumption~\ref{properties_v}.\ref{not_too_sdc}. 

A similar analysis holds for the quantities for type $\theta_3$; but in order to relate the inverse hazard rates in the discrete first-order conditions, we now make use of reverse Bayesianism (Lemma~\ref{Karni_Viero} (ii) in the appendix). This leads to $\hat{q}^{\theta_{3}}_{\Theta_P} \geq \hat{q}^{\theta_{3}}_{\bar{\Theta}}$. Moreover, by similar arguments, $\hat{q}^{\theta_{3}}_{\Theta_P} > \hat{q}^{\theta_{3}}_{\bar{\Theta}}$ is implied by $p_{\bar{\Theta}}(\theta_{1}) > p_{\bar{\Theta}}(\theta_{3})$.  

{\bf Support $\{\theta_1, \theta_2\}$:} This case is identical to its high-cost counterpart. The discrete first-order conditions are: 
$$\Delta^{+} v\left(\hat{q}^{ \theta_{1}}_{\bar{\Theta}}\right) \leq 1 \leq \Delta^{-} v\left(\hat{q}^{\theta_{1}}_{\bar{\Theta}}\right) ;  \quad \Delta^{+} v\left(\hat{q}^{\theta_{2}}_{\bar{\Theta}}\right) \leq 2 + \frac{p_{\bar{\Theta}}(\theta_{1})}{p_{\bar{\Theta}}(\theta_{2})} \leq \Delta^{-} v\left(\hat{q}^{\theta_{2}}_{\bar{\Theta}} \right).$$ By the same arguments in the case of support $\bar{\Theta}$, we have $\hat{q}^{\theta_{2}}_{\Theta_P} \geq \hat{q}^{\theta_{2}}_{\bar{\Theta}}$, with strict inequality if $p_{\bar{\Theta}}(\theta_{1}) > p_{\bar{\Theta}}(\theta_{2})$.

{\bf Support $\{\theta_1\}$:}  This type is offered the first-best contract: $$\Delta^{+} v\left(\hat{q}^{\theta_{1}}_{\bar{\Theta}}\right) \leq 1 \leq \Delta^{-} v\left(\hat{q}^{\theta_{1}}_{\bar{\Theta}} \right).$$ Such a contract satisfies this type's participation constraint due to the round-up rent, $\frac{1}{100} \hat{q}^{\theta_1}_{\bar{\Theta}}$. Of course, the principal may include further redundant contracts that are not a best response for type $\theta_1$.

\noindent {\bf Level 2, agent:} As in Section \ref{3_high}, $R_P^1$ imposes no restrictions on the agent's belief of whether he will be offered a better menu of contracts upon raising the principal's awareness of $\theta_1$. Moreover, no matter what menu he is presented by the principal, he chooses a best response. Thus, $R_A^2 = R_A^1$. 

\bigskip 

\noindent {\bf Level 3, principal:} Since $R_A^2 = R_A^1$ we have $B_P^3 = B_P^2$. Thus, $R_P^3 = R_P^2$. 

\noindent {\bf Level 3, agent:} At level 3, the agent is certain of the principal's strategies in $R_P^{2, \Theta}$. In game tree $\Theta_P$, the only disclosure action of any type $\theta \in \Theta_P$ is $s_{A1}(\theta, \Theta_P) = \Theta_P$. We focus next on game tree $\bar{\Theta}$. 

{\bf Type $\theta_3$:} The agent forms full support beliefs over the principal’s strategies in $R_P^2$. These strategies at $\bar{\Theta}$ are best responses to beliefs with various supports. We discuss the cases in turn:

{\bf Support $\bar{\Theta}$:} From $R_P^2$, we know that $\hat{q}_{\Theta_P}^{\theta_3} \geq \hat{q}_{\bar{\Theta}}^{\theta_3}$, with strict inequality if $p_{\bar{\Theta}}(\theta_1) > p_{\bar{\Theta}}(\theta_3)$. Type $\theta_3$ only earns his round up rent: 
$$u_A(\hat{q}_{\Theta_P}^{\theta_3}, \hat{t}_{\Theta_P}^{\theta_3}, \theta_3) = \frac{1}{100} \hat{q}_{\Theta_P}^{\theta_3} \geq \frac{1}{100} \hat{q}_{\bar{\Theta}}^{\theta_3} = u_A(\hat{q}_{\bar{\Theta}}^{\theta_3}, \hat{t}_{\bar{\Theta}}^{\theta_3}, \theta_3),$$ with strict inequality if $p_{\bar{\Theta}}(\theta_1) > p_{\bar{\Theta}}(\theta_3)$. Thus, the only weakly sequentially rational action for $\theta_3$ at information set $(\theta_3, \bar{\Theta})$ is to remain quiet about $\theta_1$. 

{\bf Support $\{\theta_1, \theta_2\}$:} In such a case, $\theta_3$ picks his outside option over any optimal contract offered by the principal. We have: $$u_A(\hat{q}_{\bar{\Theta}}^{\theta_2}, \hat{t}_{\bar{\Theta}}^{\theta_2}, \theta_2) = \frac{1}{100} \hat{q}_{\bar{\Theta}}^{\theta_2} > 0 > \frac{1}{100} \hat{q}_{\bar{\Theta}}^{\theta_2} - \hat{q}_{\bar{\Theta}}^{\theta_2} = u_A(\hat{q}_{\bar{\Theta}}^{\theta_2}, \hat{t}_{\bar{\Theta}}^{\theta_2}, \theta_3),$$ and similarly for the contract for $\theta_{1}$.

Any redundant contracts would also violate the participation constraint of $\theta_3$. Let $(t_r, q_r)$ denote such a redundant contract. We know that: 
$$u_A(q_r, t_r, \theta_2) = t_r - \theta_2 q_r < \lceil \theta_2 \rceil \hat{q}_{\bar{\Theta}}^{\theta_2} - \theta_2 \hat{q}_{\bar{\Theta}}^{\theta_2} = \frac{1}{100} \hat{q}_{\bar{\Theta}}^{\theta_2} = u_A(\hat{q}_{\bar{\Theta}}^{\theta_2}, \hat{t}_{\bar{\Theta}}^{\theta_2}, \theta_2).$$ Note that $\frac{1}{100} \hat{q}_{\bar{\Theta}}^{\theta_2} < 1$, by the assumptions on the contract grid. It now follows that if $q_r > 0$, then $u_A(q_r, t_r, \theta_3) = t_r - \theta_3 q_r < 0$ because $\theta_3 q_r - \theta_2 q_r \geq 1$. On the other hand, if there were a redundant contract with $q_r = 0$ and $t_r > 0$, then $t_r \geq 1 > \frac{1}{100} \hat{q}_{\bar{\Theta}}^{\theta_2} > 0$ by the requirement of the grid. But this contradicts the fact that $(q_r, t_r)$ is not a best response choice from the menu for $\theta_2$. 

Since $\theta_3$ earns positive payoffs from non-disclosure, we conclude that non-disclosure of $\bar{\Theta}$ is the only weakly sequentially rational message for $\theta_3$ in this case.




{\bf Support $\{\theta_1\}$:} The analysis is analogous to the case for support $\{\theta_1, \theta_2\}$. Once again, only non-disclosure of $\bar{\Theta}$ is weakly sequentially rational for $\theta_3$ in this case. 

We conclude that not raising awareness of $\bar{\Theta}$ is the only weak sequentially rational action for type $\theta_3$ in game tree $\bar{\Theta}$. 

{\bf Type $\theta_2$:} We discuss every case of the support in turn:

{\bf Support $\bar{\Theta}$:} This case is analogous to type $\theta_3$'s case. It is only weak sequential rational for $\theta_2$ at information set $(\theta_2, \bar{\Theta})$ to not raise awareness of $\theta_1$, and instead just disclose $\Theta_P$. 

{\bf Support $\{\theta_1, \theta_2\}$:} We know from $R_P^2$ that in this case $\hat{q}_{\Theta_P}^{\theta_2} \geq \hat{q}_{\bar{\Theta}}^{\theta_2}$, with strict inequality for some of the principal's strategies in $R_P^2$. We have: 
$$u_A(\hat{q}_{\Theta_P}^{\theta_2}, \hat{t}_{\Theta_P}^{\theta_2}, \theta_2) = \hat{q}_{\Theta_P}^{\theta_3} + \frac{1}{100} \hat{q}_{\Theta_P}^{\theta_2} > \frac{1}{100} \hat{q}_{\bar{\Theta}}^{\theta_2} = u_A(\hat{q}_{\bar{\Theta}}^{\theta_2}, \hat{t}_{\bar{\Theta}}^{\theta_2}, \theta_2).$$ Thus, the only weak sequentially rational action at $(\theta_2, \bar{\Theta})$ is not to raise awareness of $\theta_1$.

{\bf Support $\{\theta_1\}$:} In such a case, $\theta_2$ picks his outside option over any of the unique optimal contracts offered by the principal. We have: $$u_A(\hat{q}_{\bar{\Theta}}^{\theta_1}, \hat{t}_{\bar{\Theta}}^{\theta_1}, \theta_1) = \frac{1}{100} \hat{q}_{\bar{\Theta}}^{\theta_1} > 0 >  \frac{1}{100} \hat{q}_{\bar{\Theta}}^{\theta_1} - \hat{q}_{\bar{\Theta}}^{\theta_1} = u_A(\hat{q}_{\bar{\Theta}}^{\theta_1}, \hat{t}_{\bar{\Theta}}^{\theta_1}, \theta_2).$$ 

It remains to show that $\theta_2$ will not pick any redundant contract possibly offered by the principal, but rather keeps the principal unaware of $\theta_1$---picking the contract offered to him by the unaware principal. Note that any redundant contract offered by the principal must yield to $\theta_1$ a lower profit than $u_A(\hat{q}_{\bar{\Theta}}^{\theta_1}, \hat{t}_{\bar{\Theta}}^{\theta_1}, \theta_1) = \frac{1}{100} \hat{q}_{\bar{\Theta}}^{\theta_1} < 1$, where the last inequality follows from the assumptions on the grid of quantities. The profit to $\theta_2$ from picking the redundant contract must be even lower, as his marginal costs are higher. On the other hand, $\theta_2$ earns a profit larger than $1$ by picking his optimal contract when the principal remains unaware of $\theta_{1}$, namely, $u_A(\hat{q}_{\Theta_P}^{\theta_2}, \hat{t}_{\Theta_P}^{\theta_2}, \theta_2) = \hat{q}_{\Theta_P}^{\theta_3} + \frac{1}{100} q_{\Theta_P}^{\theta_2}$ whenever $\hat{q}_{\Theta_P}^{\theta_3} > 0$. 

We conclude that reporting $\Theta_{P}$ in this case is also the only weakly sequentially rational action for $\theta_2$ in the upper tree $\bar{\Theta}$. 

{\bf Type $\theta_1$:} We consider the possible supports of the principal's beliefs in turn:

{\bf Support $\bar{\Theta}$:} For type $\theta_1$ in game tree $\bar{\Theta}$, both disclosure of his type and non-disclosure are level-3 $\Delta$-O prudent rationalizable. His profit from disclosure is: $$u_A\left(\hat{q}^{\theta_1}_{\bar{\Theta}}, \hat{t}^{\theta_1}_{\bar{\Theta}}, \theta_1\right) = \frac{1}{100} \hat{q}^{\theta_1}_{\bar{\Theta}} + \hat{q}^{\theta_2}_{\bar{\Theta}} + \hat{q}^{\theta_3}_{\bar{\Theta}}.$$ 

In case of non-disclosure, type $\theta_{1}$ may pick a ``redundant'' contract. Recall that these contracts are indeed redundant insofar as types $\theta_2$ and $\theta_3$, the types that the principal is aware of, are concerned; but they may be appealing to type $\theta_{1}$. Assume for the moment that $\theta_1$ chooses only among the unique optimal contracts included in $s_P(\Theta_P)$. In this case, $\theta_1$ bunches with type $\theta_2$ and picks contract $(\hat{q}^{\theta_2}_{\Theta_P}, \hat{t}^{\theta_2}_{\Theta_P}) \in s_P(\Theta_P)$. To see this, note that: $$u_A(\hat{q}_{\Theta_P}^{\theta_2}, \hat{t}_{\Theta_P}^{\theta_2}, \theta_1) = \frac{1}{100} \hat{q}^{\theta_2}_{\Theta_P} + \hat{q}^{\theta_2}_{\Theta_P} + \hat{q}^{\theta_3}_{\Theta_P}  \geq \frac{1}{100} \hat{q}^{\theta_3}_{\Theta_P} + 2 \hat{q}^{\theta_3}_{\Theta_P} = u_A(\hat{q}^{\theta_3}_{\Theta_P}, \hat{t}^{\theta_3}_{\Theta_P}, \theta_1)$$ follows from monotonicity, $\hat{q}^{\theta_2}_{\Theta_P} > \hat{q}^{\theta_3}_{\Theta_P}$, observed earlier. 

To compare now the payoff from disclosure, $u_A(\hat{q}^{\theta_1}_{\bar{\Theta}}, \hat{t}^{\theta_1}_{\bar{\Theta}}, \theta_1)$, with the payoff from non-disclosure, $u_A(\hat{q}_{\Theta_P}^{\theta_2}, \hat{t}_{\Theta_P}^{\theta_2}, \theta_1)$, recall that $\hat{q}^{\theta_2}_{\bar{\Theta}} \leq \hat{q}^{\theta_2}_{\Theta_{P}}$ and $\hat{q}^{\theta_3}_{\bar{\Theta}} \leq \hat{q}^{\theta_3}_{\Theta_{P}}$, with strict inequality for some of the principal's strategies $s_P \in R_P^2$. We claim that $\hat{q}^1_{\bar{\Theta}} > \hat{q}^2_{\Theta_P}$. We prove this by contradiction. Consider first the case $\hat{q}^1_{\bar{\Theta}} < \hat{q}^2_{\Theta_P}$. The discrete first-order conditions, $$\Delta^+ v(\hat{q}^1_{\bar{\Theta}}) \leq 1 \leq \Delta^- v(\hat{q}^1_{\bar{\Theta}}) \quad \mbox{ and } \quad \Delta^+ v(\hat{q}^2_{\Theta_P}) \leq 2 \leq \Delta^- v(\hat{q}^2_{\Theta_P}),$$ imply: 
$$\Delta^+ v(\hat{q}^{\theta_1}_{\bar{\Theta}})  - \Delta^- v(\hat{q}^{\theta_2}_{\Theta_P}) \leq -1.$$ In fact, Assumption~\ref{properties_v}.\ref{efficient_unique} implies:  
$$\Delta^+ v(\hat{q}^{\theta_1}_{\bar{\Theta}})  - \Delta^- v(\hat{q}^{\theta_2}_{\Theta_P}) < -1.$$
If $q^{\theta_1}_{\bar{\Theta}} + 1 = \hat{q}^{\theta_2}_{\Theta_P}$, then $\Delta^+ v(\hat{q}^{\theta_1}_{\bar{\Theta}})  = \Delta^- v(\hat{q}^{\theta_2}_{\Theta_P})$ by the definition of discrete derivatives, yielding a contradiction. If $q^{\theta_1}_{\bar{\Theta}} + 1 < \hat{q}^{\theta_2}_{\Theta_P}$, then $\Delta^+ v(\hat{q}^{\theta_1}_{\bar{\Theta}})  > \Delta^- v(\hat{q}^{\theta_2}_{\Theta_P})$ by Francetich and Schipper (2025, Lemma A7 (iv)), another contradiction. Finally, consider the case $\hat{q}^{\theta_1}_{\bar{\Theta}} = \hat{q}^{\theta_2}_{\Theta_P}$. Then the discrete first-order conditions coupled with Assumption~\ref{properties_v}.\ref{efficient_unique} stated above implies $$\Delta^+ v(\hat{q}^{\theta_1}_{\bar{\Theta}})  - \Delta^- v(\hat{q}^{\theta_1}_{\bar{\Theta}}) < -1,$$ which contradicts Assumption~\ref{properties_v}.\ref{not_too_sdc}.

To sum up, if type $\theta_1$ in game tree $\bar{\Theta}$ believes with a sufficiently large probability that the principal uses a strategy that upon becoming aware of $\theta_1$ puts a sufficiently large probability on $\theta_1$ compared to $\theta_2$ and $\theta_3$ such that $\hat{q}^{\theta_2}_{\Theta_P} + \hat{q}^{\theta_3}_{\Theta_P} - q^{\theta_2}_{\bar{\Theta}} - q^{\theta_3}_{\bar{\Theta}}$ becomes larger than $\frac{1}{100}(\hat{q}^{\theta_1}_{\bar{\Theta}} - \hat{q}^{\theta_2}_{\Theta_P})$, then non-disclosure is the weak sequential best response for $\theta_1$. We reach the same conclusion if any redundant contracts for types $\theta_{2}$ and $\theta_{3}$ are appealing to this type in tree $\Theta_{P}$. Otherwise, disclosure is a weakly sequential best response. 

{\bf Support $\{\theta_1, \theta_2\}$:} The analysis is similar to the prior case, except that $\theta_1$ does not benefit from information rents due to the relevance of $\theta_3$ in game tree $\bar{\Theta}$ in the prior case. Non-disclosure of $\bar{\Theta}$ is weak sequentially rational, $$u_A(\hat{q}_{\Theta_P}^{\theta_2}, \hat{t}_{\Theta_P}^{\theta_2}, \theta_1) = \hat{q}_{\Theta_P}^{\theta_2} + \frac{1}{100} \hat{q}_{\Theta_P}^{\theta_2} + \hat{q}_{\Theta_P}^{\theta_3} > + \frac{1}{100} \hat{q}_{\bar{\Theta}}^{\theta_1} + \hat{q}_{\bar{\Theta}}^{\theta_2} = u_A(\hat{q}_{\bar{\Theta}}^{\theta_1}, \hat{t}_{\bar{\Theta}}^{\theta_1}, \theta_1),$$ if and only if: 
$$\hat{q}_{\Theta_P}^{\theta_2} + \hat{q}_{\Theta_P}^{\theta_3} - \hat{q}_{\bar{\Theta}}^{\theta_2} > \frac{1}{100}( \hat{q}_{\bar{\Theta}}^{\theta_1} - \hat{q}_{\Theta_P}^{\theta_2}).$$ We know from arguments like in the prior case that $\hat{q}_{\Theta_P}^{\theta_2}  -  \hat{q}_{\bar{\Theta}}^{\theta_2} \geq 0$ (with strict inequality for some of the principal's strategies) and $\hat{q}_{\bar{\Theta}}^{\theta_1} - \hat{q}_{\Theta_P}^{\theta_2} > 0$. Moreover $\frac{1}{100}( \hat{q}_{\bar{\Theta}}^{\theta_1} - \hat{q}_{\Theta_P}^{\theta_2}) < 1$. Thus, non-disclosure is the only weakly sequentially rational action of $\theta_1$ if $\hat{q}_{\Theta_P}^{\theta_3} > 0$. Note that ``redundant'' contracts in game tree $\bar{\Theta}$ must be even worse than $(\hat{q}_{\bar{\Theta}}^{\theta_1}, \hat{t}_{\bar{\Theta}}^{\theta_1})$.

{\bf Support $\{\theta_1\}$:} We have:
$$u_A(\hat{q}_{\Theta_P}^{\theta_2}, \hat{t}_{\Theta_P}^{\theta_2}, \theta_1) = \hat{q}_{\Theta_P}^{\theta_2} + \frac{1}{100} \hat{q}_{\Theta_P}^{\theta_2} + \hat{q}_{\Theta_P}^{\theta_3} > \frac{1}{100} \hat{q}_{\bar{\Theta}}^{\theta_1} = u_A(\hat{q}_{\bar{\Theta}}^{\theta_3}, \hat{t}_{\bar{\Theta}}^{\theta_3}, \theta_1).$$ Thus, the only weak sequentially rational action of $\theta_1$ in this case is to keep the principal unaware and communicate just $\Theta_P$.

We summarize that both raising awareness of himself and keeping the principal unaware are weak sequentially rational actions of $\theta_1$. 

\bigskip 

\noindent {\bf Level 4, principal:} At level-4, the principal at information sets $\Theta_P$ and $\bar{\Theta}$ is certain of $R_A^{3, \Theta_P}$ and $R_A^{3, \bar{\Theta}}$, respectively. Thus, at information set $\bar{\Theta}$, she is certain that only type $\theta_1$ could have raised her awareness of $\bar{\Theta}$. Upon receiving message $\bar{\Theta}$, she offers this type's first-best contract---leaving him only with his round-up rent. Any other contract included in $s_P(\bar{\Theta})$ is not a best response for type $\theta_1$. 

Upon hearing $\Theta_P$, no further restrictions on strategies are implied at level 4. 

\noindent {\bf Level 4, agent:} Since $R_P^3 = R_P^2$ we have $B_A^4 = B_A^3$. Thus, $R_A^4 = R_A^3$.

\bigskip 

\noindent {\bf Level 5, principal:} Since $R_A^4 = R_A^3$ we have $B_P^5 = B_P^4$. Thus, $R_P^5 = R_P^4$.

\noindent {\bf Level 5, agent:} Types $\theta_{2}$ and $\theta_{3}$ of the agent behave as at level 4. As for type $\theta_{1}$, he now anticipates that raising the principal's awareness will expose him to the first-best contract with binding participation constraint, while keeping the principal unaware allows him to benefit from information rents: 
$$u_A\left(\hat{q}^{\theta_{1}}_{\bar{\Theta}}, \hat{t}^{\theta_{1}}_{\bar{\Theta}}, \theta_{1}\right) = \frac{1}{100} \hat{q}_{\bar{\Theta}}^{\theta_1} < 1 < \frac{1}{100} \hat{q}_{\Theta_P}^{\theta_2} + \hat{q}_{\Theta_P}^{\theta_2} + \hat{q}_{\Theta_P}^{\theta_3} = u_A\left(\hat{q}^{\theta_{2}}_{\Theta_P}, \hat{t}^{\theta_{2}}_{\Theta_P}, \theta_{1}\right),$$ where the first inequality follows from assumptions on the grid. 

We conclude that for any $\theta \in \bar{\Theta}$ and any $s_{A}\in R^{5}_{A}(\Delta)$, we have $s_{A1}(\theta_{1}, \bar{\Theta}) = \Theta_P$.

\bigskip 

\noindent {\bf Level 6, principal:} At information set $\bar{\Theta}$, the principal realizes that none of the types she is now aware of would have raised here awareness when using a strategy in $R_A^5$. Thus, by the best rationalizability principle (i.e., Battigalli, 1996) embodied in our solution concept, the principal must select a strategy in $R_P^5$. Therefore, $R_P^6 = R_P^5$. 

\noindent {\bf Level 6, agent:} Since $R_P^5 = R_P^4$ we have $B_A^6 = B_A^5$. Thus, $R_A^6 = R_A^5$.

\bigskip 

\noindent No further reduction is possible. 

\bigskip 

We conclude that when the principal is unaware only of low-cost types, in any $\Delta$-O prudent rationalizable outcome, no type of the agent raises her awareness.

\subsection{Any Finite Number of Marginal Cost Types\label{m_low}}

Theorem \ref{m_low_theorem} below extends the analysis to an arbitrary number of finite types. 

\begin{theo}\label{m_low_theorem} Consider the case when the principal is unaware only of low-cost types: $\min(\Theta_{P})>\min(\bar{\Theta})$ and $\max(\Theta_{P})=\max(\bar{\Theta})$. 
\begin{itemize}
\item[(i)] 
For any $s_P \in R_P^{\infty}$, $s_P(\Theta_P) \supseteq M$, where $M$ is a menu of contracts uniquely optimal for a full-support, log-concave $p \in \Delta(\Theta_P)$.
\item[(ii)] 
For every $s_{A}\in R^{\infty}_{A}$ and $\Theta \in \mathcal{T}$, $s_{A1}(\theta, \Theta) = \Theta_P$.
\item[(iii)] 
(Bunching at the top) For every $s_{A} \in R^{\infty}_{A}, s_P \in R_{P}^{\infty}$, $\Theta \supset \Theta_P$, and $\theta \in \Theta \setminus \Theta_P$, 
$$s_{A2}\left(\theta, \Theta, s_{A1}(\Theta, \theta), s_{P}(s_{A1}(\theta, \Theta))\right) = \left(\hat{q}^{\min(\Theta_{P})}_{\Theta_{P}}, \hat{t}^{\min(\Theta_{P})}_{\Theta_{P}}\right),$$ where $\left(\hat{q}^{\min(\Theta_{P})}_{\Theta_{P}}, \hat{t}^{\min(\Theta_{P})}_{\Theta_{P}}\right) \in s_{P}(s_{A1}(\theta, \Theta))$ is the optimal contract designed for $\min(\Theta_P)$.   
\end{itemize}
\end{theo}

\noindent \textsc{Proof. } Just like in the proof of the prior theorem, we proceed by induction on the levels of the rationalizability procedure.

\bigskip 

\noindent {\bf Level 1:} This level is identical to level 1 in the proof of Theorem \ref{m_high_theorem}, so details are omitted.

\bigskip 

\noindent {\bf Level 2, principal:} With any $\beta_P \in B_P^2$ and at any information set $\Theta \in \mathcal{T}$, the principal is certain of $R_A^{1, \Theta}$. Since $R_A^{1, \Theta}$ imposes no restrictions on first-stage disclosure by the agent, she can infer nothing about the type of the agent. Yet, she is certain that any type of the agent in $\Theta$ observes incentive compatibility and participation constraints when choosing $\bm{c} \in M \cup \{\bm{o}\}$ for any $M \in \mathcal{M}$.  

Recall the set $\text{Supp}(\Theta) := \{\text{supp}(\text{marg}_{\Theta} \beta_P(\Theta)):\beta_P \in B_P^2\}$. For any strategy $s_P \in R_P^2$, there must exist $\beta_P \in B_P^2$ and $O \subseteq B(\beta_P) \cap B_P^2$ such that for all $\Theta \in \mathcal{T}$, $s_P(\Theta) \in \bigcap_{\beta'_P \in O} BR_P(\beta_P', \Theta)$. Since $O \subseteq B(\beta_P)$, we must have for all $\Theta \in \mathcal{T}$, $\text{supp}\left(p_{\Theta}\right) = \text{supp}\left(p'_{\Theta}\right) \in \text{Supp}(\Theta)$. 

For any two $\Theta, \Theta' \in \mathcal{T}$ with $\Theta' \in \mathcal{T}(\Theta)$ and any type $\theta \in \text{supp}\left(p_{\Theta'}\right) \cap \text{supp}\left(p_{\Theta}\right)$, if $(q^\theta_{\Theta}, t^\theta_{\Theta})$ is the contract picked by type $\theta$ in game tree associated with $\Theta$ from $s_P(\Theta)$ and $(q^\theta_{\Theta'}, t^\theta_{\Theta'})$ is the contract picked by type $\theta$ in game tree associated with $\Theta'$ from $s_P(\Theta')$, then $q^\theta_{\Theta'} \geq q^\theta_{\Theta}$. Moreover, for any $s_P \in R_P^2$ and $\theta\in\text{supp}(p_{\Theta})$ such that $\sum_{\tilde{\theta}<\theta}p_{\Theta}(\tilde{\theta}) > p_{\Theta}(\theta)$, the inequality for the quantities holds as a strict inequality. This follows from Lemma~\ref{Karni_Viero} (ii) in the appendix and Francetich and Schipper (2025, Lemma 9) because: (i.) $\{\min (\text{supp}(p_{\Theta'})), ..., \theta\} \subseteq \text{supp}\left(p_{\Theta}\right)$, i.e., by monotone supports and log-concavity, all terms included in the sum of the numerator of the inverse hazard rate for $\Theta'$ are also included in the sum making up the numerator of the inverse hazard rate for $\Theta$; (ii.) $p_{\Theta}$ and $p_{\Theta'}$ satisfy reverse Bayesianism; and (iii.) the optimal contracts in $s_P(\Theta)$ and $s_P(\Theta')$ are unique for each type. 

\noindent {\bf Level 2, agent:} At level 2 of the procedure, the agent with any belief system is certain at every one of his information sets $(\theta, \Theta)$ for $\theta \in \Theta \in \mathcal{T}$ that the principal follows a strategy in $R_P^{1, \Theta}$. This imposes no restrictions on his belief of whether or not he is presented with a better menu of contracts upon raising the principal's awareness from $\Theta_P$ to $\Theta' \in \mathcal{T}(\Theta)$. Moreover, no matter what menu he is presented by the principal at his second-stage information sets allowed by his own strategy, he chooses a best response. Thus, $R_A^2 = R_A^1$. 

\bigskip 

\noindent {\bf Level 3, principal:} Since $R_A^2 = R_A^1$ we have $B_P^3 = B_P^2$. Thus, $R_P^3 = R_P^2$. 

\noindent {\bf Level 3, agent:} We show that, for any $\theta \in \Theta_P$, $s_A \in R_A^3$, and $\Theta\in\mathcal{T}$, we have $s_{A1}(\theta, \Theta) = \Theta_P$. 

Consider $\beta_{P}\in B_{P}^{3}$ such that $\theta \in \text{supp}(p_{\Theta})$. We show that: 
\begin{align} 
u_A(\hat{q}_{\Theta_P}^{\theta}, \hat{t}_{\Theta_P}^{\theta}, \theta) & > u_A(\hat{q}_{\Theta}^{\theta}, \hat{t}_{\Theta}^{\theta}, \theta) \nonumber
\\ 
\hat{t}_{\Theta_P}^\theta - \theta \hat{q}_{\Theta_P}^{\theta} & > \hat{t}_{\Theta}^\theta - \theta \hat{q}_{\Theta}^{\theta} \nonumber
\\
\lceil \theta \rceil \hat{q}_{\Theta_P}^{\theta} + \sum_{\theta' > \theta} \hat{q}_{\Theta_P}^{\theta'} - \theta \hat{q}_{\Theta_P}^{\theta} & >  \lceil \theta \rceil \hat{q}_{\Theta}^{\theta} + \sum_{\theta' > \theta, \theta' \in \text{supp}(p_{\Theta})} \hat{q}_{\Theta}^{\theta'} - \theta \hat{q}_{\Theta}^{\theta} \nonumber
\\
\frac{1}{\gamma} (\hat{q}_{\Theta_{P}}^{\theta} - \hat{q}_{\Theta}^{\theta}) & > \sum_{\theta' > \theta, \,\theta' \in \text{supp}(p_{\Theta})} \hat{q}_{\Theta}^{\theta'} - \sum_{\theta' > \theta} \hat{q}_{\Theta_P}^{\theta'}.\label{summies3}
\end{align} 
The l.h.s. in \eqref{summies3} is weakly positive, and strictly positive for some of the principal's strategies in $R_P^2$ by Lemma~\ref{Karni_Viero} (ii) in the appendix. For the r.h.s., we split the analysis into two cases.

If $\theta \in \text{supp}(p_{\Theta})$, we conclude from Lemma~\ref{Karni_Viero} (ii) in the appendix that the r.h.s. in \eqref{summies3} is weakly smaller than zero: All terms in the first sum have corresponding terms in the second sum, and the corresponding quantities across awareness levels satisfy weak monotonicity \`{a} la Lemma~\ref{Karni_Viero} (ii) in the appendix.  

Now consider the case $\theta \notin \text{supp}(p_{\Theta})$. In such a case, $\theta > \max (\text{supp}(p_{\Theta}))$ by log-concavity and monotone supports. When raising awareness of $\Theta$, $\theta$ is better off taking the outside option than any level-$2$ $\Delta$-O prudent rationalizable contract subsequently offered by the principal. This is clear for contracts that are optimal for the types in the support of the principal's marginal beliefs. With respect to redundant contracts, consider the least efficient type in the principal's support of the marginal belief upon becoming aware of $\Theta$, $\theta' := \max(\text{supp}(p_{\Theta} ))$. For any redundant contract $(q^r, t^r)$, we have: 
$$t^r - \theta' q^r < \hat{t}_{\Theta}^{\theta'}- \theta' \hat{q}_{\Theta}^{\theta'}.$$ Since $\theta'$ is held to his outside option, the r.h.s. is equal to $\frac{1}{\gamma} \hat{q}_{\Theta}^{\theta'}$, which is smaller than 1. Since $\theta > \theta'$, we must have $t^r - \theta q^r < t^r - \theta' q^r$. In particular, if $q^r > 0$, then $t^r - \theta q^r < 0$. We conclude that $\theta$ is better off taking the outside option than any redundant contract that the principal may offer with any strategy in $R_P^2$ upon becoming aware of $\Theta$. 

Note that when $\theta$ takes the outside option, he earns zero payoff. In contrast, he can ensure himself a positive payoff by disclosing just $\Theta_P$ and picking the contract optimal for him in $s_{P}(\Theta_{P})\in R^{2,\Theta_{P}}_{P}$.  

No other strategies in $R_A^2$ of the agent can be excluded at level 3. 

\bigskip 

\noindent {\bf Level 4, principal:} There are no further changes to the principal's actions following the uninformative message $\Theta_{P}$. Regarding any other message $\Theta \in \mathcal{T}$, the principal is now certain with any $\beta_P \in B_P^4$ that she faces none of the types in $\Theta_P$: $\text{supp}(p_{\Theta}) \cap \Theta_P = \emptyset$. Define $\text{Supp}'(\Theta) := \{X \in \text{Supp}(\Theta) : X \cap \Theta_P = \emptyset\}$. For any $s_P \in R_P^4$ and information set $\Theta$, $s_P(\Theta)\supset M$, where $M$ is the menu of optimal contracts for some $\beta_P \in B_P^4$ (as discussed at level 2), one for each type in $\text{supp}(p_{\Theta}) \in \text{Supp}'(\Theta)$; the remaining contracts, if any, are redundant contracts. 

\noindent {\bf Level 4, agent:} Since $R_P^3 = R_P^2$ we have $B_A^4 = B_A^3$. Thus, $R_A^4 = R_A^3$.

\bigskip 

\noindent {\bf Level 5, principal:} Since $R_A^4 = R_A^3$ we have $B_P^5 = B_P^4$. Thus, $R_P^5 = R_P^4$.

\noindent {\bf Level 5, agent:} 
Take any $\Theta \in \mathcal{T}$, $\theta \in\Theta\setminus \Theta_P$, and $s_A \in R_A^5$. Let $\mathcal{T}_{\theta}$ denote the subset of messages with $\theta$ as the lowest type: $$\mathcal{T}_{\theta}:=\left\{\Theta'\in\mathcal{T}:\min(\Theta')=\theta\right\}.$$
We show that, for any $\Theta'\in\mathcal{T}_{\theta}$ and any $\Theta''\in\mathcal{T}$ such that $\Theta'\subset\Theta''$, $s_{A1}(\theta, \Theta') \neq \Theta''$. In words, no type in $\Theta\setminus\Theta_{P}$ would send a message that includes cost types lower than his.

Assume first $\theta \in \text{supp}(p_{\Theta'}) \cap \text{supp}(p_{\Theta''})$. We show that:
\begin{align} 
u_A(\hat{q}_{\Theta'}^{\theta}, \hat{t}_{\Theta'}^{\theta}, \theta) & > u_A(\hat{q}_{\Theta''}^{\theta}, \hat{t}_{\Theta''}^{\theta}, \theta) \nonumber
\\
\lceil \theta \rceil \hat{q}_{\Theta'}^{\theta} + \sum_{\tilde{\theta} > \theta,\,\tilde{\theta}\in \text{supp}(p_{\Theta'})} \hat{q}_{\Theta'}^{\tilde{\theta}} - \theta \hat{q}_{\Theta'}^{\theta} & > \lceil \theta \rceil \hat{q}_{\Theta''}^{\theta} + \sum_{\tilde{\theta} > \theta,\,\tilde{\theta}\in \text{supp}(p_{\Theta''})} \hat{q}_{\Theta''}^{\tilde{\theta}} - \theta \hat{q}_{\Theta''}^{\theta}\nonumber
\\
\frac{1}{\gamma} (\hat{q}_{\Theta'}^{\theta} - \hat{q}_{\Theta''}^{\theta}) & > \sum_{\tilde{\theta} > \theta,\,\tilde{\theta}\in \text{supp}(p_{\Theta''})} \hat{q}_{\Theta''}^{\theta} - \sum_{\tilde{\theta} > \theta,\,\tilde{\theta} \in \text{supp}(p_{\Theta'})} \hat{q}_{\Theta'}^{\theta}.\label{summies4}
\end{align} 
We have that the l.h.s. of \eqref{summies4} is weakly larger zero by Lemma~\ref{Karni_Viero} (ii) in the appendix, because of monotone supports and log-concavity. (It is also strictly smaller than 1.) The r.h.s. is strictly smaller than zero if $\max(\text{supp}(p_{\Theta'})) \geq \max(\text{supp}(p_{\Theta''}))$ because of two reasons. First, any type that has a term in the first sum also has a corresponding term in the second one. Note that $\max(\text{supp}(p_{\Theta'})) \geq \max(\text{supp}(p_{\Theta''}))$ is implied by any strategy $s_P \in R_P^4$, again by monotone supports and log-concavity. Second, for all types with terms in both sums, their term in the second one is weakly larger than their counterpart in the first one (Lemma~\ref{Karni_Viero} (ii), monotone supports, and log-concavity).

If $\theta \in \text{supp}(p_{\Theta'})$ but $\theta\notin\text{supp}(p_{\Theta''})$, log-concavity and monotone supports imply that type $\theta$ would choose his outside option in $s_{P}(\Theta'')$ for any $s_{P}\in R^{4,\Theta''}_{P}$. Finally, $\theta \notin \text{supp}(p_{\Theta'})$ but $\theta\in\text{supp}(p_{\Theta''})$, this type obtains higher rents by bunching with type $\min(\text{supp}(p_{\Theta'}))$ than in the contract designed for him in $s_{P}(\Theta'')$ for any $s_{P}\in R^{4,\Theta''}_{P}$.

\bigskip 
  
\noindent {\bf Level 6, principal:} At information set $\Theta \in \mathcal{T}$ with $\Theta \neq \Theta_P$, the principal is now certain with any $\beta_P \in B_P^6$ that she faces type $\min(\Theta)$. Consequently, with any $s_P \in R_P^6$, upon receiving message $\Theta \neq \Theta_P$, she offers the first best-contract for type $\min(\Theta)$ potentially along redundant contracts that are strictly worse to type $\min(\Theta)$. 

\noindent {\bf Level 6, agent:} Since $R_P^5 = R_P^4$ we have $B_A^6 = B_A^5$. Thus, $R_A^6 = R_A^5$.

\bigskip 

\noindent {\bf Level 7, principal:} Since $R_A^6 = R_A^5$ we have $B_P^7 = B_P^6$. Thus, $R_P^7 = R_P^6$.

\noindent {\bf Level 7, agent:} We show that for any $\Theta \in \mathcal{T}$, $\theta \in \Theta$, and $s_A \in R_A^7$, $s_A(\theta, \Theta) = \Theta_P$. For $\Theta_P$ and $\theta \in \Theta_P$ this follows from $R_A^7 \subseteq R_A^3$. For any other type $\theta \notin \Theta_P$, he compares his profit from disclosing $\Theta$ with $\theta = \min(\Theta)$ with his profit from disclosing just $\Theta_P$: 
\begin{align*} 
u_A(\hat{q}_{\Theta}^{\theta}, \hat{t}_{\Theta}^{\theta}, \theta) & < u_A(\hat{q}_{\Theta_P}^{\min(\Theta_P)}, \hat{t}_{\Theta_P}^{\min(\Theta_P}), \theta) 
\\
\frac{1}{\gamma}\hat{q}_{\Theta}^{\theta} & < \lceil \min \Theta_P \rceil \hat{q}_{\Theta_P}^{\min(\Theta_P)} + \sum_{\theta' > \min(\Theta_P)} \hat{q}_{\Theta_P}^{\theta'} - \theta \hat{q}_{\Theta_P}^{\min(\Theta_P)}. 
\end{align*} 
The l.h.s. in the last line is $\theta$'s payoff from the first-best contract, which is strictly below 1. The r.h.s. is the payoff to $\theta$ from bunching with $\min(\Theta_P)$ in a menu offered by the principal when nothing is revealed to the principal. This payoff is clearly larger than 1 because of the information rents. The principal may also offer redundant contracts along with the first-best contract upon becoming aware of $\Theta$. However, by definition, the former must be worse to $\theta$ than the latter.  

\bigskip 

\noindent No further reduction of strategies occurs at higher levels. \hfill $\Box$ 

\bigskip 

Point (i) in Theorem \ref{m_low_theorem} states that, being kept in the dark, the principal offers a menu of ``standard equilibrium'' contracts that are uniquely optimal given a full-support log-concave $p \in \Delta(\Theta_P)$, possibly along with redundant contracts.

Point (iii) is straightforward: Whenever the principal is not made aware of all possible types, the low-cost types that she remains unaware of will bunch with the lowest type on the principal's radar when it comes to choosing a contract.

Last but not least, point (ii) relates to disclosure. It states that, in any $\Delta$-O prudent rationalizable outcome, \textit{none} of the types raise the principal's awareness. The proof of the theorem shows that, if the principal receives a message $\Theta \in \mathcal{T}$ such that $\Theta \supset \Theta_{P}$, she reasons that she is in fact facing type $\min(\Theta)$. Thus, if a type $\theta \in \bar{\Theta} \setminus \Theta$ raises the principal's awareness to $\{\theta, \theta + 1, \ldots, \max(\Theta_{P})\}$, he is virtually revealing himself. Consequently, the principal would offer the first-best contract to him, leaving him only with the round-up rent. This is strictly worse to him than benefiting from the information rents obtained by bunching with type $\min(\Theta_P)$ when keeping the principal in the dark.


\section{Discussion~\label{discussion}}

\subsection{Further Observations}

The results in Theorems \ref{m_high_theorem} and \ref{m_low_theorem} can be captured in a simple punch line: \textit{``Bad news are sometimes shared, good news are always withheld.''} More precisely, the principal is kept in the dark about efficient types, but inefficient types make themselves seen. This is reminiscent of the negativity bias in social psychology (Rozin and Royzman, 2001) according to which negative events are more salient than positive ones. In our case, the principal may develop endogenously a negativity bias about marginal costs of the agent. 

While the intuition for the agent's incentives to reveal higher or to hide lower cost types is fairly robust, especially for high-cost types who would otherwise not be served, it relies on the comparison of quantities across awareness levels:
\begin{itemize}   

\item No cost type of which the principal is initially aware wants to make her aware of lower cost types because they will be awarded a smaller quantity (there's additional distortion) and no extra rents.

\item Low-cost types of which the principal is initially unaware keep the principal in the dark and bunch with the lowest type she is aware of. Raising awareness about himself would leave him without information rents because the principal would infer the type perfectly. 

\item High-cost types that the principal is unaware of would not be served unless they speak up. Thus, they have a strict incentive to raise the principal's awareness and earn at least their round-up rent. 

\item Under unawareness of high-cost types only, with three types, the lower cost types---which the principal is aware of---also want to raise awareness about the highest cost type. The lowest type continues to be awarded the first-best quantity, while reverse Bayesianism implies that the middle type is also awarded the same quantity as before; but now, both types enjoy (additional) information rents.

\item However, with more than three types, a type the principal is initially aware of who raises her awareness about higher costs may not be served anymore with a menu containing a contract dedicated to him.  If the subsequent menu caters only to the highest of the new types, it is possible for it to be less profitable than the ``default'' contract offered by the unaware principal.

\end{itemize}
The intuition also relies on log-concavity; bunching in the contracts would cloud the comparison of quantities across awareness levels, especially if raising the principal's awareness causes some types that the principal was already aware of to become bunched. Reverse Bayesianism preserves log-concavity and prevents bunching of those types. Of course, it does not prevent bunching of types the principal remains unaware of---the efficient ones. 

Since we use a rationalizability concept, we can dispense with the assumption of a common prior over types. Although the contract menus---and therefore the agent's profit---depend on the specific belief that the principal holds, the comparison of quantities across awareness levels---and therefore the agent's incentive whether or not to raise the principal's awareness---apply to any $\Delta$-O prudent rationalizable belief. Intuitively:
\begin{itemize}

\item When the principal is initially unaware of high-cost types only, said types need not know the precise belief the principal will form upon becoming aware to know that they will be offered an acceptable contract, as opposed to not being served. With three types, when the lower-cost types raise the principal's awareness of the highest-cost types, the former need not be certain about the principal's updated belief to infer that he will be awarded the same quantity plus an extra information-rent term.

\item If the agent's types that the principal is initially aware of were to raise the principal's awareness about lower cost types, he will be offered the first-best quantity for the lowest type reported. At best, such a contract will take away all of his information rents; at worst, it will be unacceptable.\footnote{In the best-case scenario, even if the quantity awarded after essentially revealing himself is higher, recall that the round-up rent is always less than 1 and so information rents dominate.} 

\end{itemize}

\subsection{Unawareness versus zero-probability events}

We interpret our model as one of changing the awareness level of the principal. How is this different from the principal assigning zero probability to some types? 

If the principal assigns zero probability to some type $\theta$, then she assigns probability 1 to the complement of $\theta$; that is, she is certain that the agent is not of type $\theta$. Hence, receiving a message like ``Have you considered that the agent could have type $\theta$?'' (a question) or ``The agent could be of type $\theta$ or not'' (a tautology) does not contain inherent information and should not change the principal's probabilistic assessment of $\theta$, especially if she is absolutely certain that the agent's type is not $\theta$ (as implied by assigning zero probability to it). Yet, if she were unaware of $\theta$, she realizes upon receiving such a message that she did not consider that the agent's type could be $\theta$, and hence may reevaluate her probabilistic assessment of types. Thus, we find the interpretation of unawareness much more compelling for the implications of belief change we study in this paper.

We refer to Heifetz, Meier, and Schipper (2013) for a comparison of zero probability and unawareness in games with unawareness in the extensive form, and to Schipper (2013) for a test of zero probability versus unawareness with choice experiments.

\subsection{Awareness of Unawareness}

Our principal is initially aware only of $\Theta_P$ and unaware of types in $\bar{\Theta} \setminus \Theta_P$. When the agent raises her awareness of further types, she must realize that initially she missed some types. That is, at least at this point, she must be aware that she can be unaware of some types. For instance, she can reason that counterfactually the agent could have disclosed \textit{fewer} types to her, in which case she would be aware of less types than she is now. By induction, she may now suspect that the agent may have not disclosed all types to her in which case she is still unaware of some types. However, she cannot conceive of the specific types that agent may have chosen to omit. This is especially compelling in an context in which we interpret the unidimensional types as a score of multidimensional types (Asker and Cantillon, 2006; Bajari, Houghton, and Tadelis, 2014) and awareness is about cost dimensions. A feature of awareness of unawareness (e.g., Schipper, 2022) is that the principal could be uncertain whether she considered all cost-dimensions, but she cannot know that she missed a particular dimension. In our model, there is no action the principal can take to mitigate her awareness of unawareness such as an action to ask experts (i.e., other agents) or the design of complicated mechanisms that may or may not be able reveal the entire awareness of the agent. That is, while we acknowledge that the agent may be aware of her unawareness, there is nothing in our model where it becomes behaviorally relevant. The best the principal can do is to take into account all types that she is aware of. In many settings, principals such as regulators, CEOs etc. have to justify their actions, which sometimes requires them to testify what exactly they had taken or not taken into account. Requesting money for something unspecific usually does not go down well with legislators. Even if they face awareness of unawareness and do their own investigations on what they may miss, at the end of the day they can only take into account events of which they are or were made aware.

This leads us to avenues for further research. Recently, Pram and Schipper (2025) introduced \textit{dynamic elaboration VCG mechanisms} for efficient implementation at the pooled awareness level. These mechanisms allow for awareness of unawareness and succeed in incentivizing the pooling the awareness among all participants. The question is whether these dynamic direct elaboration mechanisms can be used to also implement optimally from the principal's point of view. While for efficient implementation Pram and Schipper (2025) can focus on belief-free conditional dominant strategies, the extension to optimal implementation must go beyond the belief-free approach and is likely to involve our combination of reverse Bayesianism and monotone inverse hazard rates. Such an approach should not only extend the screening problem to awareness of unawareness but the entire area of optimal auction design to unawareness. If the principal is an auctioneer seeking to maximize expected revenue, how can she design mechanisms that encourage bidders to raise awareness of events relevant to the valuation of the object? Should she allow agent to raise awareness in public or in private? Etc. These questions are beyond the scope of the current paper. Yet, we are confident that the methods of the current paper, in particular, the combination log-concavity and reverse Bayesianism, will be instrumental for making progress on these questions. 

Another promising avenue for further research may be to allow for disclosure of both awareness and verifiable information. Disclosure of information (without awareness) has recently been studied by Pram (2021) and Ali, Lewis, and Vasserman (2022). Yet, another avenue for future research is to test the predictions of Theorems~\ref{m_high_theorem} and~\ref{m_low_theorem} in experiments.

\appendix 

\setcounter{lem}{0}
\renewcommand{\thelem}{\Alph{section}\arabic{lem}}
\setcounter{prop}{0}
\renewcommand{\theprop}{\Alph{section}\arabic{prop}}
\setcounter{cor}{0}
\renewcommand{\thecor}{\Alph{section}\arabic{cor}}


\appendix

\section{Reverse Bayesian Updates and Log-Concavity}\label{reverse_logconc}

In this appendix, we present the main properties of reverse Bayesianism and log-concavity for our analysis. We compare different awareness levels for the principal, represented by different type spaces. For any $\Theta \in \mathcal{T}$, let $\kappa_{\Theta}^{(1)} > \cdots > \kappa_{\Theta}^{(|\Theta|)}$ be the relabeling of types in $\Theta$ from highest to lowest.

In what follows, fix $\Theta\in\mathcal{T}$. The same type $\theta\in\Theta$ may have different places in the ordering of $\Theta$ vs. $\bar{\Theta}$; for every $i \in \{1, \ldots, |\Theta|\}$, let $j(i)$ be the subscript $j(i) \in \{1, \ldots, m\}$ such that: $$\kappa^{(i)}_{\Theta} = \kappa^{(j(i))}_{\bar{\Theta}}.$$
Notice that $j(\cdot)$ is strictly increasing. Moreover, $j(i) = i$ for all $i = 1, ..., |\Theta|$ if $\max (\Theta) = \max (\bar{\Theta})$. 

\begin{lem}\label{RB} Let $\Theta \in \mathcal{T}$, let $p_{\bar{\Theta}}$ be a probability measure with full-support on $\bar{\Theta}$, and let $p_{\Theta}$ be the conditional probability measure of $p_{\bar{\Theta}}$ on $\Theta$. 
\begin{itemize}
\item[(i)] For any $i = 1, ..., |\Theta| - 1$ and $h \leq |\Theta| - i$, 
$$\frac{p^{i + h}_{\Theta}}{p^i_{\Theta}} = \frac{p^{j(i + h)}_{\bar{\Theta}}}{p^{j(i)}_{\bar{\Theta}}};$$
\item[(ii)] If $p_{\bar{\Theta}}$ is log-concave, then $p_{\Theta}$ is also log-concave.
\item[(iii)] For any $i = 1, ..., |\Theta| - 1$ and $k \leq |\Theta| - i$, 
$$\frac{\sum_{h = 1}^{k} p^{i + h}_{\Theta}}{p^i_{\Theta}} = \frac{\sum_{h = 1}^{k} p^{j(i + h)}_{\bar{\Theta}}}{p^{j(i)}_{\bar{\Theta}}}.$$
\end{itemize}
\end{lem}

\noindent \textsc{Proof. } (i) For any $i = 1, ..., |\Theta| - 1$ and $h \leq |\Theta| - i$,
\begin{eqnarray*} \frac{p^{i + h}_{\Theta}}{p^i_{\Theta}} = \frac{\frac{p^{j(i + h)}_{\bar{\Theta}}}{\sum_{\ell = 1}^{n} p^{j(\ell)}_{\bar{\Theta}}}}{\frac{p^{j(i)}_{\bar{\Theta}}}{\sum_{\ell = 1}^{n} p^{j(\ell)}_{\bar{\Theta}}}}
= \frac{p^{j(i + h)}_{\bar{\Theta}}}{p^{j(i)}_{\bar{\Theta}}}. 
\end{eqnarray*} 

(ii)  Since $p_{\bar{\Theta}}$ is log-concave, for all $i = 2, \ldots, |\Theta| - 1$,
\begin{eqnarray*} \frac{p^i_{\Theta}}{p^{i-1}_{\Theta}} = \frac{p^{j(i)}_{\bar{\Theta}}}{p^{j(i-1)}_{\bar{\Theta}}}
& \geq & \frac{p^{j(i+1)}_{\bar{\Theta}}}{p^{j(i)}_{\bar{\Theta}}}
= \frac{p^{i+1}_{\Theta}}{p^{i}_{\Theta}},
\end{eqnarray*} where the first and last equality follows from (i). 

(iii) For any $i = 1, ..., |\Theta| - 1$ and $h \leq |\Theta| - i$, we have by (i) 
$$\frac{p^{i + h}_{\Theta}}{p^i_{\Theta}} = \frac{p^{j(i + h)}_{\bar{\Theta}}}{p^{j(i)}_{\bar{\Theta}}},$$ which implies 
$$p^{i + h}_{\Theta} p^{j(i)}_{\bar{\Theta}} - p^i_{\Theta} p^{j(i + h)}_{\bar{\Theta}} = 0.$$ For any $k \leq |\Theta| - i$, 
\begin{eqnarray*} \sum_{h = 1}^k \left( p^{i + h}_{\Theta} p^{j(i)}_{\bar{\Theta}} - p^i_{\Theta} p^{j(i + h)}_{\bar{\Theta}} \right) & = & 0 \\
p^{j(i)}_{\bar{\Theta}} \sum_{h = 1}^k p^{i + h}_{\Theta}  - p^i_{\Theta} \sum_{h = 1}^k p^{j(i + h)}_{\bar{\Theta}}  & = & 0 \\
\frac{\sum_{h = 1}^{k} p^{i + h}_{\Theta}}{p^i_{\Theta}} & = & \frac{\sum_{h = 1}^{k} p^{j(i + h)}_{\bar{\Theta}}}{p^{j(i)}_{\bar{\Theta}}},
\end{eqnarray*}

as desired. \hfill $\Box$\\ 

These observations generalize to reverse Bayesianism. 

\begin{lem}\label{realRB} For any $\Theta \in \mathcal{T}$, let $p_{\Theta}$ and $p_{\bar{\Theta}}$ be probability measures (not necessarily full support) on $\Theta$ and $\bar{\Theta}$, respectively. Assume that they satisfy reverse Bayesianism.
\begin{itemize}
\item[(i)] For any $i = 1, ..., |\Theta| - 1$ and $h \leq |\Theta| - i$ such that $\kappa_{\Theta}^i, \kappa_{\Theta}^{i + h} \in \text{supp}(p_{\Theta}) \cap \text{supp}(p_{\bar{\Theta}})$, 
$$\frac{p^{i + h}_{\Theta}}{p^i_{\Theta}} = \frac{p^{j(i + h)}_{\bar{\Theta}}}{p^{j(i)}_{\bar{\Theta}}}.$$
\item[(ii)] If $p_{\bar{\Theta}}$ is log-concave, then $p_{\Theta}$ is log-concave for any $i = 1, ..., |\Theta| - 1$ and $h \leq |\Theta| - i$ such that $\kappa_{\Theta}^i, \kappa_{\Theta}^{i + h} \in \text{supp}(p_{\Theta}) \cap \text{supp}(p_{\bar{\Theta}})$.
\item[(iii)] For any $i = 1, ..., |\Theta| - 1$ and $k \leq |\Theta| - i$ such that $\kappa_{\Theta}^i, \kappa_{\Theta}^{i + h} \in \text{supp}(p_{\Theta}) \cap \text{supp}(p_{\bar{\Theta}})$, 
$$\frac{\sum_{h = 1}^{k} p^{i + h}_{\Theta}}{p^i_{\Theta}} = \frac{\sum_{h = 1}^{k} p^{j(i + h)}_{\bar{\Theta}}}{p^{j(i)}_{\bar{\Theta}}}.$$
\end{itemize}
\end{lem}

\noindent \textsc{Proof. } (i) follows directly from the definition of reverse Bayesianism. (ii) follows directly from (i). (iii) follows by the same arguments as in the proof of Lemma~\ref{RB} (iii) in the appendix. \hfill $\Box$ \\

The following lemma is at the core of our disclosure results: 

\begin{lem}\label{Karni_Viero} For any $\Theta \in \mathcal{T}$, let $p_{\Theta}$ and $p_{\bar{\Theta}}$ be probability measures (not necessarily full support) on $\Theta$ and $\bar{\Theta}$, respectively. Assume that they satisfy reverse Bayesianism and log-concavity. For $i = 1, \ldots, |\Theta|$, let $\hat{Q}_{\Theta}^i$ denote the set of solutions for agent $i$ of the principal's optimization problem under $p_{\Theta}$ if $p_{\Theta}^{i} > 0$ (see Francetich and Schipper, 2025). Similarly, let $\hat{Q}_{\bar{\Theta}}^{j(i)}$ denote the set of solutions for agent $j(i)$ of the principal's optimization problems under $p_{\bar{\Theta}}$ if $p_{\bar{\Theta}}^{j(i)} > 0$. Recall that $\hat{Q}_{\Theta}^i$ and $\hat{Q}_{\bar{\Theta}}^{j(i)}$ are at least singletons and at most doubletons with adjacent members.
\begin{itemize}
\item[(i)] If $\min(\text{supp}(p_{\bar{\Theta}})) = \min(\text{supp}(p_{\Theta}))$, then for all $i = 1,\ldots, |\Theta|$ with $\kappa_{\Theta}^{(i)} \in \text{supp}(p_{\Theta}) \cap \text{supp}(p_{\bar{\Theta}})$, $\hat{Q}_{\Theta}^i=\hat{Q}_{\bar{\Theta}}^{j(i)}$.
\item[(ii)]If $\min(\text{supp}(p_{\bar{\Theta}})) < \min(\text{supp}(p_{\Theta}))$, then for all $i = 1,\ldots, |\Theta|$ with $\kappa_{\Theta}^{(i)} \in \text{supp}(p_{\Theta}) \cap \text{supp}(p_{\bar{\Theta}})$, $q^i_{\Theta} \in \hat{Q}_{\Theta}^i$, $q_{\bar{\Theta}}^{j(i)} \in \hat{Q}_{\bar{\Theta}}^{j(i)}$, $q^i_{\Theta} \geq q^{j(i)}_{\bar{\Theta}}$. In this case, $q^i_{\Theta} > q^{j(i)}_{\bar{\Theta}}$ if $\sum_{h = j(|\Theta|) + 1}^{m} p_{\bar{\Theta}}^h > p_{\bar{\Theta}}^{j(i)}$. 
\end{itemize}
\end{lem} 

\noindent \textsc{Proof. } (i) By log-concavity, all types between $\min(\text{supp}(p_{\bar{\Theta}}))$ and $\kappa_{\Theta}^{(i)}$ must be in the support of both $p_{\bar{\Theta}}$ and $p_{\Theta}$. Lemma~\ref{realRB} (iii) implies, in particular, that inverse hazard rates are equal: 
$$\frac{\sum_{h > i}^{|\Theta|} p^h_{\Theta}}{p^i_{\Theta}} = \frac{\sum_{h > j(i)}^{|\bar{\Theta}|} p^{h}_{\bar{\Theta}}}{p^{j(i)}_{\bar{\Theta}}} \mbox{ for } i = 1, ..., n - 1.$$ To see this note the sum $\sum_{h > i}^{|\Theta|} p^h_{\Theta}$ has as many non-zero terms as the sum $\sum_{h > j(i)}^{|\bar{\Theta}|} p^{h}_{\bar{\Theta}}$. 

It follows now that the principal's F.O.C.s for $i = 1,\ldots, |\Theta|$ with $\kappa_{\Theta}^{(i)} \in \text{supp}(p_{\Theta}) \cap \text{supp}(p_{\bar{\Theta}})$ are the same across awareness levels.

(ii) Since $v$ is discrete strictly concave (Assumption~\ref{properties_v}.\ref{sdc}), by Lemmata A5 (2.) and A7 (iii) in Francetich and Schipper (2025), $q_{\Theta}^i \geq q_{\bar{\Theta}}^{j(i)}$ is implied by $\Delta^{+} v(q_{\Theta}^i) < \Delta^{-} v(q_{\bar{\Theta}}^{j(i)})$.
From the principal's F.O.C.s, we have: 
$$\Delta^+ v(q^i_{\Theta}) \leq \lceil \kappa^{(i)}_{\Theta}  \rceil + \frac{\sum_{h = i + 1}^{|\Theta|} p_{\Theta}^h}{p_{\Theta}^i}$$ and 
$$\lceil \kappa^{(j(i))}_{\bar{\Theta}} \rceil + \frac{\sum_{h = j(i) + 1}^{|\bar{\Theta}|} p_{\bar{\Theta}}^h}{p_{\bar{\Theta}}^{j(i)}} \leq \Delta^- v(q_{\bar{\Theta}}^{j(i)}).$$ 

We show that:  
\begin{eqnarray*} \lceil \kappa^{(i)}_{\Theta}  \rceil + \frac{\sum_{h = i + 1}^{|\Theta|} p_{\Theta}^h}{p_{\Theta}^i} & < & \lceil \kappa^{(j(i))}_{\bar{\Theta}} \rceil + \frac{\sum_{h = j(i) + 1}^{|\bar{\Theta}|} p_{\bar{\Theta}}^h}{p_{\bar{\Theta}}^{j(i)}}\\
\frac{\sum_{h = i + 1}^{|\Theta|} p^{h}_{\Theta}}{p^{i}_{\Theta}} & < & \frac{\sum_{h = j(i) + 1}^{|\bar{\Theta}|} p^{h}_{\bar{\Theta}}}{p^{j(i)}_{\bar{\Theta}}} \\
\frac{\sum_{h = i + 1}^{|\Theta|} p^{h}_{\Theta}}{p^{i}_{\Theta}}  & < & \frac{\sum_{h = j(i) + 1}^{j(|\Theta|)} p^{h}_{\bar{\Theta}}}{p^{j(i)}_{\bar{\Theta}}} + \frac{\sum_{h = j(|\Theta|) + 1}^{|\bar{\Theta}|} p^{h}_{\bar{\Theta}}}{p^{j(i)}_{\bar{\Theta}}} \\
\frac{\sum_{h = i + 1}^{|\Theta|} p^{h}_{\Theta}}{p^{i}_{\Theta}} & < & \frac{\sum_{h = i + 1}^{|\Theta|} p^{h}_{\Theta}}{p^{i}_{\Theta}} + \frac{\sum_{h = j(|\Theta|) + 1}^{|\bar{\Theta}|} p^{h}_{\bar{\Theta}}}{p^{j(i)}_{\bar{\Theta}}} \\
0 & < & \frac{\sum_{h = j(|\Theta|) + 1}^{|\bar{\Theta}|} p^{h}_{\bar{\Theta}}}{p^{j(i)}_{\bar{\Theta}}},
\end{eqnarray*} where the second last line follows from Lemma~\ref{realRB} (iii). This proves $\Delta^{+} v(q_{\Theta}^i) < \Delta^{-} v(q_{\bar{\Theta}}^{j(i)})$ and thus $q^i_{\Theta} \geq q^{j(i)}_{\bar{\Theta}}$. 

To prove that $\sum_{h = j(|\Theta|) + 1}^{|\bar{\Theta}|} p_{\bar{\Theta}}^h > p_{\bar{\Theta}}^{j(i)}$ implies $q^i_{\Theta} > q^{j(i)}_{\bar{\Theta}}$ for $i = 1, ..., |\Theta|$ with $\kappa_{\Theta}^{(i)} \in \text{supp}(p_{\Theta}) \cap \text{supp}(p_{\bar{\Theta}})$, suppose to the contrary that $q^i_{\Theta} = q^{j(i)}_{\bar{\Theta}}$. Then $\Delta^{+} v(q_{\Theta}^i) - \Delta^{-} v(q_{\bar{\Theta}}^{j(i)}) = \Delta^+ \Delta^- v(q_{\Theta}^i) \geq -1$, where the first equality is implied by $q^i_{\Theta} = q^{j(i)}_{\bar{\Theta}}$ and the last inequality is required by Assumption~\ref{properties_v}.\ref{not_too_sdc}. From the discrete first-order conditions discussed just above, we get: $$\Delta^{+} v(q_{\Theta}^i) - \Delta^{-} v(q_{\bar{\Theta}}^{j(i)}) \leq - \frac{\sum_{h = j(|\Theta|) + 1}^{|\bar{\Theta}|} p_{\bar{\Theta}}^h}{p_{\bar{\Theta}}^{j(i)}}.$$ Observe that: $$\Delta^{+} v(q_{\Theta}^i) - \Delta^{-} v(q_{\Theta}^{i}) \leq - \frac{\sum_{h = j(|\Theta|) + 1}^{|\bar{\Theta}|} p_{\bar{\Theta}}^h}{p_{\bar{\Theta}}^{j(i)}} < -1,$$ where the l.h.s. follows from $q^i_{\Theta} = q^{j(i)}_{\bar{\Theta}}$ and the right-hand inequality from $\sum_{h = j(|\Theta|) + 1}^{|\bar{\Theta}|} p_{\bar{\Theta}}^h > p_{\bar{\Theta}}^{j(i)}$, a contradiction.\hfill $\Box$\\

Note that in case (ii) we have $q^i_{\Theta} > q^{j(i)}_{\bar{\Theta}}$ if the principal assigns sufficiently large probability (i.e., larger than probability $p_{\bar{\Theta}}^{j(i)}$) to the types that she just became aware of.


\end{document}